\documentclass[a4paper,11pt]{article}
\usepackage{jcappub}
\usepackage{aas_macros}
\usepackage{amsmath}
\usepackage{physics}
\usepackage[capitalize]{cleveref}
\usepackage{eivars}
\usepackage{mrv}
\usepackage{modelnames}
\usepackage{longtable}

\makeatletter
\gdef\@fpheader{\mbox{}}
\makeatother

\newlength{\fullw}
\newlength{\halfw}
\newlength{\threefigw}
\newlength{\twofigw}
\newlength{\onefigw}
\newlength{\bigfigw}
\newlength{\roww}
\newcommand{\nnow}{287}
\newcommand{\nold}{193}

\newcommand{\nignore}{19}

\newcommand{\nprev}{\the\numexpr \nold - \nignore \relax}
\newcommand{\nadd}{\the\numexpr \nnow - \nprev \relax}

\newenvironment{priortable}[1]
               {
                \ifnum0=#1
               \fi
               \ifnum1=#1
                  \begin{longtable}{{|l||c|}}
                    \endhead \hline 
                    Name & $\cte{1}$-prior  \\
                    \hline  \endfirsthead
               \fi
               \ifnum2=#1
                  \begin{longtable}{{|l||c||c|}}
                    \endhead \hline 
                    Name & $\cte{1}$-prior  & $\cte{2}$-prior  \\
                    \hline  \endfirsthead
               \fi
               \ifnum3=#1
                  \begin{longtable}{{|l||c||c|c|}}
                    \endhead \hline 
                    Name & $\cte{1}$-prior  & $\cte{2}$-prior & $\cte{3}$-prior   \\
                    \hline \endfirsthead
               \fi
               }
               {
                 \hline
                 \end{longtable} 
               }

\newcommand{\BAYASPIC}{\texttt{Bayaspic}}
\newcommand{\POLYCHORD}{\texttt{PolyChord}}
\newcommand{\MULTINEST}{\texttt{MultiNest}}
\newcommand{\INFDISTBAYES}{\texttt{infdistbayes}}
\newcommand{\ANESTHETIC}{\texttt{anesthetic}}
\newcommand{\COBAYA}{\texttt{Cobaya}}
\newcommand{\CAMSPEC}{\texttt{CamSpec}}
\newcommand{\PRNPIPE}{\texttt{PR4/NPIPE}}
\newcommand{\PLIK}{\texttt{Plik}}
\newcommand{\CLIK}{\texttt{clik}}
\newcommand{\SIMALL}{\texttt{SimAll}}

\newcommand{\dirac}[1]{\delta\negthinspace\left(#1\right)}

\newcommand{\cte}[1]{c_{#1}}
\newcommand{\Ader}[1]{A_\uder^{#1}}
\newcommand{\efolds}{$e$-folds}
\newcommand{\efold}{$e$-fold}

\newcommand{\epsast}[1]{\epsstar{#1}}
\newcommand{\Hast}{\Hstar}
\newcommand{\kast}{\kstar}
\newcommand{\calPz}{\calP_\zeta}
\newcommand{\calPh}{\calP_h}
\newcommand{\vareps}[1]{\varepsilon_{#1}}

\newcommand{\btheta}{\boldsymbol{\theta}}
\newcommand{\bthetainf}{\btheta_{\uinf}}
\newcommand{\bthetareh}{\btheta_{\ureh}}
\newcommand{\bthetastd}{\btheta_{\us}}
\newcommand{\bdata}{\boldsymbol{D}}
\newcommand{\bvareps}{\boldsymbol{\varepsilon}}

\newcommand{\OmegaCDM}{\Omega_\udm}
\newcommand{\OmegaB}{\Omega_\ub}

\newcommand{\thetaMC}{\theta_{\usssMC}}

\newcommand{\BayesFactor}[2]{B^{#1}_{#2}}
\newcommand{\Bref}[1]{\BayesFactor{#1}{\usssSRTHIRD}}
\newcommand{\calMref}{\calM_{\usssSRTHIRD}}
\newcommand{\calMmax}{\calM_{\ubest}}
\newcommand{\Bbest}[1]{\BayesFactor{#1}{\ubest}}

\newcommand{\evid}[2]{\calE\negthinspace\left(#1|#2\right)}
\newcommand{\prior}[1]{\pi\negthinspace\left(#1\right)}
\newcommand{\prob}[1]{P\negthinspace\left(#1\right)}

\newcommand{\post}[2]{P\negthinspace\left(#1|#2\right)}
\newcommand{\calLeff}{\calL_\ueff}
\newcommand{\calLeffmin}{\calL_\ueff^{\min}}
\newcommand{\calLeffmax}{\calL_\ueff^{\max}}
\newcommand{\likeff}[2]{\calLeff\negthinspace\left(#1|#2\right)}
\newcommand{\likeffb}[2]{\calLeff\negthinspace\left[#1|#2\right]}

\newcommand{\Ztol}{Z_{\utol}}

\newcommand{\DKL}[1]{D_{\usssKL}^{#1}}
\newcommand{\DKLreh}{\DKL{\ureh}}
\newcommand{\DKLtot}{\DKL{}}
\newcommand{\calI}{\mathcal{I}}

\begin{document}

\title{Cosmic Inflation at the Crossroads}

\author[a]{J\'er\^ome Martin,}
\author[b,a]{Christophe Ringeval}
\author[c,a]{and Vincent Vennin}

\affiliation[a]{Institut d'Astrophysique de Paris, 98bis boulevard
  Arago, 75014 Paris, France}

\affiliation[b]{Cosmology, Universe and Relativity at Louvain (CURL),
  Institute of Mathematics and Physics, University of Louvain, 2
  Chemin du Cyclotron, 1348 Louvain-la-Neuve, Belgium}

\affiliation[c]{Laboratoire de Physique de l'\'Ecole Normale
  Sup\'erieure, ENS, CNRS, Universit\'e PSL, Sorbonne Universit\'e,
  Universit\'e Paris Cit\'e, 75005 Paris, France}

\emailAdd{jmartin@iap.fr}
\emailAdd{christophe.ringeval@uclouvain.be}
\emailAdd{vincent.vennin@ens.fr}

\date{today}

\abstract{The capability of Cosmic Inflation to explain the latest
  Cosmic Microwave Background and Baryonic Acoustic Oscillation data
  is assessed by performing Bayesian model comparison within the
  landscape of nearly three-hundred models of single-field slow-roll
  inflation. We present the first Bayesian data analysis based on the
  third-order slow-roll primordial power spectra. In particular, the
  fourth Hubble-flow function $\vareps{4}$ remains unbounded while the
  third function verifies, at two-sigma, $\vareps{3}\in[-0.4,0.5]$,
  which is perfectly compatible with the slow-roll predictions for the
  running of the spectral index. We also observe some residual excess
  of $B$-modes within the BICEP/Keck data favoring, at a
  non-statistically significant level, non-vanishing primordial tensor
  modes: $\log(\vareps{1}) > -3.9$, at $68\%$ confidence level. Then,
  for $\nnow$ models of single-field inflation, we compute the Bayesian
  evidence, the Bayesian dimensionality and the marginalized
  posteriors of all the models' parameters, including the ones
  associated with the reheating era. The average information gain on
  the reheating parameter $\Rreh$ reaches $1.3 \pm 0.18$ bits, which
  is more than a factor two improvement compared to the first Planck
  data release. As such, inflationary model predictions cannot meet
  data accuracy without specifying, or marginalizing over, the
  reheating kinematics. We also find that more than $40\%$ of the
  scenarios are now strongly disfavored, which shows that the
  constraining power of cosmological data is winning against the
  increase of the number of proposed models. In addition, about $20\%$
  of all models have evidences within the most probable region and are
  all favored according to the Jeffreys' scale of Bayesian evidences.}



\maketitle

\newpage

\section{Introduction}
\safeguard
\label{sec:intro}

In the last decades, the quest for understanding the physical
conditions that prevailed in the very early universe, prior to the
radiation dominated epoch of the standard hot Big Bang model, has been
revolutionized twice.

The first breakthrough was the advent of a scenario, cosmic inflation,
that, for the first time, was able to offer a consistent and
convincing description of the universe in its early stages of
evolution. Based on the principles of General Relativity and Quantum
Mechanics, cosmic inflation can explain the puzzles of the hot Big
Bang model and, in addition, provides a mechanism for the origin of
the structures observed in the universe~\cite{Starobinsky:1979ty,
  Starobinsky:1980te, Guth:1980zm, Linde:1981mu, Albrecht:1982wi,
  Linde:1983gd, Mukhanov:1981xt, Mukhanov:1982nu, Starobinsky:1982ee,
  Guth:1982ec, Hawking:1982cz, Bardeen:1983qw}. Moreover, it has led
to a prediction, namely that the spectral index of the scalar power
spectrum should be close to one but, crucially, not exactly one, that
has now been confirmed at a high statistical confidence
level~\cite{2013ApJS..208...20B,Planck:2013oqw,Planck:2018jri}.

Obviously, the theory has not escaped critical evaluation, eliciting a
range of criticisms.  A detailed discussion of these criticisms, with
a possible answer for each of them, can (for instance) be found in
\Refa{Chowdhury:2019otk}. One of the most popular fault-finding made
against inflation, which, if true, would be especially relevant in the
context of the present paper, is that it would not be a falsifiable
theory. It is based on the idea that inflation, if viewed as a
paradigm, can always be modified to account for new data. Indeed,
there exist different incarnations of inflation and, if a particular
version turns out to be incompatible with some data set, then there is
always the possibility to invoke another, possibly more complicated,
realization. For instance, if some new data show that the scalar
primordial power spectrum contains superimposed oscillations, then
single-field slow-roll inflation starting in the Bunch-Davis vacuum
would be ruled out. But a non slow-roll model, with a potential having
a feature at the required scale~\cite{Avila:2013ela}, or some
trans-Planckian
initial~\cite{Martin:2000xs,Martin:2003sg,Martin:2004iv,Martin:2004yi}
conditions, could explain the new data. If, in addition to the
presence of oscillations, one also discovers that, say,
non-Gaussianities are non-negligible, then one could further modify
the scenario, for example by introducing additional fields, or
non-minimal kinetic terms. A priori, this process has no end, hence
the naive claim that inflation would be non falsifiable. This
argument, however, is not specific to inflation. It could be made, for
instance, against quantum field theory (QFT). Indeed, QFT also comes
in different realizations and one could imagine changing the gauge
group or adding new fields each time new particles are
discovered. Clearly, it is meaningless to try ruling out the paradigm
of QFT \emph{per se} but, nevertheless, each of its incarnations is of
course falsifiable, as beautifully exemplified by the establishment of
the standard model of particle physics. And so it is for inflation.

Another example of a popular criticism against inflation that has
turned out to be inadequate concerns the initial conditions. For a
long time, it was believed that inflation can start only if,
initially, the universe is already quite homogeneous, thus questioning
the relevance of a scenario which is precisely supposed to dynamically
achieve this property. However, a series of recent works have
demonstrated that, on the contrary, even initially strongly
inhomogeneous situations can lead to a phase of
inflation~\cite{Kurki-Suonio:1993lzy, East:2015ggf, Clough:2017efm,
  Aurrekoetxea:2019fhr, Joana:2020rxm, Joana:2022uwc}. The jury might
be still out in the sense that the most general case has yet to be
treated; there are indeed always some assumptions about the type of
inhomogeneities initially present~\cite{Garfinkle:2023vzf}. This is
because, technically, the subject is not easy, requiring to carry out
full General Relativity simulations. It is, however, fair to say that
the trend seems to be in favor of inflation since configurations that
were previously believed to prevent inflation have in fact been shown
to be perfectly compatible with this theory.

The fact that, despite many attempts, the numerous criticisms against
inflation have not really threatened its conceptual attraction and
explanatory power, together, of course, with its adequation with the
data, are probably the reasons why inflation remains considered by a
vast majority of practitioner cosmologists as the leading theory for
understanding the very early universe.

The second game changer, which occurred at the same time our
understanding of inflation was gradually improved and deepened
(namely, roughly speaking, in the last 30 or 40 years), is the fact
that high-accuracy cosmological data became progressively
available. As briefly mentioned before, this has allowed us to confirm
that vanilla inflation is a good description of the early universe and
that its emblematic prediction $0<\left \vert \nS-1\right\vert \ll 1$
is verified. Recall that this prediction was made before the data were
accurate enough to confirm it; it is, therefore, not a postdiction but
a genuine prediction. The high-accuracy data now concern different
observables but the role played by the Cosmic Microwave Background
(CMB) anisotropy measurements has been crucial. It started with the
first detection by the NASA COBE satellite~\cite{COBE:1992syq},
followed by the first measurement of multiple Doppler peaks by the
WMAP satellite~\cite{WMAP:2003elm} and, more recently, culminated with
the publication of the final results obtained by the ESA
cosmic-variance limited mission Planck~\cite{Planck:2013oqw}. Other
results obtained either by
balloon-born~\cite{deOliveira-Costa:1998zkk, Boomerang:2000efg,
  Hanany:2000qf, Henrot-Versille:2003int, Reichborn_Kjennerud_2010,
  SPIDER:2017xxz} and/or ground based~\cite{Ruhl:1995jy,
  Netterfield:1996nb, Reichardt:2008ay, Rubino-Martin:2008duq,
  Reichardt_2009, Fowler_2010, Chiang_2010, POLARBEAR:2013oat}
telescopes should not be forgotten and the fact that the CMB is
polarized~\cite{Kovac:2002fg} is, and will be, of major importance.

The concomitant nature of the two major aforementioned developments
suggests that one has reached a particular stage in the history of the
subject. In other words, cosmic inflation is at the crossroads and,
before new observations become available, it is the right time to assess
where do we stand and to study what we have finally learned about
inflation. This will also allow us to understand what is to be
expected when new data become available, both from CMB (CMB-S4,
LiteBIRD, \etc)~\cite{CMB-S4:2022ght,Sugai_2020} and large-scale
structure (EUCLID, SKA, \etc)
surveys~\cite{laureijs2011euclid, Weltman:2018zrl}. In this paper, we
are interested in the simplest incarnation of inflation (namely,
single-field slow-roll inflation with minimal kinetic terms,
see~\Sec{sec:post-planck-landscape}) and aim at assessing its
compatibility with recent astrophysical and cosmological data.

In order to carry out this task, we have developed new and efficient
analysis methods that are presented in \cref{sec:analysis} whereas new
results concerning the slow-roll power spectra are presented in
\cref{sec:srposteriors}. In \cref{sec:evidences}, pipelining these
tools with basic machine-learning methods, we perform Bayesian model
comparison in the space of $\nnow$ models of inflation. We complement
Bayesian evidences with the computation of various other Bayesian
quantities, such as the Bayesian dimensionality and the information
gain on the reheating, to determine which inflationary scenarios are
the most probable given the data. We also present the posterior
probability distributions for the parameters associated with the best
models in the appendix (all data will be made available
online~\cite{urlci}). Our conclusion are presented in
\cref{sec:bests}.

\section{Post-Planck landscape of single-field inflation}
\label{sec:post-planck-landscape}

As described in the introduction, cosmic inflation comes in various
incarnations. However, a major piece of information inferred from the
Planck data is that one of these incarnations, as a matter of fact the
simplest one, based on a slowly rolling scalar field, with minimal
kinetic term, turns out to be favored. This conclusion is supported,
for instance, by the absence of Non-Gaussianities, of non-adiabatic
modes or of non-trivial features in the power spectra of primordial
fluctuations. From an effective field theory (EFT) perspective, one
may wonder why the phenomenology of inflation appears single field
while most of its embedding feature various high-energy degrees of
freedom (this may be explained by volume selection effects, as
recently shown in \Refa{Tokeshi:2023swe}). But according to the
principle of parsimony, our task is to focus on the simplest
``vanilla'' models, since more complicated versions of inflation are,
for the moment, not required to understand the data

Another important piece of information recently obtained is that, for
the vanilla scenarios, the reheating stage that follows inflation
itself is, to some extent, already constrained. These constraints are,
however, not very tight and, moreover, depend on the shape of inflaton
potential considered. Overall, Planck 2015 and BICEP2/KECK yield to an
information gain of $0.82 \pm 0.13$ bits on the reheating history of
slow-roll models~\cite{Martin:2016oyk}. The Planck 2013 data were
giving a gain of, typically, half a bit.

Finally, there also exist constraints on the shape of the inflaton
potential itself. Models can be organized according to the
Schwarz–Terrero-Escalante (STE)
classification~\cite{Schwarz:2004tz}. There are three main categories
depending on the behavior of the Hubble-flow functions (defined in
\cref{sec:analysis}): the first category corresponds to models with
$\eps2 > 2\eps1 >0$, namely models for which the kinetic energy
increases during inflation as well as the ratio of the kinetic energy
to the total energy; the second category represents potentials where
$0 < \eps2 < 2\eps1$, that is to say where the kinetic energy
decreases while the ratio of the kinetic energy to the total energy
still increases. Finally, the third category is such that both
quantities decrease during inflation. As shown in
\Refa{Martin:2013nzq}, Planck data have clearly favored models in the
first category. A Bayesian analysis (model selection) further reveals
that plateau models, exemplified by the Starobinsky model (but it is
important to keep in mind that the performances of many other models
are comparable), are the most effective explanation of the data. It is
truly remarkable that cosmological data can lead to constraints on
physical phenomena that could possible occur at energies as high as
$10^{16}\, \GeV$.

This brief status of the art rests on models and methods presented in
the first edition of {\EI}~\cite{EIoriginal}, published in $2013$. In
this article, we aim at presenting an updated status report given the
peculiar situation of the cosmic inflation theory which, as already
pointed out in the introduction, is today at something of a
crossroads, midway between the old and new generations of cosmological
measurements.

In order to carry out this task, new methods must be designed in order
to take maximum advantage of the recent theoretical and observational
developments. On the theoretical front, since the first edition of
{\EI}~\cite{EIoriginal}, several new models of inflation have been put
forward. As a consequence, in order to have an up-to-date description
of the landscape of models, a second edition of {\EI}, including these
new scenarios, has recently been published~\cite{EIopiparous}. The new
potential functions in the new edition (here ordered alphabetically)
are: Axion Hilltop Inflation (AHI), Cubicly Corrected Starobinsky
Inflation (CCSI), Double Exponential Inflation (DEI), Dual Inflation
(DI), Fibre Inflation (FI), Generalized Double Well Inflation (GDWI),
Hyperbolic Inflation (HBI), Hybrid Natural Inflation (HNI),
Non-Renormalizable Corrected Loop Inflation (NCLI), N-Formalism
Inflation (NFI), Non-Minimal Large Field Inflation (NMLFI), Pure
Arctan Inflation (PAI), Radiatively Corrected Inflection Point
Inflation (RCIPI), Radiatively Corrected Large Field Inflation
(RCLFI), String Axion Inflation I (SAII), String Axion Inflation II
(SAIII), Super-conformal Alpha Attractor A Inflation (SAAI),
Super-conformal Alpha Attractor B Inflation (SABI), Super-conformal
Alpha Attractor T Inflation (SATI), Symmetry Breaking K\"ahler
Inflation (SBKI), S-Dual Inflation (SDI), Smeared Higgs Inflation
(SHI), T-Model Inflation (TMI), Mukhanov Inflation (VFMI). With these
new scenarios, the total number of models included has increased by
$\nadd$ to reach $\nnow$. It is important to remark that a model of
inflation is not only characterized by a potential function but also
by the prior distributions of the parameters appearing in each
potential. In other words, two different inflationary models can share
a same potential with different priors for its parameters, for
instance if these models come from different high-energy
embedding. As such, there can be more models than potential
functions.

\section{Analysis methods}
\label{sec:analysis}

In this section, we describe the theoretical grounds and analysis
workflow used to perform Bayesian model comparison and parameter
estimation in the space of the single-field slow-roll models of
inflation. The basic idea is to compress all the available information
associated with the cosmological data into an small set of parameters
for which inflationary predictions can be quickly, and accurately,
made.  For slow-roll inflation, these parameters are the Hubble-flow
functions, encoding the drift of the Hubble parameter $H$,
\begin{equation}
\eps{i+1}(N) \equiv \dfrac{\ud \ln \left|\eps{i}\right|}{\ud N}\,, \quad
\eps{0}(N) = \dfrac{\Mp}{H}\,,
\label{eq:epsHdef}
\end{equation}
evaluated at a peculiar {\efold} number $\Nstar$ during inflation. This
number is defined as the {\efold} at which an observable pivot
wavenumber $\kstar$ crosses the Hubble radius during inflation. More
precisely, it is the {\efold} such that $\kstar \eta(\Nstar)=-1$ where
$\eta$ is conformal time (this coincides with Hubble crossing, $k=aH$,
at leading order in slow roll only, but it is taken as a generic
definition). The actual value of $\kstar$ is an observer choice, and,
in the following, we choose $\kstar=0.05\,\Mpc^{-1}$, which lies in
the middle of the observed range of modes. Once $\kstar$ is set,
$\Nstar$ is no longer a free parameter: it is completely fixed by both
the inflationary model parameters and the kinematics of the reheating
era. In the following, we give some details explaining why the
Hubble-flow functions are best-suited to parameterize the inflationary
dynamics, how to properly deal with the reheating era and how one can
extract the posterior probability distributions for the inflationary
theoretical parameters associated with a given model.

\subsection{Data compression using the slow-roll power spectra}

During inflation, it is possible to analytically solve the evolution
of linear perturbations in General Relativity by performing a
functional expansion in terms of the Hubble flow functions
$\eps{i}(N)$~\cite{Stewart:1993bc, Liddle:1994dx, Nakamura:1996da,
  Gong:2001he, Hoffman:2000ue, Schwarz:2001vv, Leach:2002ar,
  Schwarz:2004tz, Martin:2013uma, Jimenez:2013xwa}. The sole
assumption in deriving such an analytical solution is that all the
$\eps{i}$ are small and of the same order (at least above the
truncation order), which is the mere definition of slow-roll
inflation. The analytical expressions for the primordial power spectra
have been the subject of various works in the last two
decades~\cite{Martin:2002vn, Habib:2002yi, Habib:2004kc, Choe:2004zg,
  Casadio:2004ru, Casadio:2005xv, Casadio:2005em, Martin:2013uma,
  Jimenez:2013xwa}, and are currently known at third order in the
Hubble flow functions~\cite{Auclair:2022yxs}. Using the shorthand
notation $\epsstar{i} = \eps{i}(\Nstar)$ as well as $\Hstar =
H(\Nstar)$, the primordial scalar power spectrum reads {
  \allowdisplaybreaks
\begin{align}
    \calPz(k) &= \dfrac{\Hast^2}{8 \pi^2 \Mp^2 \epsast{1}}\bigg\{
    1 - 2 (C + 1) \epsast{1} - C \epsast{2}
    + \qty(\frac{\pi^2}{2} + 2 C^2 + 2 C - 3) \epsast{1}^2 \nonumber \\ &
    + \qty(\frac{7 \pi^2}{12} + C^2 - C - 6) \epsast{1} \epsast{2}
    + \frac{1}{8} \qty(\pi^2 + 4 C^2 - 8) \epsast{2}^2
    + \frac{1}{24} (\pi^2 - 12 C^2) \epsast{2} \epsast{3} \nonumber \\ &
    - \frac{1}{24} \qty[4 C^3 + 3 \qty(\pi^2 - 8) C + 14 \zeta(3) - 16]
    \qty(8 \epsast{1}^3 + \epsast{2}^3) \nonumber\\
    & + \frac{1}{12} \qty[13 \pi^2 - 8 (\pi^2 - 9) C + 36 C^2 - 84 \zeta(3)] \epsast{1}^2 \epsast{2}
    \nonumber \\ & - \frac{1}{24} \qty[8 C^3 - 15 \pi^2 + 6 (\pi^2 -
      4) C - 12 C^2 + 100 \zeta(3) + 16] \epsast{1} \epsast{2}^2
    \nonumber \\
    & + \frac{1}{24} \qty[\pi^2 C - 4 C^3 - 8 \zeta(3) + 16] \qty(\epsast{2}
    \epsast{3}^2 + \epsast{2} \epsast{3} \epsast{4})
     + \frac{1}{24} \qty[12 C^3 + \qty(5 \pi^2 - 48) C] \epsast{2}^2
     \epsast{3} \nonumber \\
     & + \frac{1}{12} \qty[8 C^3 + \pi^2 + 6 \qty(\pi^2 - 12) C - 12
       C^2 - 8 \zeta(3) - 8] \epsast{1} \epsast{2} \epsast{3}
     \nonumber \\
    & + \ln\qty(\frac{k}{\kast})  \bigg[
        - 2 \epsast{1}
        - \epsast{2}
        + 2 (2 C + 1) \epsast{1}^2
        + (2 C - 1) \epsast{1} \epsast{2}
        + C \epsast{2}^2
        - C \epsast{2} \epsast{3}
        \nonumber \\ & - \frac{1}{8} \qty(\pi^2 + 4 C^2 - 8)
        (8 \epsast{1}^3 + \epsast{2}^3) - \frac{2}{3} \qty(\pi^2 - 9 C - 9)
        \epsast{1}^2 \epsast{2}
        - \frac{1}{4} \qty(\pi^2 + 4 C^2 - 4 C - 4) \epsast{1} \epsast{2}^2
        \nonumber \\ & + \frac{1}{2} \qty(\pi^2 + 4 C^2 - 4 C - 12) \epsast{1}
        \epsast{2} \epsast{3}  \nonumber \\ & + \frac{1}{24} \qty(\pi^2 - 12 C^2)
        \left(\epsast{2} \epsast{3}^2 + \epsast{2} \epsast{3} \epsast{4}\right)
        + \frac{1}{24} \qty(5 \pi^2 + 36 C^2 - 48) \epsast{2}^2 \epsast{3}
    \bigg] \nonumber \\
    & + \frac{1}{2}  \ln^2\qty(\frac{k}{\kast}) \bigg[
        4\epsast{1}^2
        + 2 \epsast{1} \epsast{2}
        + \epsast{2}^2
        - \epsast{2} \epsast{3}
        + 6 \epsast{1}^2 \epsast{2}
        - (2 C - 1) \qty(\epsast{1} \epsast{2}^2 - 2 \epsast{1}
        \epsast{2} \epsast{3}) \nonumber \\
        & - C \left(8\epsast{1}^3
        + \epsast{2}^3
        - 3\epsast{2}^2 \epsast{3}
        + \epsast{2} \epsast{3}^2
        + \epsast{2} \epsast{3} \epsast{4}\right)
    \bigg] \nonumber \\
    & + \frac{1}{6} \ln^3\qty(\frac{k}{\kast}) \qty(
        - 8 \epsast{1}^3
        - 2 \epsast{1} \epsast{2}^2
        + 4 \epsast{1} \epsast{2} \epsast{3}
        - \epsast{2}^3
        + 3 \epsast{2}^2 \epsast{3}
        - \epsast{2} \epsast{3}^2
        - \epsast{2} \epsast{3} \epsast{4}
    ) 
    \bigg\},
    \label{eq:calPzsr3ast}
\end{align}
}
while the primordial power spectrum for gravitational waves is given
by
\begin{equation}
\begin{aligned}
    \calPh(k) & = \dfrac{2 \Hast^2}{\pi^2 \Mp^2} \bigg\{
        1 - 2 \qty(C + 1) \epsast{1}
        + \frac{1}{2} \qty(\pi^2 + 4 C^2 + 4 C - 6) \epsast{1}^2
        \\ & + \frac{1}{12} (\pi^2 - 12 C^2 - 24 C - 24) \epsast{1} \epsast{2}
        -\frac{1}{3} \qty[4 C^3 + 3 (\pi^2 - 8) C + 14 \zeta(3) - 16] \epsast{1}^3
        \\ & + \frac{1}{12} \qty[24 C^3 + 13 \pi^2 + 2 (5 \pi^2 - 36) C + 36 C^2 - 96]
        \epsast{1}^2 \epsast{2} \\
        & - \frac{1}{12} \qty[4 C^3 - \pi^2 - (\pi^2 - 24) C + 12 C^2 + 8 \zeta(3) + 8]
        \qty(\epsast{1} \epsast{2}^2 + \epsast{1} \epsast{2} \epsast{3}) \\
        & + \ln \qty(\frac{k}{\kast})\bigg[
            - 2 \epsast{1}
            + 2 (2 C + 1) \epsast{1}^2
            - 2 (C + 1) \epsast{1} \epsast{2}
            - \qty(\pi^2 + 4 C^2 - 8) \epsast{1}^3
            \\& + \frac{1}{6} \qty(5 \pi^2 + 36 C^2 + 36 C - 36)
            \epsast{1}^2 \epsast{2} + \frac{1}{12} \qty(\pi^2 - 12 C^2 - 24 C - 24)
            \qty(\epsast{1} \epsast{2}^2 + \epsast{1} \epsast{2} \epsast{3})
        \bigg]  \\
        & + \ln^2 \qty(\frac{k}{\kast}) \bigg[
            2 \epsast{1}^2
            - \epsast{1}\epsast{2}
            - 4 C \epsast{1}^3
            + 3 (2 C + 1) \epsast{1}^2\epsast{2}
            - \qty(C + 1) \qty(\epsast{1} \epsast{2}^2
            + \epsast{1} \epsast{2} \epsast{3})
        \bigg]  \\
        & + \frac{1}{3}\ln^3 \qty(\frac{k}{\kast}) \qty(
            - 4 \epsast{1}^3
            + 6 \epsast{1}^2 \epsast{2}
            - \epsast{1} \epsast{2}^2
            - \epsast{1} \epsast{2} \epsast{3})        
    \bigg\}.
  \end{aligned}
\label{eq:calPhsr3ast}
\end{equation}
Some irrational numbers appear in
\cref{eq:calPzsr3ast,eq:calPhsr3ast}, there is $\pi$, a constant $C$
related to the Euler constant $\gammaE$ by
\begin{equation}
C \equiv \gammaE + \ln 2 - 2 \simeq -0.729637,
\end{equation}
and the Riemann Zeta function $\zeta(3) \simeq 1.20206$. As explicit
in these expressions, once $\kstar$ is chosen, the $k$-dependence of
the tensor and scalar primordial power spectra is completely fixed by
the parameters
$(\epsstar{0},\epsstar{1},\epsstar{2},\epsstar{3},\epsstar{4})$. As
discussed in \Refs{Leach:2002ar, Leach:2002dw}, performing data
analysis using \cref{eq:calPzsr3ast,eq:calPhsr3ast} is not the same as
postulating a power-law shape for the primordial power
spectra. Indeed, the latter option can be viewed as adding an infinite
number of extra-terms that are not consistent with the actual
slow-roll dynamics. Still, \cref{eq:calPzsr3ast,eq:calPhsr3ast} are
approximate, but that approximation is under control and missing terms
are of order $\order{\epsstar{}^4}$. As we will see later on, this is
more than enough to match the present data accuracy. Defining
\begin{equation}
\label{eq:Pstar:def}
  \Pstar \equiv \calPz(\kstar),
\end{equation}
one can also trade the parameter $\epsstar{0}$ for $\Pstar$, which is
better suited for data analysis. Indeed, the amplitude of the CMB
anisotropies is a well-measured quantity, which is in direct
correspondence with $\Pstar$. Notice from \cref{eq:calPzsr3ast} that
recovering $\epsstar{0}$ from $\Pstar$ requires the knowledge of all the
other $\epsstar{i}$.

The observable predictions for the whole class of single-field
slow-roll models can thus be encoded within the small parameter set
$(\Pstar,\epsstar{i})$ by using \cref{eq:calPzsr3ast,eq:calPhsr3ast}
as functional forms for the primordial power spectra. As a
consequence, for any data set, one could add the slow-roll parameters
into the inference problem and extract their posterior probability
distribution. This is done in \Sec{sec:srposteriors} for the latest
cosmological data.

Given a specific inflationary model $\calM$, within the slow-roll
class, what remains to be done is to compute the actual values of the
$(\Pstar,\epsstar{i})$ parameters. We now turn to this question.

\subsection{Model predictions}

For the minimally coupled single-field inflationary models, with a
canonical kinetic term, the potential $V(\phi)$, complemented with the
Friedmann-Lema\^{\i}tre equations, completely fixes the dynamics. In
particular, one can define an infinite hierarchy of potential-related
functions, denoted $\epsV{i}(\phi)$, defined as~\cite{Liddle:2003py,
  Vennin:2014xta}
\begin{equation}
\epsV{n+1}(\phi) \equiv \sqrt{2 \epsV{1}} \Mp \, \dfrac{\ud \ln
  \qty|\epsV{n}|}{\ud \phi}\,, \quad \epsV{0}(\phi) = \sqrt{\dfrac{3\Mp^4}{V}}\,.
\label{eq:epsVdef}
\end{equation}
The functions $\epsV{i}(\phi)$ can be straightforwardly derived from
the functional shape of the potential and are, sometimes, also
referred to as the ``slow-roll parameters''. The reason underlying
such a confusing naming being that the Hubble-flow and $\epsV{i}$
hierarchies match at leading order in an expansion in Hubble flow
functions. More simply stated, one has $\epsV{i} = \eps{i} +
\order{\eps{}^2}$ for all $i$~\cite{Liddle:1994dx}. However, if one
wants to consistently make predictions at a given order in the
Hubble-flow functions, say order three as to match the accuracy at
which \cref{eq:calPzsr3ast,eq:calPhsr3ast} have been derived, one has
to use the following mapping
{\allowdisplaybreaks
\begin{align}
    \eps{1} & = \epsV{1} - \dfrac{1}{3} \epsV{1}\epsV{2} -\dfrac{1}{9}
    \epsV{1}^2 \epsV{2} + \dfrac{5}{36} \epsV{1} \epsV{2}^2 +
    \dfrac{1}{9} \epsV{1} \epsV{2} \epsV{3}, \nonumber \\
    \eps{2} & = \epsV{2} - \dfrac{1}{6} \epsV{2}^2 - \dfrac{1}{3}
    \epsV{2} \epsV{3} - \dfrac{1}{6} \epsV{1} \epsV{2}^2 + \dfrac{1}{18}
    \epsV{2}^3 - \dfrac{1}{9} \epsV{1} \epsV{2} \epsV{3} + \dfrac{5}{18}
    \epsV{2}^2 \epsV{3} + \dfrac{1}{9} \epsV{2}\epsV{3}^2 + \dfrac{1}{9}
    \epsV{2} \epsV{2}\epsV{4},\nonumber \\
    \eps{3} & = \epsV{3} - \dfrac{1}{3} \epsV{2} \epsV{3} -
    \dfrac{1}{3} \epsV{3}\epsV{4} - \dfrac{1}{6} \epsV{1} \epsV{2}^2
    -\dfrac{1}{3} \epsV{1}\epsV{2}\epsV{3} + \dfrac{1}{6}
    \epsV{2}^2\epsV{3} + \dfrac{5}{18} \epsV{2} \epsV{3}^2 -
    \dfrac{1}{9} \epsV{1}\epsV{3}\epsV{4} \nonumber \\ & +\dfrac{5}{18} \epsV{2}
    \epsV{3}\epsV{4} + \dfrac{1}{9} \epsV{3}^2 \epsV{4} + \dfrac{1}{9}
    \epsV{3} \epsV{4}^2 + \dfrac{1}{9} \epsV{3} \epsV{4}
    \epsV{5},\nonumber \\
    \eps{4} & = \epsV{4} - \dfrac{1}{3} \epsV{2} \epsV{3} -
    \dfrac{1}{6} \epsV{2}\epsV{4} - \dfrac{1}{3} \epsV{4} \epsV{5} + \cdots.
\label{eq:epsVtoH}
\end{align}
}
These expressions have been derived from the Friedmann-Lema\^{\i}tre
equations, using \cref{eq:epsHdef,eq:epsVdef} and assuming a field
trajectory $\phi(N)$ on shell, i.e., solving the Klein-Gordon
equation, which reads~\cite{Chowdhury:2019otk}
\begin{equation}
\dfrac{2}{6 -\Gamma^2} \dfrac{\ud \Gamma}{\ud N} + \Gamma = -
\Mp \dfrac{\ud \ln (V/\Mp^4)}{\ud \phi}\,,
\label{eq:KGefolds}
\end{equation}
where we have defined $\Gamma=\Mp^{-1}\ud \phi/\ud N$. From
\cref{eq:epsVtoH}, one sees that one would need up to $\epsV{5}$ to
determine $\eps{4}$ at second order in the Hubble-flow
functions. However, because $\epsstar{4}$ is always multiplied by
$\epsstar{2} \epsstar{3}$ in \cref{eq:calPzsr3ast}, maintaining
third-order accuracy for $\calP(k)$ actually requires $\epsstar{4}$ to
be known at leading order only. Hence, we have omitted here the
highest-order terms.

In practice, all the $\epsV{i}(\phi)$ functions can be
straightforwardly obtained by taking the successive derivatives of
$V(\phi)$ according to \cref{eq:epsVdef}. The field trajectory
$\phi(N)$, or, equivalently, the relation giving the number of
{\efolds} $N$ given a field value can also be obtained within the
slow-roll framework. Indeed, \cref{eq:KGefolds} is similar to the
equation of motion of a relativistic particle with a constant friction
term and accelerated by an external force deriving from the potential
$\ln (V/\Mp^4)$. After a transient regime, the field reaches its ``terminal
velocity'' and settles into the solution
\begin{equation}
\Gamma \simeq - \Mp \dfrac{\ud \ln (V/\Mp^4)}{\ud \phi}\,,
\label{eq:GammaSR}
\end{equation}
which is the ``slow-roll'' regime, an attractor. On the attactor, and up
to a constant term, one gets from \cref{eq:GammaSR},
\begin{equation}
N = -\frac{1}{\Mp^2}\int^{\phi} \dfrac{V(\psi)}{V'(\psi)} \ud \psi.
\label{eq:nufunc}
\end{equation}
In some regimes, as for instance in ultra slow-roll, or, at the end of
inflation, \cref{eq:GammaSR} may become inaccurate thereby requiring a
numerical integration of \cref{eq:KGefolds}.

Given a specific inflationary model, \crefrange{eq:epsVdef}{eq:nufunc}
allow us to determine, at the wanted precision, all the Hubble flow
functions $\eps{i}(N)$ given a potential $V(\phi$). This potential is
uniquely specified by some intrinsic model parameters that will be
referred to as $\bthetainf$. In practice, the determination of the
$\eps{i}(N)$ functions given $\bthetainf$ is automated by a modern
Fortran library {\ASPIC}~\cite{2018ascl.soft06031M}, supporting all
the inflationary models discussed in the ``Opiparous Edition'' of the
{\EI} paper~\cite{EIopiparous}. Let us stress that various motivated
models are non-minimally coupled to gravity. For these, various
extensions of the method discussed in this section have been
implemented, they are discussed at length in \Refa{EIopiparous} and
they have been numerically implemented in the {\ASPIC} code. In
summary, the slow-roll approximation, and its extensions, allow us to
determine all the necessary $\eps{i}(N)$ functions given the
theoretical parameters $\bthetainf$.

\subsection{One reheating parameter to rule them all}
\label{sec:rehparam}

During inflation and the subsequent cosmological era, the wavelengths
of cosmological perturbations are stretched by the expansion of the
universe. As such, the correspondence between a mode $k/\azero$
observed today and its blue-shifted value $k/a$ during inflation
requires to know the complete history of the universe in between. As
mentioned above, the value of $\Nstar$, to be used as an argument of
the $\eps{i}(\Nstar)$ in \cref{eq:calPzsr3ast,eq:calPhsr3ast}, is
defined by $\kstar \eta(\Nstar)=-1$. As shown in
\Refs{Martin:2010kz,Ringeval:2013hfa,Martin:2014nya,Martin:2016iqo,
  Martin:2016oyk}, this condition is equivalent, at leading order in
the Hubble-flow functions, to the following algebraic equation
determining the value of $\Nstar$
\begin{equation}
\Delta\Nstar \equiv \Nstar - \Nend = -\ln\Rrad + \Nzero + \dfrac{1}{4}
\ln\left[\dfrac{9}{\epsstar{1} \left(3-\eps{1\uend}\right)}
  \dfrac{\Vend}{\Vstar}\right] -
\dfrac{1}{4} \ln \left(\dfrac{\Hstar^2}{\Mp^2 \epsstar{1}}\right),
\label{eq:DeltaNstarRrad}
\end{equation}
where $\Vstar = V[\phi(\Nstar)]$ and $\Vend = V[\phi(\Nend)]$, the
index ``end'' denoting the end of inflation. Using \cref{eq:nufunc},
this equation can also be viewed as an algebraic equation on the field
value $\phistar$. The quantity $\Nzero$ is the typical number of
{\efolds} of decelerated expansion and reads
\begin{equation}
\Nzero \equiv \ln \left[\dfrac{k_*/a_0}{\left(3 \rdofreh \Omega_\urad \Hzero^2 \right)^{1/4}}\right],
\end{equation}
where $\rdofreh$ is a measure of the change of number of entropic and
energetic relativistic degrees of freedom between the beginning of the
radiation era and today~\cite{Ringeval:2013hfa}. The parameter $\Rrad$
in \cref{eq:DeltaNstarRrad} is the so-called \emph{reheating
parameter} defined as
\begin{equation}
\Rrad \equiv
\dfrac{\aend}{\areh}\left(\dfrac{\rhoend}{\rhoreh}\right)^{1/4}.
\label{eq:defRrad}
\end{equation}
It encodes all the relevant kinematic effects the reheating may induce
on the conserved super-Hubble cosmological
perturbations~\cite{Martin:2006rs, Ringeval:2007am}. In
\cref{eq:defRrad}, $\rhoend$ is the energy density of the universe at
the end of inflation and $\rhoreh$ stands for the energy density at
the beginning of the radiation era (i.e., the end of the reheating
era). Clearly, one must have $\rhoreh \le \rhoend$, and, in order not to spoil
Big-Bang Nucleosynthesis (BBN), we should have $\rhoreh > \rhonuc$. Using
energy conservation during the reheating era, one can also show
that~\cite{EIopiparous}
\begin{equation}
\ln \Rrad = \dfrac{\Delta\Nreh}{4}\left(-1 + 3\wrehbar\right) =
\dfrac{1 - 3\wrehbar}{12\left(1+ \wrehbar\right)}
\ln\left(\dfrac{\rhoreh}{\rhoend}\right),
\label{eq:Rradwrho}
\end{equation}
where $\Delta\Nreh$ is the duration of reheating in {\efolds} and
$\wrehbar$ is the average equation of state parameter of the universe
during that period. These other forms of $\Rrad$ show that a
radiation-like reheating era, having $\wrehbar=1/3$, and not
necessarily thermalized, is indistinguishable from an instantaneous
reheating from the point of view of the cosmological
perturbations. Such a statement should make clear that specifying only
$\Delta\Nreh$ is not enough to make inflationary predictions. As
explicit in \cref{eq:DeltaNstarRrad}, whichever representation of the
reheating era is chosen $\bthetareh=(\Delta\Nreh,\wrehbar)$,
$\bthetareh=(\wrehbar,\rhoreh)$, or others, the overall stretching
induced on the cosmological perturbations will always be given by one
number: $\Rrad$. As such, the simplest motivated way to parameterize
the reheating era is, a priori, to use the parameter $\bthetareh = \Rrad$.

Recapping that $\Nstar$ is the parameter fixing the actual values of
the $\epsstar{i}\equiv \eps{i}(\Nstar)$, one should expect any
cosmological data to be sensitive to $\Delta\Nstar$. However, the last
term of \cref{eq:DeltaNstarRrad} is, at leading order in the Hubble
flow functions, equal to $8\pi^2\Pstar$, the amplitude of the scalar
spectrum which is a very well measured quantity. As such,
\cref{eq:DeltaNstarRrad} shows that there is a strong degeneracy
between $\Rrad$, $\Pstar$ and $\Delta\Nstar$ in inferring their
posterior probability distribution from any cosmological data
sets. This issue was discussed, and solved, in \Refs{Martin:2006rs,
  Ringeval:2007am} by introducing the \emph{rescaled reheating
parameter} $\Rreh$ defined by
\begin{equation}
\ln\Rreh \equiv \ln\Rrad + \dfrac{1}{4}\ln\left(\dfrac{\rhoend}{\Mp^4}\right).
\label{eq:Rrehdef}
\end{equation}
Indeed, from the Friedmann-Lema\^{\i}tre equation, one has
\begin{equation}
\rhoend = \dfrac{3 \Vend}{3-\epsoneend} = \dfrac{\Vend}{\Vstar}\dfrac{3
  \Vstar}{3-\epsoneend} = 3\Mp^2  \Hstar^2 \dfrac{\Vend}{\Vstar}
\dfrac{3-\epsstar{1}}{3-\epsoneend}\,,
\end{equation}
One always has $\epsstar{1}\ll 1$ and \cref{eq:Rrehdef} can be
inverted as
\begin{equation}
\ln\Rrad = \ln\Rreh - \dfrac{1}{4}\ln\left(\dfrac{\Hstar^2}{\Mp^2
  \epsstar{1}}\right) - \dfrac{1}{4}\ln\left(\dfrac{9\Vend}{\Vstar}
\dfrac{\epsstar{1}}{3-\epsoneend} \right).
\label{eq:RradLO}
\end{equation}
Plugging \cref{eq:RradLO} into \cref{eq:DeltaNstarRrad}, the terms in
$\Hstar^2/(\Mp^2 \epsstar{1})$ cancel out and one finally gets
\begin{equation}
\Delta\Nstar = -\ln \Rreh + \Nzero + \dfrac{1}{2} \ln
\left(\dfrac{9}{3 - \epsoneend} \dfrac{\Vend}{\Vstar}\right).
\label{eq:DeltaNstarRreh}
\end{equation}
As a result, performing cosmological data analysis with the rescaled
reheating parameter $\Rreh$ completely suppresses the degeneracies
between the kinematics of the reheating and the amplitude of the
scalar perturbations $\Pstar$. Even though the previous calculation
has been made at leading order in the Hubble flow functions, it is
possible to show that $\Rreh$, as defined in \cref{eq:Rrehdef}, allows
one to decouple $\Delta\Nstar$ and $\calP(\kstar)$ out of the
slow-roll assumptions, i.e., assuming only linear perturbations within
General Relativity~\cite{Ringeval:2007am}.  For this reason, the
simplest and optimal choice for parameterizing the reheating is to use
$\bthetareh = \Rreh$. Let us notice that, given an inflationary model,
the quantity $\rhoend$ is determined by the parameters
$\bthetainf$. As such, \cref{eq:Rrehdef} shows that, at fixed
$\bthetainf$, $\Rrad$ and $\Rreh$ are deterministically related. Their
posterior probability distributions will however differ and will be
correlated to the ones associated with $\bthetainf$.

\subsection{Posteriors and evidences}

All inflationary scenarios considered in this work are assumed to be
followed by a reheating era, possibly instantaneous, and which ends
when the standard Friedmann-Lema\^{\i}tre radiation era starts. In
this picture, we are comparing to data a consistent embedding of each
inflationary model within the standard cosmological framework. This
implies that the theoretical parameters of a model $\calM$ are made of
the pure inflationary model parameters $\bthetainf$, the reheating
parameters $\bthetareh$, all the standard cosmological and
astrophysical parameters $\bthetastd$, possibly complemented by some
instrumental nuisance parameters.

Given some cosmological data set $\bdata$, let us denote the
posterior probability distribution of the parameters
$\{\bthetainf,\bthetareh,\bthetastd\}$, within the hypothesis that
inflation has occurred according to model $\calM$, by
$\post{\bthetainf,\bthetareh,\bthetastd}{\bdata,\calM}$. This
multidimensional probability distribution can be determined by
sampling methods, given a likelihood functional
$\post{\bdata}{\bthetainf,\bthetareh,\bthetastd,\calM}$, the prior
probability distributions for all parameters
$\prior{\bthetainf,\bthetareh,\bthetastd} =
\post{\bthetainf,\bthetareh,\bthetastd}{\calM}$, and by using the
Bayes' theorem
\begin{equation}
\post{\bthetainf,\bthetareh,\bthetastd}{\bdata,\calM} =
\dfrac{\post{\bdata}{\bthetainf,\bthetareh,\bthetastd,\calM}
  \prior{\bthetainf,\bthetareh,\bthetastd}}{\evid{\bdata}{\calM}}\,.
\label{eq:bayesth}
\end{equation}
The normalization constant $\evid{\bdata}{\calM}$ is the Bayesian
evidence, or, strictly speaking, the model likelihood: the probability
of having observed the data $\bdata$ given the hypothesis that
inflation occurred according to model $\calM$. From \cref{eq:bayesth},
it reads
\begin{equation}
\evid{\bdata}{\calM} = \int
\post{\bdata}{\bthetainf,\bthetareh,\bthetastd,\calM}\prior{\bthetainf,\bthetareh,\bthetastd}
\, \ud\bthetainf \ud \bthetareh \ud \bthetastd.
\label{eq:eviddef}
\end{equation}
The Bayesian evidence allows us to determine the probability of a
given model to explain the data. Indeed, using again the Bayes'
theorem one has
\begin{equation}
\post{\calM}{\bdata} = \dfrac{\evid{\bdata}{\calM} \prob{\calM}}{\prob{\bdata}}\,,
\end{equation}
where $\prob{\bdata}$ is a model-free normalization factor and
$\prob{\calM}$ is the prior probability of model $\calM$. In order to
compare models $\calM_1$ and $\calM_2$, one can evaluate their
probability ratio to explain the data, namely
\begin{equation}
\dfrac{\post{\calM_1}{\bdata}}{\post{\calM_2}{\bdata}} =
\BayesFactor{1}{2} \,\dfrac{\prob{\calM_1}}{\prob{\calM_2}}\,,
\end{equation}
where we have introduced the ratio of their evidences, namely, the
Bayes factor
\begin{equation}
\BayesFactor{1}{2} = \dfrac{\evid{\bdata}{\calM_1}}{\evid{\bdata}{\calM_2}}\,.
\label{eq:bayesfactor}
\end{equation}
Without any prejudice on which models, $\calM_1$ or $\calM_2$, is
better suited to explain the data, one has $\prob{\calM_1}=\prob{\calM_2}$ and
\cref{eq:bayesfactor} shows that the Bayesian evidences are the
quantities of interest to determine what are the best models~\cite{Martin:2010hh}.

In practice, \cref{eq:eviddef} shows that determining
$\calE(\bdata|\calM)$ requires to compute a multidimensional integral
of the likelihood multiplied by the prior over all the
parameters. This can be performed by using nested-sampling
methods~\cite{Mukherjee:2005wg, Feroz:2007kg, Feroz:2008xx,Handley:2015fda,
  2015MNRAS.453.4384H}. However, this is computationally time
consuming, especially if the likelihood evaluation is slow, and the
task becomes extremely challenging when the number of models $\calM$
is large, as in the present case.

\subsection{Fast nested sampling with machine learning}

In order to speed-up the computation of evidences and of the posterior
probability distribution for all the primordial parameters
$\bthetainf$ and $\bthetareh$ associated with each model $\calM$, we
have applied the method discussed in \Refa{Ringeval:2013lea} to the
third-order power spectra of \cref{eq:calPzsr3ast,eq:calPhsr3ast}.

It consists in determining, and machine-learning, an effective
likelihood $\calLeff$ for the primordial parameters constructed over
some model-free Hubble-flow functions, say $\vareps{i}$, and such that
\begin{equation}
\likeff{\bdata}{\vareps{0},\vareps{1},\vareps{2},\vareps{3}} \propto \int
\post{\bdata}{\bthetastd,\vareps{0},\vareps{1},\vareps{2},\vareps{3},\vareps{4}}
\prior{\vareps{4}} \prior{\bthetastd} \ud \vareps{4} \ud \bthetastd.
\label{eq:Leffdef}
\end{equation}
The full likelihood
$\post{\bdata}{\bthetastd,\vareps{0},\vareps{1},\vareps{2},\vareps{3},\vareps{4}}$
is associated with the primordial power spectra of
\cref{eq:calPzsr3ast,eq:calPhsr3ast} where all the $\epsstar{i}$ are
systematically replaced with the model-free $\vareps{i}$.

Determining $\calLeff$ requires to marginalize a full likelihood over
many parameters, the $\bthetastd$ and $\vareps{4}$ here. This is done
in \Sec{sec:srposteriors} by performing an exploration of the full
parameter space using Markov-Chains-Monte-Carlo (MCMC)
methods~\cite{Lewis:2002ah,Torrado:2020dgo}.

Let us prove here that \cref{eq:Leffdef} is indeed sufficient to make
Bayesian inference in the space of primordial parameters
$\{\bthetainf,\bthetareh\}$. Within a given inflationary model, say
$\calM$, there is a deterministic relationship
$\epsstar{i}(\bthetainf,\bthetareh)$ between the values of the
Hubble-flow functions and the model parameters. Notice the presence of
the reheating parameters for the reasons explained in
\Sec{sec:rehparam}. One can therefore add to the inference problem a
new set of hyperparameters, the
$\bvareps=\{\vareps{0},\vareps{1},\dots\}$, whose values are precisely
defined to be the functional forms
$\epsstar{i}(\bthetainf,\bthetareh)$ for $i \geq 1$ and
$\varepsilon_0=P_*$ when one is concerned with a particular model
$\calM$, but are unspecified otherwise. Explicitly, the
hyperparameters are deterministic and satisfy
\begin{equation}
\post{\bvareps}{\bthetainf,\bthetareh,\calM} \equiv
\dirac{\vareps{0} - \Pstar} \prod_i \dirac{\vareps{i}-\epsstar{i}}.
\label{eq:hyperdef}
\end{equation}

If we don't specify the model, it is still possible to perform data
analysis based on the slow-roll power spectra of
\cref{eq:calPzsr3ast,eq:calPhsr3ast}, now parameterized by the
$\bvareps$ instead of the $\epsstar{i}$, still with
$\vareps{0}=\Pstar$, and we will refer to this model-free situation as
hypothesis $I$. From MCMC exploration, one can determine the full
posterior distribution
$\post{\bthetastd,\bvareps}{\bdata,I}$, under
the model-free hypothesis $I$.

The posterior for the primordial parameters associated with $\calM$ is
obtained by marginalizing over all the other parameters and reads
\begin{equation}
  \post{\bthetainf,\bthetareh}{\bdata,\calM} = \int
  \post{\bthetainf,\bthetareh,\bthetastd,\bvareps}{\bdata,\calM,I} \ud
  \bvareps \ud \bthetastd.
\label{eq:postprim}
\end{equation}
Using the Bayes' theorem, the right-hand side can be rewritten as
\begin{equation}
\post{\bthetainf,\bthetareh,\bthetastd,\bvareps}{\bdata,\calM,I} =
\dfrac{\post{\bthetainf,\bthetareh,\bthetastd,\bvareps}{\calM,I}
  \post{\bdata}{\bthetainf,\bthetareh,\bthetastd,\bvareps,\calM,I}}{\post{\bdata}{\calM,I}}\,,
\label{eq:postbayes}
\end{equation}
which can be further simplified using the product rule:
\begin{equation}
\post{\bthetainf,\bthetareh,\bthetastd,\bvareps}{\calM,I} =
\post{\bthetainf,\bthetareh,\bthetastd}{\calM,I} \post{\bvareps}{\bthetainf,\bthetareh,\bthetastd,\calM,I}.
\end{equation}
Assuming that $I$ is generic enough to contain all models, for
inflation occurring according to model $\calM \subset I$, the last term is
conditioned by $\calM$ only and one can use
\cref{eq:hyperdef}. Moreover, assuming that
$\{\bthetainf,\bthetareh\}$ and $\{\bthetastd\}$ are independent, one
gets
\begin{equation}
\post{\bthetainf,\bthetareh,\bthetastd,\bvareps}{\calM,I} = 
\post{\bthetainf,\bthetareh}{\calM} \post{\bthetastd}{I} \dirac{\vareps{0} - \Pstar}
\prod_i \dirac{\vareps{i}-\epsstar{i}}.
\end{equation}
Plugging this expression into \cref{eq:postbayes}, and provided that
the full likelihood has no explicit dependence on
$\{\bthetainf,\bthetareh\}$, \cref{eq:postprim} becomes
\begin{equation}
\post{\bthetainf,\bthetareh}{\bdata,\calM} =
\dfrac{\likeffb{\bdata}{\Pstar(\bthetainf,\bthetareh),\epsstar{i}(\bthetainf,\bthetareh)}
  \prior{\bthetainf,\bthetareh}}{\evid{\bdata}{\calM}}\,,
\label{eq:posteff}
\end{equation}
where
\begin{equation}
\likeff{\bdata}{\bvareps} \equiv \dfrac{\prob{D}}{\evid{\bdata}{I}} \int
\post{\bdata}{\bthetastd,\bvareps} \post{\bthetastd}{I} \ud \bthetastd.
\label{eq:Leff}
\end{equation}
Once $I$ is set, the overall factor will be a constant for all $\calM$
and will disappear in the Bayes' factors of \cref{eq:bayesfactor}. As
announced, we have dramatically reduced the dimensionality of the
likelihood. The determination of $\calLeff$ is complex, but, once done
and fitted, Bayesian data analysis for any model $\calM$ can be
performed with \cref{eq:posteff} only. Provided the machine learned
$\calLeff$ is fast and accurate, considerable speed-up can be achieved
in the determination of the posteriors
$\post{\bthetainf,\bthetareh}{\bdata,\calM}$ and evidences
$\evid{\bdata}{\calM}$.

An important remark is that hyperparameters to which the data are not
sensitive can be marginalized over instead of being kept in the
effective likelihood. Conversely, if one of the primordial parameters,
say $\theta_1$, would enter explicitly in the full likelihood (i.e.,
not only via the $\epsilon_{i*}$ functions), it could be kept as a new
parameter in $\calLeff$ thereby ensuring no information loss. As shown
in the next section, the current data are not sensitive to the values
of $\vareps{4}$, and even though our full data analysis is based on
third-order power spectra, and includes $\vareps{4}$, the effective
likelihood in \cref{eq:Leffdef} does not include $\vareps{4}$
anymore. Let us stress that defining $\calLeff$ by marginalizing over
$\vareps{4}$, as opposed to drop $\vareps{4}$ from the very
beginning, ensures the robustness of our results with respect to all
possible third-order slow-roll corrections.

\subsection{Workflow}

As detailed in the previous sections, the analysis workflow is
bootstrapped by a complete data analysis of the cosmological data
using the third-order slow-roll primordial power spectra of
\cref{eq:calPzsr3ast,eq:calPhsr3ast}. This step, described in
\Sec{sec:srposteriors}, allows us to determine $\calLeff$, which is
then fitted over the four-dimensional space
$\{\Pstar,\vareps{1},\vareps{2},\vareps{3}\}$ using machine-learning
methods.

Next, as presented in \Sec{sec:evidences}, for all models $\calM$
under scrutiny, we determine the evidence $\evid{\bdata}{\calM}$, the
information gain and other relevant statistical measures, by nested
sampling methods running over the machine-learned $\calLeff$.

\section{Third-order slow-roll data analysis}
\label{sec:srposteriors}

In this section, we present novel results coming from a full data
analysis based on the third-order slow-roll primordial power spectra
and using the latest CMB and BAO data. Some technical details are
given as this analysis constitutes the first attempt to constrain
$\vareps{4}$ while providing more robust bounds on $\vareps{1}$,
$\vareps{2}$ and $\vareps{3}$ as they are marginalized over $\vareps{4}$.

The computation of the full posterior probability distribution
$\post{\bthetastd,\Pstar,\vareps{1},\vareps{2},\vareps{3},\vareps{4}}{\bdata,I}$
has been achieved using a modified version of the public Fortran code
{\COSMOMC}~\cite{Lewis:2002ah}. It has been interfaced with a modified
version of the {\CAMB} code~\cite{Lewis:1999bs}, in which we have
replaced the initial power spectrum module by a new slow-roll module
encoding the third-order slow-roll power spectra of
\cref{eq:calPzsr3ast,eq:calPhsr3ast}.

The last public version of the {\COSMOMC} code dates back from 2022
and we had to update various parts of the code to include newest
data. Our choice of using {\COSMOMC} instead of more recent and,
possibly, more user-friendly Monte-Carlo-Markov-Chains samplers, such
as the {\COBAYA} Python code~\cite{Torrado:2020dgo}, has been
motivated by the needs to robustly run massively parallel MCMC to get
tens of million of samples over more than fifty dimensions. These
unusually large numbers are indeed needed to properly sample the
four-dimensional likelihood
$\likeff{\bdata}{\Pstar,\vareps{1},\vareps{2},\vareps{3}}$, which
is fed to a machine-learning algorithm afterwards. The
four-dimensional samples to determine $\calLeff$ have been computed by
marginalizing the full posterior on all but the four dimensions
$\{\Pstar,\vareps{1},\vareps{2},\vareps{3}\}$. For this purpose, we
have coded and used $N$-dimensional marginalization function within
the {\GETDIST} Python code~\cite{Lewis:2019xzd, Martin:2013nzq,
  Martin:2014lra}.

\subsection{Data sets}
\label{sec:data}
The cosmological data sets $\bdata$ used are the latest Planck
satellite data~\cite{Aghanim:2019ame}, the BICEP/Keck array $B$-mode
polarization 2021 measurements~\cite{BICEP:2021xfz}, the South Pole
Telescope third generation $TE$ and $EE$ angular power
spectra~\cite{SPT-3G:2021eoc}, and a compilation of Baryonic Acoustic
Oscillations data from the Sloan Digital Sky Survey
IV~\cite{Dawson:2015wdb, SDSS:2017yll}.

In more details, for the Planck data, we have used the 2020
post-legacy release with the small angular scales $TT$, $TE$ and $EE$
power spectra extracted from the {\PRNPIPE} maps~\cite{Planck:2020olo,
  Tristram:2020wbi}. The likelihood is the {\CAMSPEC}
code~\cite{Efstathiou:2019mdh}, in its Fortran version $12.6$, and
discussed in \Refa{Rosenberg:2022sdy}. For the low multipoles, we have
used the official {\SIMALL} likelihood built on the large scales $EE$
polarization measurements, also refereed to as the $\texttt{lowE}$
data~\cite{Aghanim:2019ame}. Finally, we have also included the
lensing potential measurements as described in \Refa{Planck:2018lbu}.

The BICEP/Keck $B$-mode data are from the latest release presented in
\Refa{BICEP:2021xfz}, using both the WMAP and Planck dust tracers for
map cleaning~\cite{2013ApJS..208...20B,Planck:2020olo}. The
likelihoods are public, provided by the BICEP/Keck team, and they are
ready-to-use for {\COSMOMC}.

The South Pole Telescope data are from the SPT-3G survey described in
\Refa{SPT-3G:2021wgf} and consist in high multipoles measurements for
the $TE$ and $EE$ angular power spectra. The likelihood codes for
{\COSMOMC} are provided by the SPT collaboration and have been
incorporated within the multiplatform library
{\CLIK}\footnote{\url{https://github.com/benabed/clik}}.

The Baryon Acoustic Oscillation data are the DR16 compilation, built
using multiple tracers: the Main Galaxy Sample, the Luminous Red
Galaxies, the Emission Line Galaxies, the Quasars and Lymann-$\alpha$
forests from eBOSS~\cite{SDSS:2008tqn, Bautista:2020ahg,
  Gil-Marin:2020bct, deMattia:2020fkb, Hou:2020rse, Neveux:2020voa,
  duMasdesBourboux:2020pck}. The likelihood data files are public and
provided by the SDSS collaboration. They have been used to upgrade the
BAO likelihood module provided with {\COSMOMC}.

In order to maximize the constraining power of the combined data sets,
we have chosen not to include the Atacama Cosmology Telescope
data~\cite{ACT:2020gnv, ACT:2020frw}, due to a possible small tension
with Planck~\cite{Handley:2020hdp}, in addition to some optimization
issues in calling a Python likelihood from {\COSMOMC}. Also, within
the Planck data sets, we have excluded the first thirty low multipoles
from the $TT$ channel (the so-called \texttt{lowl} likelihood) due to
their known under power compared to the higher
multipoles~\cite{Planck:2018bsf}. Notice that the low multipole data
in the $TE$ and $EE$ channels are included and they have been shown to
be, at least, as powerful as the $TT$ channel in constraining the
shape of the primordial power spectra, while not being in tension with
the small scales~\cite{Galli:2014kla}.

\subsection{Priors}
\label{eq:priors}

In addition to assume third-order slow-roll power spectra generating
both scalar and tensor perturbations, we consider a spatially flat
Friedmann-Lema\^{\i}tre model described by the usual four cosmological
parameters. They are the density parameter for baryons $\OmegaB h^2$,
for cold dark matter $\OmegaCDM h^2$ (both multiplied by $h^2$), $100
\thetaMC$, the angular size of the sound radius at last scattering
(multiplied by $100$), and the reionization optical depth $\tau$. As
already mentioned, these are complemented by the four Hubble-flow
parameters $\{\Pstar,\vareps{1},\vareps{2},\vareps{3}\}$. In addition
to these eight fundamental cosmological parameters, the data sets come
with a relatively large number of nuisance and foreground
parameters. There is little interest to enumerate them here, their
priors are already set in their associated likelihoods and are
discussed at length in the references cited in \Sec{sec:data}. In
total, the nuisance and foreground parameters amount to $49$
dimensions.

Concerning the priors for the four cosmological parameters, they have
been taken as flat distributions, wide enough to encompass their
currently preferred values $\OmegaB h^2 \in[0.05,0.1]$, $\OmegaCDM h^2
\in[0.01,0.99]$, $100\thetaMC \in[0.5,10]$, $\tau\in[0.01,0.8]$. An
additional hard prior is set on the Hubble parameter today
$\Hzero\in[40,100]$. Although we do not expect the primordial
parameters to be sensitive to the following settings, non-linear
lensing corrections have been switched on in {\CAMB} while we have let
to the default various other parameters: the number of effective
neutrinos is $3.046$ (as for the Planck analysis), the sum of their
masses is set to $0.06\,\eV$, and the Helium mass fraction is enforced
from BBN constraints.  For the primordial parameters, the pivot scale
is set at $\kstar=0.05\,\Mpc^{-1}$. The scalar amplitude, $\Pstar$, is
sampled from a flat prior on
$\ln\left(10^{10}\Pstar\right)\in[2.4,3.6]$, largely encompassing the
required amplitude to match the one of the CMB fluctuations. The order
of magnitude of the first Hubble-flow parameter is unknown, it is
positive definite and it mostly determines the tensor-to-scalar
ratio. As such, it has been sampled from a Jeffreys' prior, set as a
flat prior on $\log(\vareps{1})\in[-5,-0.7]$ (decimal logarithm). For
the three other Hubble-flow functions, we have chosen a flat prior
$\vareps{2,3,4}\in[-0,2,0.2]$ in order to enforce the slow-roll
condition that these parameters should be smaller than unity. However,
as we discuss in \Sec{sec:priortests}, this choice could be questioned
for the parameter $\vareps{3}$. Indeed, as shown in the next section,
our data sets are putting some constraints on $\vareps{3}$: its
posterior distribution is not flat in the range $[-0,2,0.2]$ but it is
not vanishing at the boundaries. This necessarily triggers some
sensitivity with respect to the prior range. For this reason, we have
performed two complete MCMC analysis, one with the aforementioned
prior choices, referred to as ``strictly slow roll'' (labeled as
``Strict SR'' in the figures), and another one with the extended prior
$\vareps{3}\in[-1,1]$, which saturates the slow-roll conditions and
referred to as ``extended slow roll'' (labeled as ``Extended SR'').

\subsection{Marginalized posteriors}
\label{eq:posteriors}

\begin{figure}
\begin{center}
\includegraphics[width=\bigfigw]{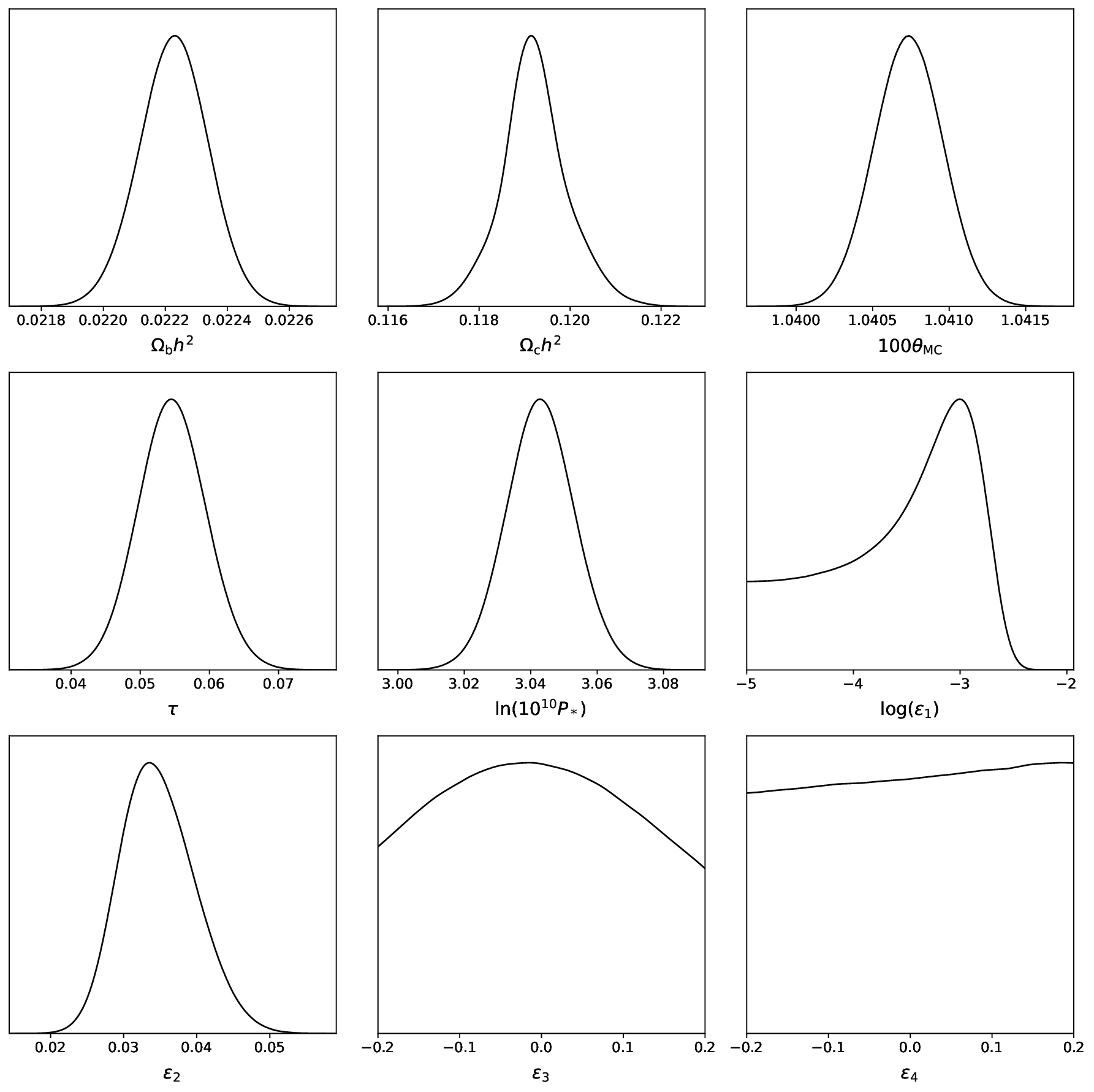}
\caption{One-dimensional marginalized posterior distributions for the
  bare cosmological parameters and the third-order Hubble flow
  parameters $\Pstar$, $\vareps{1}$, $\vareps{2}$, $\vareps{3}$ and
  $\vareps{4}$. The data sets $\bdata$ used are described in
  \Sec{sec:data} and the prior is strictly slow-roll
  $\vareps{3}\in[-0.2,0.2]$. Notice the sensitivity of the data to
  $\vareps{3}$ whereas $\vareps{4}$ is unconstrained (see also
  \Fig{fig:sr3ndlog_2D}). The peak in the posterior for
  $\log(\vareps{1})$ shows that, despite marginalization of the
  BICEP/Keck data over dust contamination, there remains in the
  current data a $B$-mode excess which is slightly favoring, at
  one-sigma, tensor modes of slow-roll inflationary origin (see
  text).}
\label{fig:sr3ndlog_1D}
\end{center}
\end{figure}

\begin{figure}
\begin{center}
\includegraphics[width=\bigfigw]{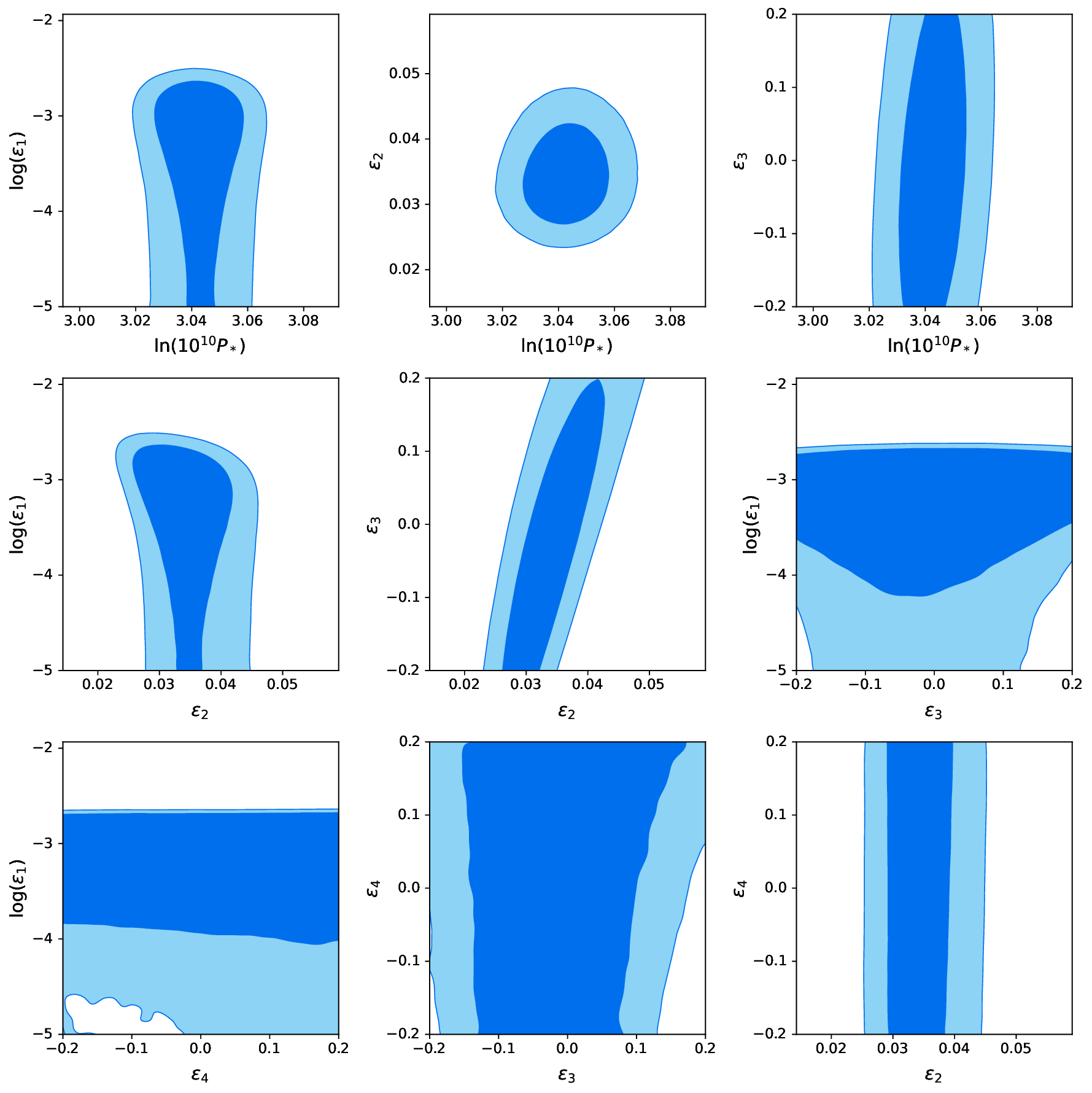}
\caption{Two-dimensional marginalized posterior distributions within the
  Hubble-flow parameter space for the strict slow-roll prior
  $\vareps{3}\in[-0.2,0.2]$ (see also \Fig{fig:sr3ndlog_1D}).}
\label{fig:sr3ndlog_2D}
\end{center}
\end{figure}

\begin{figure}
\begin{center}
\includegraphics[width=\bigfigw]{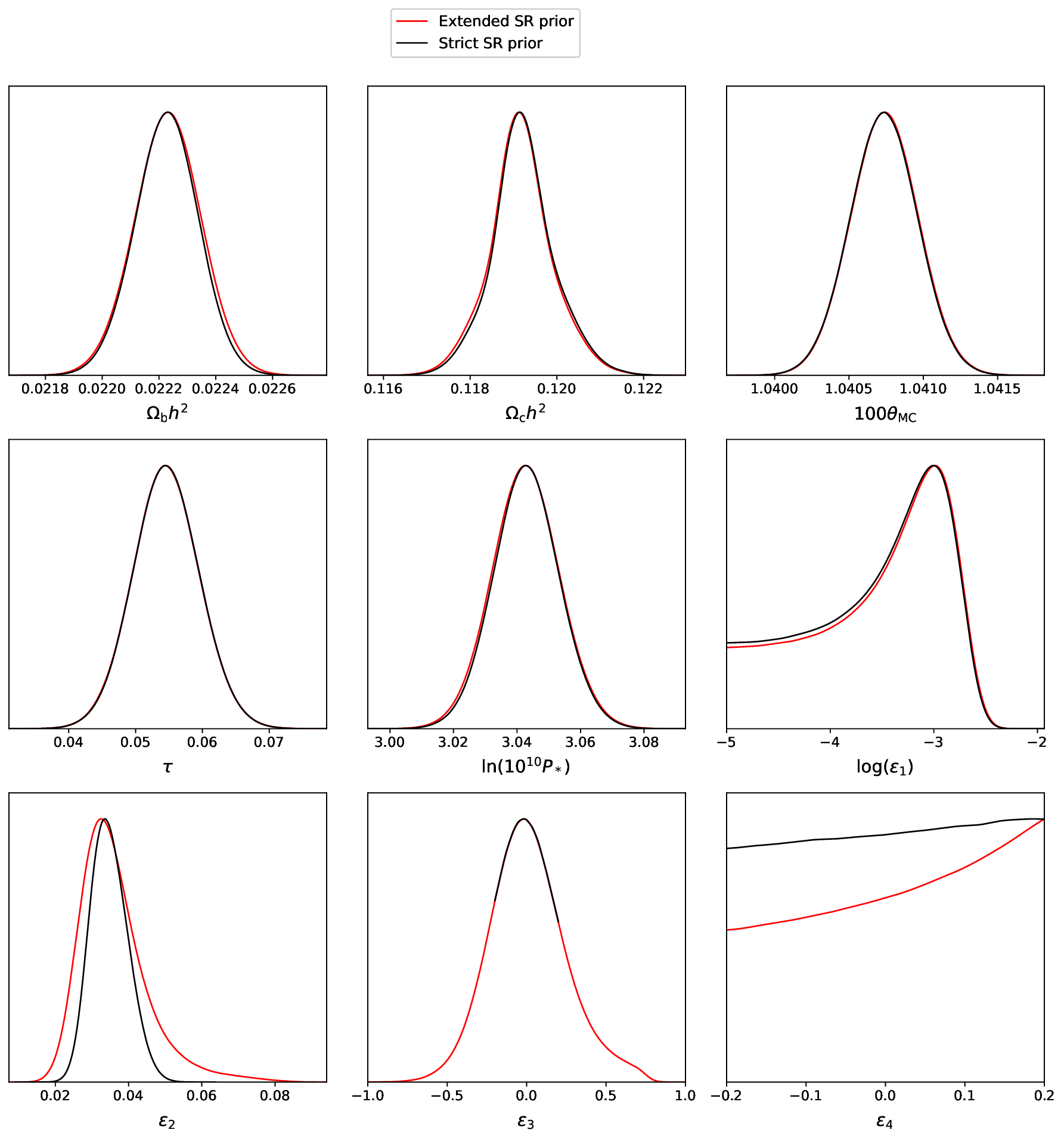}
\caption{One-dimensional marginalized posterior distributions,
  normalized to their maximum, for the bare cosmological parameters
  and the third-order Hubble-flow parameters for the extended
  slow-roll prior on $\vareps{3}\in[-1,1]$ (red). The black curves are
  the one- and two-sigma contours of \Fig{fig:sr3ndlog_1D} obtained
  under the strict prior $\vareps{3}\in[-0.2,0.2]$. The posterior of
  $\vareps{3}$ is well bounded within the prior. Notice the appearance
  of a tail in the $\vareps{2}$ posterior due to its correlations with
  $\vareps{3}$.}
\label{fig:sr3ndlogext_1D}
\end{center}
\end{figure}

\begin{figure}
\begin{center}
\includegraphics[width=\bigfigw]{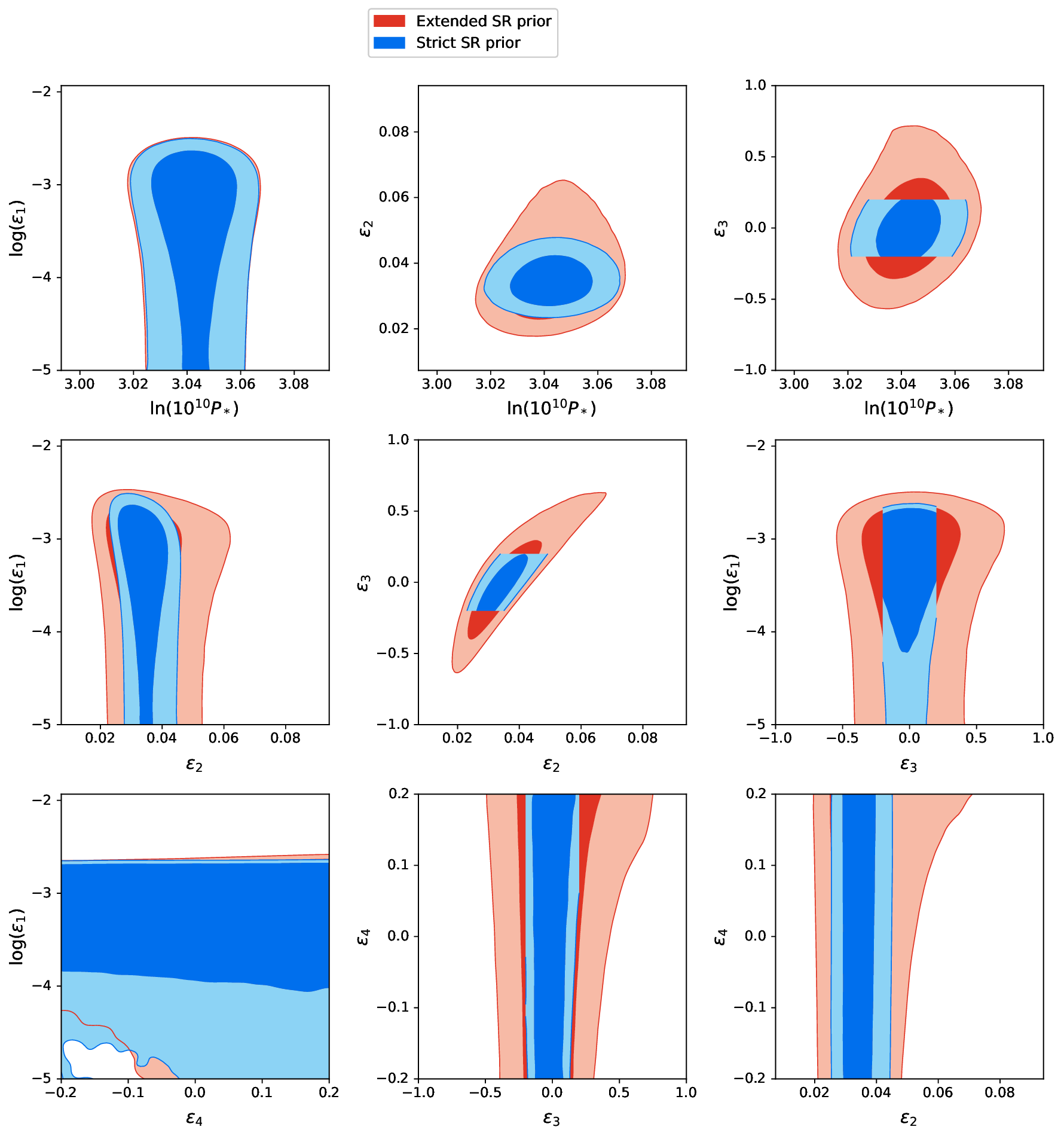}
\caption{Two-dimensional marginalized posterior distributions within
  the Hubble-flow parameter space for the extended slow-roll prior
  range $\vareps{3}\in[-1,1]$ and compared to the strict slow-roll
  posteriors (in red), see also \Fig{fig:sr3ndlog_1D}. Notice the
  wider posterior for $\vareps{2}$ and the slight bias induced on the
  posterior for $\vareps{4}$.}
\label{fig:sr3ndlogext_2D}
\end{center}
\end{figure}

\begin{figure}
\begin{center}
\includegraphics[width=\bigfigw]{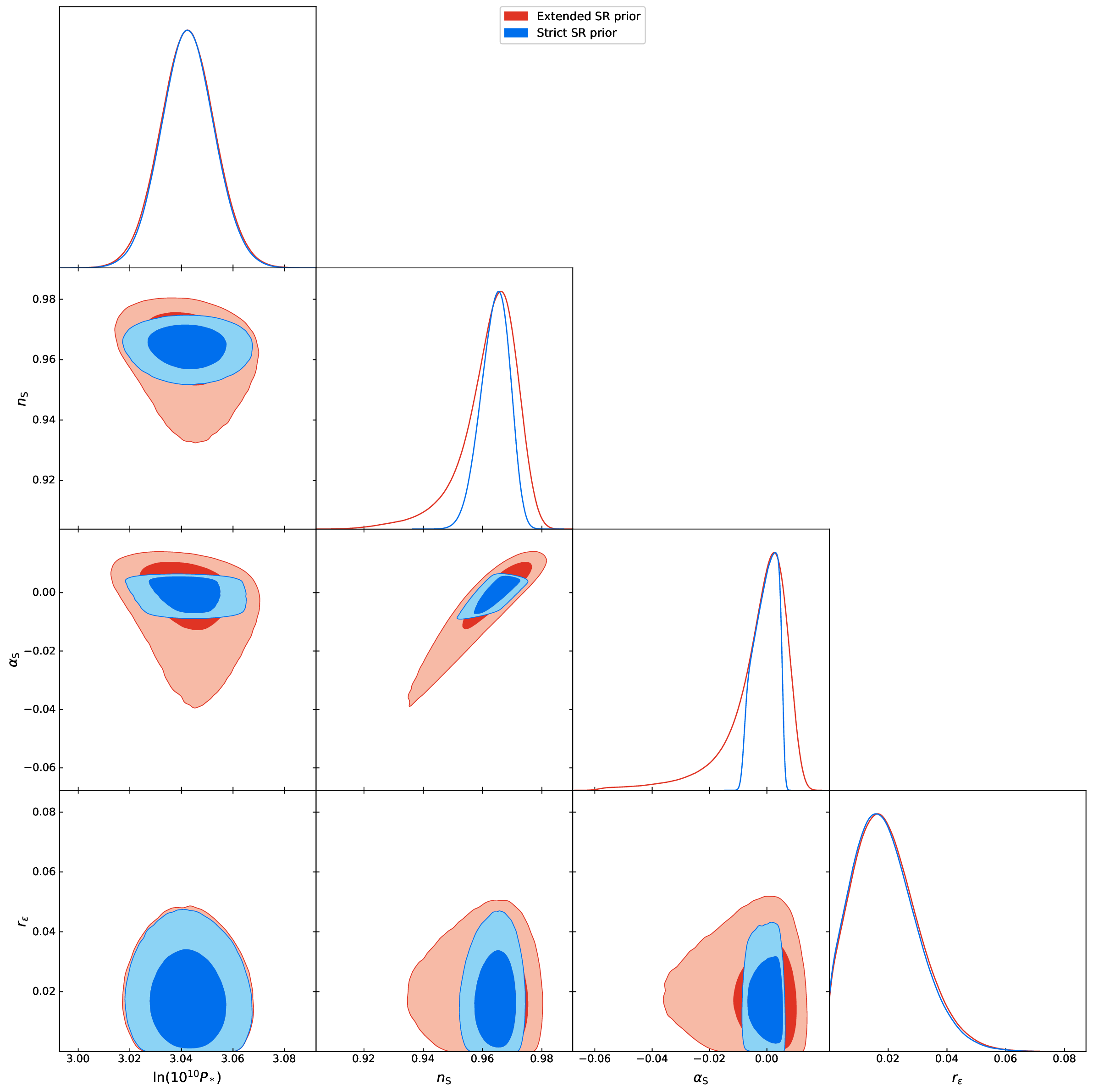}
\caption{One- and two-dimensional marginalized posterior distributions
  within the power-law parameter space $\{\Pstar,\nS,\reps,\alphaS\}$
  where $\nS$ is the spectral index, $\reps$ the tensor-to-scalar
  ratio and $\alphaS$ the running of the spectral index. These have
  been obtained by importance sampling from the Hubble-flow
  posteriors, the samples have been re-weighted to correspond to a
  flat prior on $\vareps{1}$. As such the prior \emph{is not}
  associated with a power-law shaped spectrum but with
  \cref{eq:calPzsr3ast,eq:calPhsr3ast}. The colors are the same as in
  \Fig{fig:sr3ndlogext_2D} and correspond to the strict and extended
  prior choices on $\vareps{3}$.}
\label{fig:powerlaw_2D}
\end{center}
\end{figure}

The MCMC chains has been generated using {\COSMOMC}, running in
parallel over $96$ \texttt{MPI}-processes, each processor running the
Boltzmann solver {\CAMB}, \texttt{OpenMP}-parallelized over $8$
threads (processors AMD \texttt{EPYC-7742}). The chains have been
stopped once the number of recorded samples reached $25$ million, for which the
$R$-statistics criterion of convergence (Gelman-Rubin) is as low as
$10^{-3}$~\cite{Lewis:2002ah}.

In \Fig{fig:sr3ndlog_1D}, we have plotted the marginalized posterior
probability distribution for all eight cosmological parameters,
including the Hubble-flow ones. These are obtained by marginalizing
over the other nuisance parameters, the one-dimensional posteriors of
which can be found in \cref{sec:nuisances}. \Fig{fig:sr3ndlog_2D}
shows the most interesting two-dimensional posteriors in the
Hubble-flow parameter space. These posteriors show that the fourth
Hubble flow function is unconstrained by the data, its posterior is
flat. On the contrary, the posterior for $\vareps{3}$ is peaked around
vanishing values and this shows that the current CMB and BAO data are
sensitive to a running of a spectral index typical of slow roll. As
can be checked in \Fig{fig:sr3ndlog_2D}, $\vareps{3}$ is correlated
with $\vareps{2}$, and the one- and two-dimensional posteriors are not
vanishing at the boundary of the strict slow-roll prior range
$[-0.2,0.2]$.

Let us notice the interesting posterior for $\log(\vareps{1})$ which
is clearly peaked at a value well inside our prior range. We measure,
indeed, a weak statistical preference, at one-sigma, for the
one-dimensional posterior
\begin{equation}
  \log(\vareps{1}) > -3.9, \qquad \texttt{($68\%$ CL)},
\label{eq:pgwdiscovery}
\end{equation}
moving down to $\log(\vareps{1}) > -4.9$ at $95\%$ CL. We have checked
that this excess is coming from the BICEP/Keck data and it means that,
in spite of the marginalization over dust contamination, there is
still some $B$-mode excess remaining. This effect was visible on the
posteriors presented by the Planck
collaboration~\cite{Planck:2018jri}, but with the latest data, and
using slow-roll power spectra, its significance has increased. Let us
also stress that we have chosen a Jeffreys' prior for $\vareps{1}$,
which naturally favor small values. This excess is more pronounced
when one assumes a flat prior instead. This is illustrated in
\Fig{fig:powerlaw_2D} where we have used importance sampling to
re-weight the samples according to a flat prior on $\vareps{1}$ and
computed the power law parameter posteriors. These are the spectral
index $\nS$, the running of the spectral index $\alphaS$ and the
tensor-to-scalar ratio $\reps$. The one-sigma contour is closed in all
two-dimensional posteriors involving the tensor-to-scalar ratio
$\reps$. Also, let us stress that this small residual excess is robust
to third-order slow-roll corrections, which are automatically
marginalized over when looking at the $\log(\vareps{1})$
posterior. Still, one-sigma is non-statistically significant, and, at
three-sigma, only an upper bound remains
\begin{equation}
  \log(\vareps{1}) < -2.6, \qquad \texttt{($98\%$ CL)}.
\end{equation}
Still, it will be worth paying attention to this trend in the future, since
the favored values found here will be well inside the sensitivity of
the LiteBIRD satellite~\cite{LiteBIRD:2022cnt}.

In order to quantify the effects associated with the choice of the
$\vareps{3}$ prior, we have duplicated the analysis starting from the
extended prior choice $\vareps{3}\in[-1,1]$. The marginalized
posteriors are represented in
\Figs{fig:sr3ndlogext_1D,fig:sr3ndlogext_2D} and compared to the ones
of \Figs{fig:sr3ndlog_1D,fig:sr3ndlog_2D}. The bare cosmological
parameters are unaffected, as expected, but the posterior for
$\vareps{2}$ is slightly enlarged due to the degeneracies existing
between $\vareps{2}$ and $\vareps{3}$, these ones being enhanced when
the latter take values close to unity. Similar degeneracies between
$\vareps{3}$ and $\vareps{4}$ are also responsible for the tiny
positive bias visible in the one-dimensional posterior of $\vareps{4}$
in \Fig{fig:sr3ndlogext_1D}. Let us stress that the posterior for
$\vareps{1}$ is unchanged, confirming the robustness of the peak
amplitude with respect to $\vareps{3}$. The one-sigma bound of
\cref{eq:pgwdiscovery} ends up being unaffected by the extended prior
choice\footnote{By the very nature of an improper Jeffreys' prior,
\cref{eq:pgwdiscovery} has, however, some sensitivity to the choice of
the lower bound of the $\log(\vareps{1})$ prior. Assuming much smaller
minimal values of $\log(\vareps{1})$ would decrease its
significance.}. Finally, the posterior for $\vareps{3}$ is now well
bounded within the extended prior range $[-1,1]$, which confirms that
the current data are strongly disfavoring any running which is not of
slow-roll magnitude~\cite{Planck:2018jri}. We find the mean value
$\mean{\vareps{3}} =0.015$ in the credible interval
\begin{equation}
-0.44 < \vareps{3} < 0.55, \qquad \texttt{($95\%$ CL)}.
\end{equation}

\subsection{Other consistency checks}
\label{sec:priortests}

Our results have been cross-checked using the Planck official
likelihood {\PLIK} running on the Planck 2018 legacy
data~\cite{Planck:2018lkk, Aghanim:2019ame}. The posteriors are all
compatible and we do not observe more than half a sigma degradation in
any parameter constraints by doing so. We have also cross-checked our
implementation of the BAO likelihoods for {\COSMOMC} by comparing the
posteriors to small MCMC explorations made using the Python package
{\COBAYA}~\cite{Torrado:2020dgo}.

In the following, we marginalize the $25$ million samples over $54$
dimensions to determine the four-dimensional posterior probability
distribution $\post{\Pstar,\vareps{1},\vareps{2},\vareps{3}}{\bdata}$
that will be used as the effective likelihood to perform Bayesian model
comparison in \Sec{sec:evidences}. Although not explicit in the
following, all computations have been duplicated in order to quantify
the possible effects associated with the strict and extended prior
choices on $\vareps{3}$. As mentioned above, because $\vareps{4}$ is
unconstrained, it is part of the parameter set over which we
marginalize the posteriors.

\subsection{Effective likelihood}
\label{sec:caleff}

\begin{figure}
\begin{center}
\includegraphics[width=\onefigw]{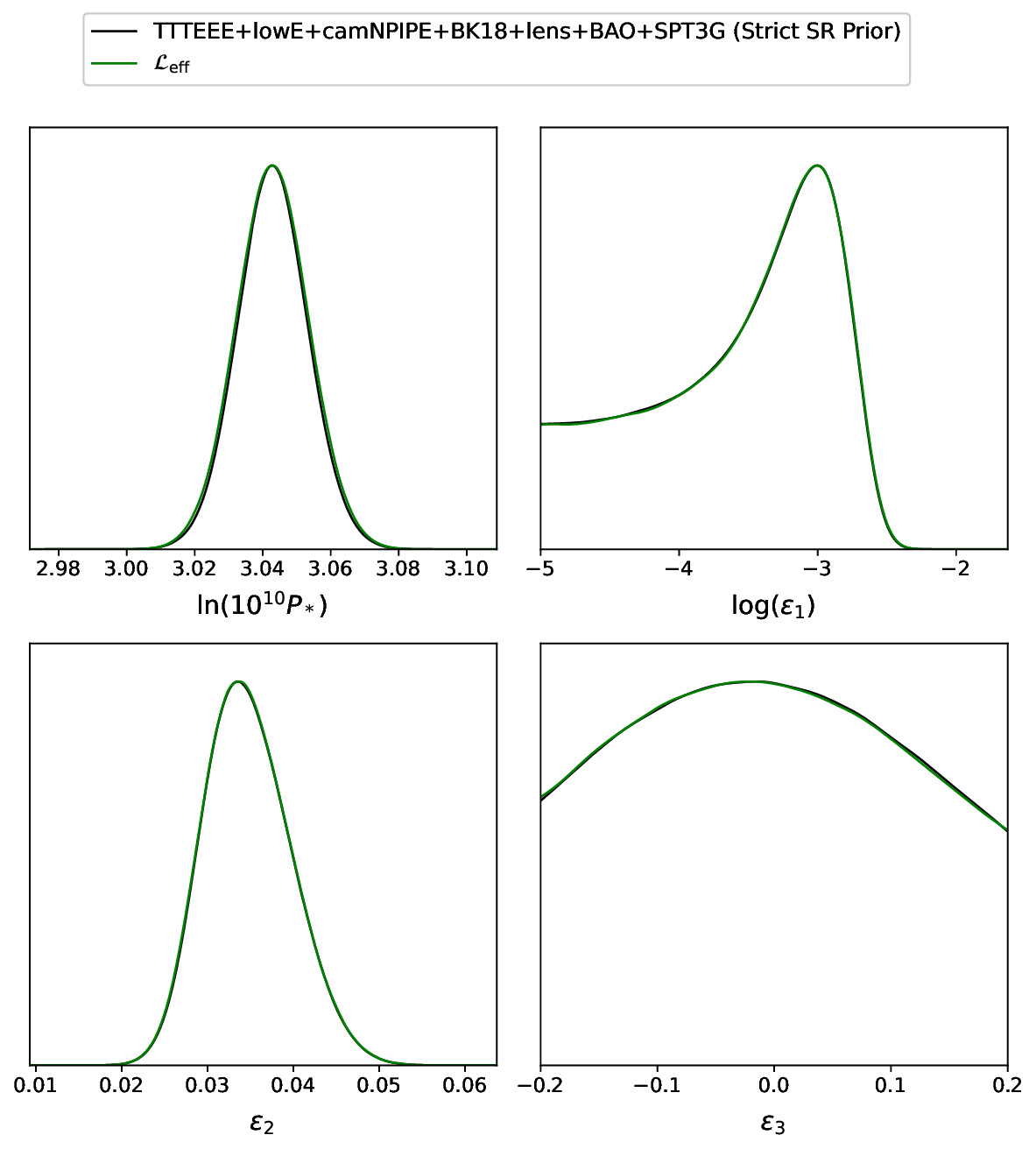}
\caption{Comparison between the one-dimensional posteriors in the
  Hubble-flow parameter space
  $\{\Pstar,\vareps{1},\vareps{2},\vareps{3}\}$ obtained by nested
  sampling over our machine-learned $\calLeff$ (green), and the exact
  ones (black) of \Fig{fig:sr3ndlog_1D}, the latter being obtained
  from marginalization of the full likelihood over the
  $54$-dimensions. The differences are barely visible, see also
  \Fig{fig:bayesinf_sr3_2D}.}
\label{fig:bayesinf_sr3_1D}
\end{center}
\end{figure}

\begin{figure}
\begin{center}
\includegraphics[width=\bigfigw]{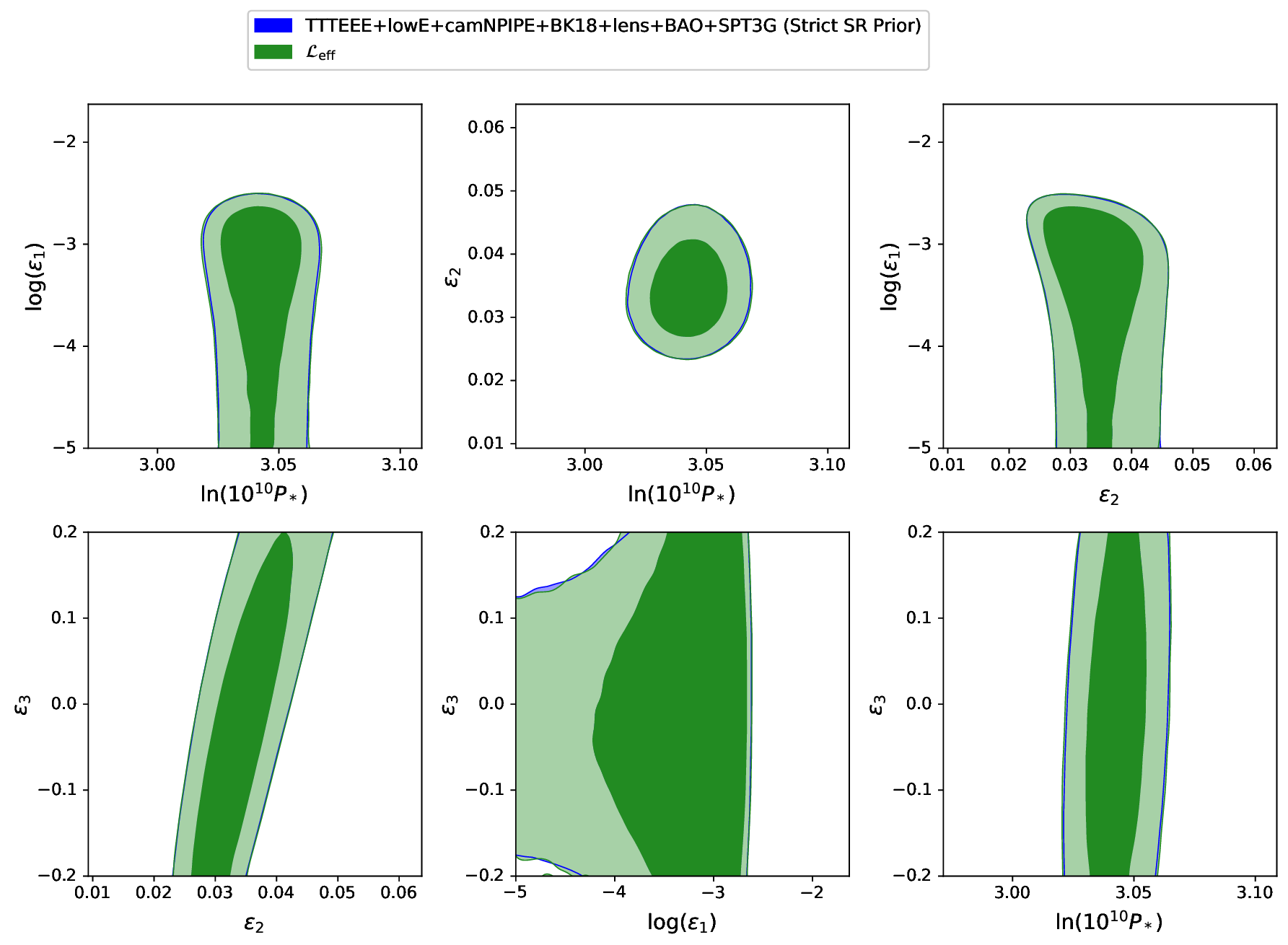}
\caption{Comparison between the two-dimensional posteriors in the
  Hubble-flow parameter space
  $\{\Pstar,\vareps{1},\vareps{2},\vareps{3}\}$ obtained by nested
  sampling over our machine-learned $\calLeff$ (green) to the exact
  ones (blue) of \Fig{fig:sr3ndlog_2D}.}
\label{fig:bayesinf_sr3_2D}
\end{center}
\end{figure}

Up to a multiplicative constant, $\calLeff \propto
\post{\Pstar,\vareps{1},\vareps{2},\vareps{3}}{\bdata}$ and its
two-dimensional slices are already represented in
\Figs{fig:sr3ndlog_2D,fig:sr3ndlogext_2D}. In order to ``fit'' this
four-dimensional function, we have first determined
$\post{\Pstar,\vareps{1},\vareps{2},\vareps{3}}{\bdata}$, binned over
$20^4$ bins, as computed by the $N$-dimensional {\GETDIST}
marginalization functions~\cite{Lewis:2019xzd}. Given $25$ millions
MCMC samples, had the bins been equally sampled (which is not the case
for MCMC), each of them would have contained typically $10^2$ samples and
this yields a typical $10\%$ error per bin. This binned posterior is
thus not very practical per se but can be efficiently used to train a
basic-machine learning algorithm to smoothly fit its four-dimensional
shape. We have used a modified quadratic Shepard's
method~\citep{Shepard:1968, Thacker:2010}, which can be viewed as a
neural network having one input layer of four nodes ($\Pstar$,
$\vareps{1}$, $\vareps{2}$, $\vareps{3}$), one internal layer of $300$
nodes, and, a single node output layer providing the value of
$\calLeff$. Determining the optimal weights has been done using a least
square fitting method running on the binned posterior.

As with any neural networks, there is no definite way to ensure the
accuracy of the fitted multivariate function, a priori. Therefore, in
order to validate that the machine-learned
$\calLeff(\Pstar,\vareps{1},\vareps{2},\vareps{3})$ is accurate enough
for posteriors and evidences, we have probed its multidimensional
shape using the nested sampling algorithm
{\POLYCHORD}~\cite{2015MNRAS.453.4384H, Handley:2015fda} (running with
$40000$ live points). It has been cross-checked with another nested
sampler, {\MULTINEST}~\cite{Feroz:2008xx}, ran with $100000$ live
points. Using nested sampling, as opposed to MCMC, allows us to probe
the tails of $\calLeff$ thereby testing if the neural network provides
a proper fit in regions not having many samples to start with. In
addition, the nested sampling gives accurate information on the
minimal values reached by $\calLeff$ and these values are also the
lower bound of the minimal evidence ratios reachable with our
effective likelihood. We find $\calLeffmin \simeq e^{-11}$ (for a
maximum likelihood $\calLeffmax = 1$). Let us mention that we
extrapolate $\calLeff$ outside its prior definition by a constant
value along the $\log(\vareps{1})<-5$ direction, realistic
inflationary models could indeed generate very low tensor-to-scalar
ratios. Moreover, within the strict slow-roll prior, we have to deal
with a non-vanishing $\calLeff$ along the $\vareps{3}$ direction. We
have made the most conservative choice of extrapolating $\calLeff$ by
a constant value for $\abs{\vareps{3}}>0.2$. For the extended prior
choice, there is no need to do any extrapolation as the likelihood
vanishes outside the prior of $\vareps{3}$.

In \Figs{fig:bayesinf_sr3_1D,fig:bayesinf_sr3_2D} we compare the one-
and two-dimensional posteriors obtained by nested-sampling the
machine-learned $\calLeff$ (green curves) to the ones of
\Figs{fig:sr3ndlog_1D,fig:sr3ndlog_2D}, the latter having being
obtained by full MCMC marginalization of the exact posteriors. There
is no visible difference. In \Sec{sec:evidences}, we use this
machine-learned $\calLeff$ in \Eq{eq:posteff} to carry out fast parameter
estimation of $\{\bthetainf,\bthetareh\}$ for all slow-roll models
$\calM$ and to compute their evidences $\evid{\bdata}{\calM}$. 

\section{Bayesian model comparison}
\label{sec:evidences}

In this section, we perform parameter estimation and model comparison
for $\nnow$ models of single-field slow-inflation along the lines
discussed in \Sec{sec:analysis}. The theoretical motivations,
derivation of the $\epsV{i}(\phi)$ functions, and various other
technical considerations for each of these models can be found in the
``Opiparous Edition'' of the {\EI} in \Refa{EIopiparous}. Let us stress
that for a given potential in \Refa{EIopiparous}, there are usually
more than one inflationary model, as various different theoretical
setups can yield the same field potential, with different values,
and hence priors, for the primordial parameters $\bthetainf$ and
$\bthetareh$. These priors are listed in \cref{sec:modelpriors}.

Let us summarize how the calculations are performed. Having at our
disposal the reheating
\crefrange{eq:DeltaNstarRrad}{eq:DeltaNstarRreh}, and the {\ASPIC}
library, we can quickly compute $\Pstar(\bthetainf,\bthetareh)$ and
$\epsstar{1,2,3}(\bthetainf,\bthetareh)$ for each of the $\nnow$
models. Plugging these numbers into the machine-learned effective
likelihood $\likeff{\bdata}{\Pstar,\vareps{1},\vareps{2},\vareps{3}}$,
it is, in principle, straightforward to derive the posteriors of the
primordial parameters $\post{\bthetainf,\bthetareh}{\calM}$ as
well as the evidence $\evid{\bdata}{\calM}$ using
\cref{eq:eviddef,eq:posteff}.

For this purpose, we have developed a modern Fortran code, named
{\BAYASPIC}, which is interfacing the machine-learned likelihood to
the {\ASPIC} library. This code is also a nested sampler which is
using {\POLYCHORD} or, as a user choice, {\MULTINEST}. The {\BAYASPIC}
code also deals with the prior choices and is in charge of scheduling
computations over the $\nnow$ models. For this purpose, it is
parallelized using the Message Passing Interface (\texttt{MPI}) and
this allows us to perform nested sampling on many models at the same
time. The Hubble-flow parameters can take quite different amounts of
computing time to be evaluated, thereby triggering some differences in
the running times for each of the inflationary models. The
fastest-to-compute models, sampled with ${\POLYCHORD}$ and $20000$
live points, are processed in about ten minutes of CPU-time. The
requested precision on their evidence is set at $\Ztol=10^{-3}$, and
this generates chains containing a few hundred thousand samples. The
slowest to compute is named Dual Inflation (see \cref{sec:diprior}), a
model for which none of the slow-roll calculations can be done
analytically, and it takes two days for the nested sampling to reach
the same precision. At the end of the day, even if straightforward,
the task is numerically challenging and the amount of data produced
under the form of nested-sampled chains is about $300\,\GB$.

These chains are then analyzed using a code we have developed, named
{\INFDISTBAYES}, coded as various Python modules and using some
functions built over the {\GETDIST} and {\ANESTHETIC}
codes~\cite{Lewis:2019xzd, Handley:2019mfs}. Analyzing multiple chains
at once is basically parallelized with a Bash script and the output of
{\INFDISTBAYES} are $\nnow$ one- and two-dimensional posteriors,
figures, and various statistical outputs, including the Bayesian
evidences and the Bayesian dimensionality~\cite{Handley:2019pqx}.

As detailed in \Sec{sec:gain}, the code {\INFDISTBAYES} is also used
to compute the information gain on the reheating parameter (or any
other parameters) directly, i.e., by integrating the marginalized
posteriors. In order to do so, one needs to know the prior
distributions $\prior{\bthetainf,\bthetareh}$, which are not all
analytically encoded. As such, to avoid any bias due to badly determined
priors, we have actually performed two nested-sampling explorations of
the $\nnow$ models: one with the actual $\calLeff$ discussed above,
and another one with a constant $\calLeff=1$ which ensures that we are
sampling the priors. In order to cross-check our results, we have also
duplicated these two runs by using {\MULTINEST} instead of
{\POLYCHORD}, and, we have again duplicated everything by starting
from another $\calLeff$ machine-learned over the Hubble-flow
posteriors associated with the extended prior on $\vareps{3}$ (see
\Sec{sec:srposteriors}). All in all, $2\,\TB$ of data are to be dealt
with.
  
\subsection{Priors and model space}
\label{sec:modelspace}
For each model $\calM$, we have to choose the prior probability
distribution for the primordial parameters $\bthetainf$ and the
reheating parameter $\bthetareh$. For the former, they are chosen
according to the underlying theoretical model presented in
{\EI}~\cite{EIopiparous}. These ones are detailed, model by model, in
\cref{sec:modelpriors} for the $\nadd$ new models added in the
``Opiparous Edition'' while the remaining $\nprev$ models have been
set with the exact same priors presented in \Refa{Martin:2013nzq} (see
Appendix A). Let us notice that, compared to \Refa{Martin:2013nzq}, we
have dropped $\nignore$ models that were either extremely fine-tuned
and strongly disfavored, or, extrapolated in disfavored slow-roll
violating regions (see \Sec{sec:modelpriors}).

As discussed in \Sec{sec:rehparam}, the reheating occurring after
inflation enforces that $\rhoreh\le\rhoend$ while we do certainly want
the mean equation of state of the universe during that period to
satisfy $-1/3 <\wrehbar < 1$, the lower bound ensuring that the
universe has really stopped inflating. Even though the energy scale at
which the reheating ended is unknown, not spoiling Big-Bang
Nucleosynthesis demands $\rhoreh > \rhonuc$ and we have set a safe
lower limit at $\rhonuc^{1/4} = 10 \, \MeV$. From \Eqs{eq:Rradwrho},
having $\rhoend < \Mp^4$, these numbers yield a maximally extended
prior for the reheating parameter $\Rrad$, which is a flat distribution
for $\ln\Rrad \in [-46,15]$. This is indeed consistent with our
complete ignorance about the order of magnitude of the values of
$\Rrad$. However, as motivated in \Sec{sec:rehparam}, it is far more
accurate to sample the reheating kinematics using the rescaled
parameter $\bthetareh=\ln \Rreh$ defined by \Eq{eq:Rrehdef}. The prior
range for $\ln \Rreh$ is also confined within $[-46,15]$ but, because
$\rhoend$ depends on $\bthetainf$, some values within this range are actually
excluded on a model-by-model basis and the actual prior of $\ln\Rreh$
is correlated with $\prior{\bthetainf}$. In practice, we start by
sampling from the flat prior distribution $\ln\Rreh\in[-46,15]$ and,
then, add a hard prior during the computation which ignores all values
$\ln\Rreh > 15 + (1/3)\ln[\rhoend(\bthetainf)/\Mp^4]$.

\subsection{Posteriors and evidences}
\label{sec:evid}

\begin{figure}
\begin{center}
\includegraphics[width=\bigfigw]{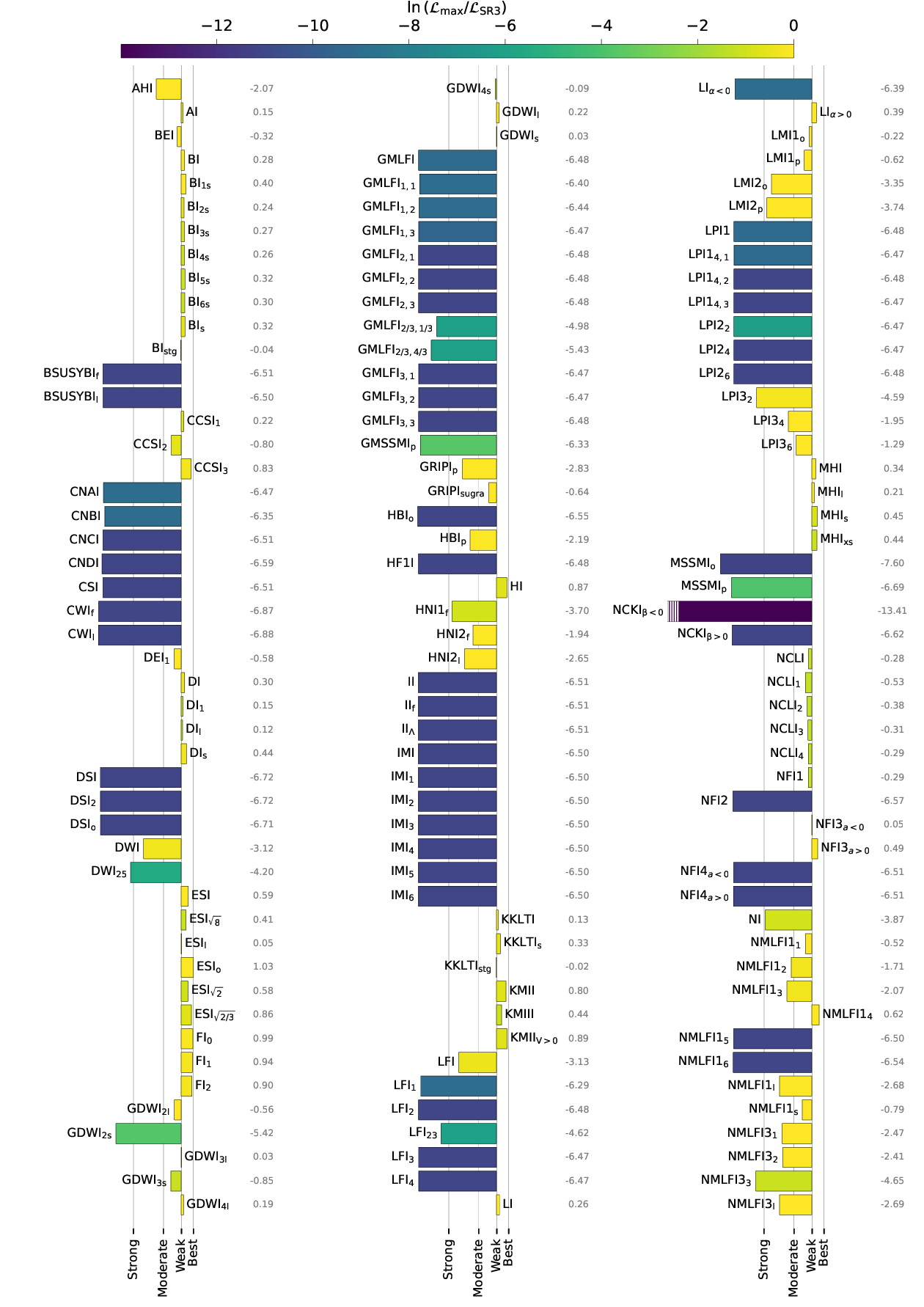}
\caption{Logarithm of the Bayes factors $\Bref{\calM}$, in reference
  to the Bayesian evidence of the slow-roll third-order power
  spectra. The length of each bar corresponds to the logarithm of the
  Bayesian evidence, whose value is also displayed in light gray. The
  color scale shows the best likelihood, again in reference to best
  fit of the slow-roll spectra. The color scale is also the best
  possible Bayes factor the model could get, would the prior be
  associated with a Dirac-distribution at the best fit. Continued in
  \Fig{fig:bestevid_2}.}
\label{fig:bestevid_1}
\end{center}
\end{figure}

\begin{figure}
\begin{center}
\includegraphics[width=\bigfigw]{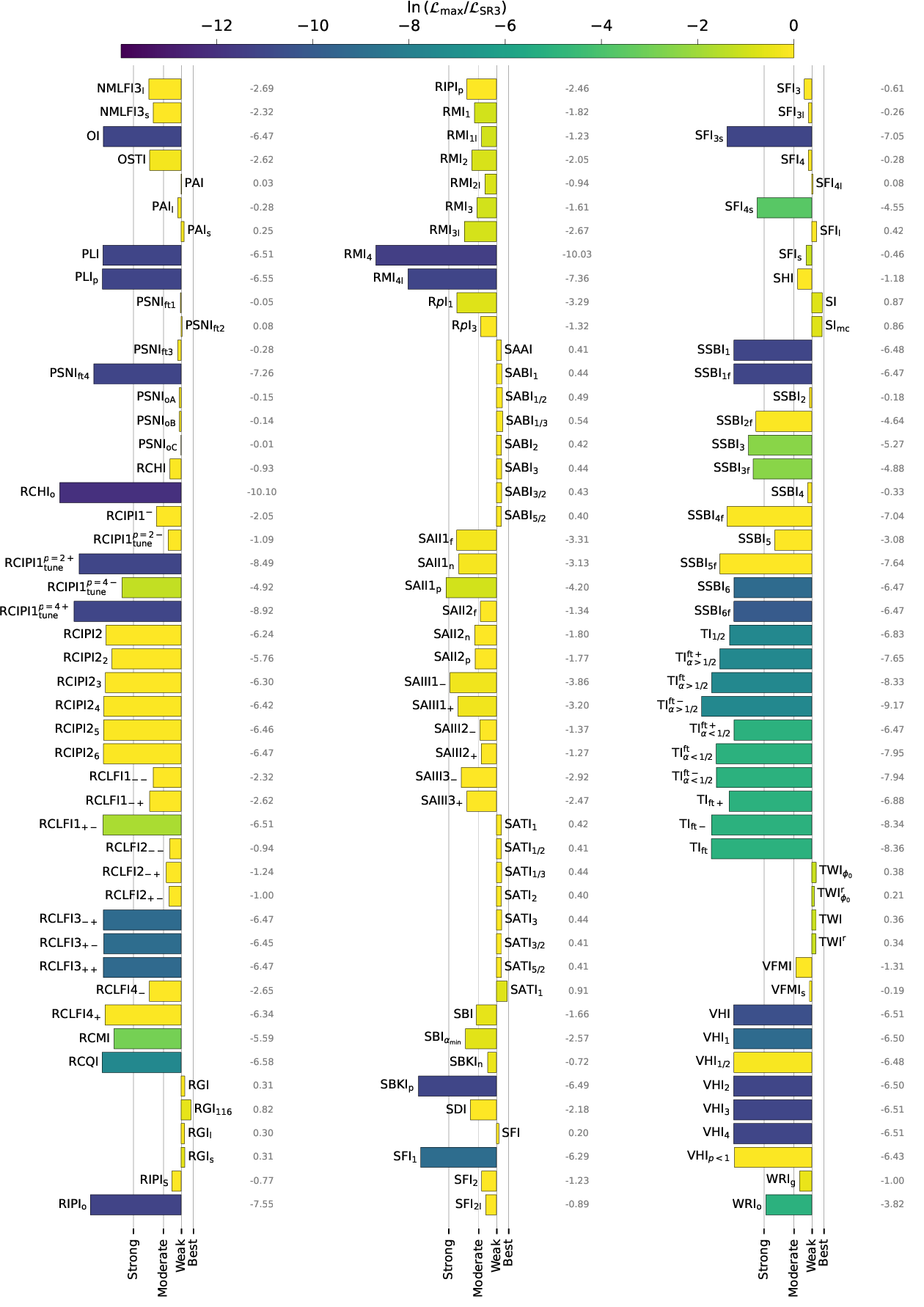}
\caption{Logarithm of the Bayes factors $\Bref{\calM}$, in
  reference to the Bayesian evidence of the slow-roll third-order
  power spectra. \Fig{fig:bestevid_1} continued.}
\label{fig:bestevid_2}
\end{center}
\end{figure}

\begin{figure}
\begin{center}
\includegraphics[width=\onefigw]{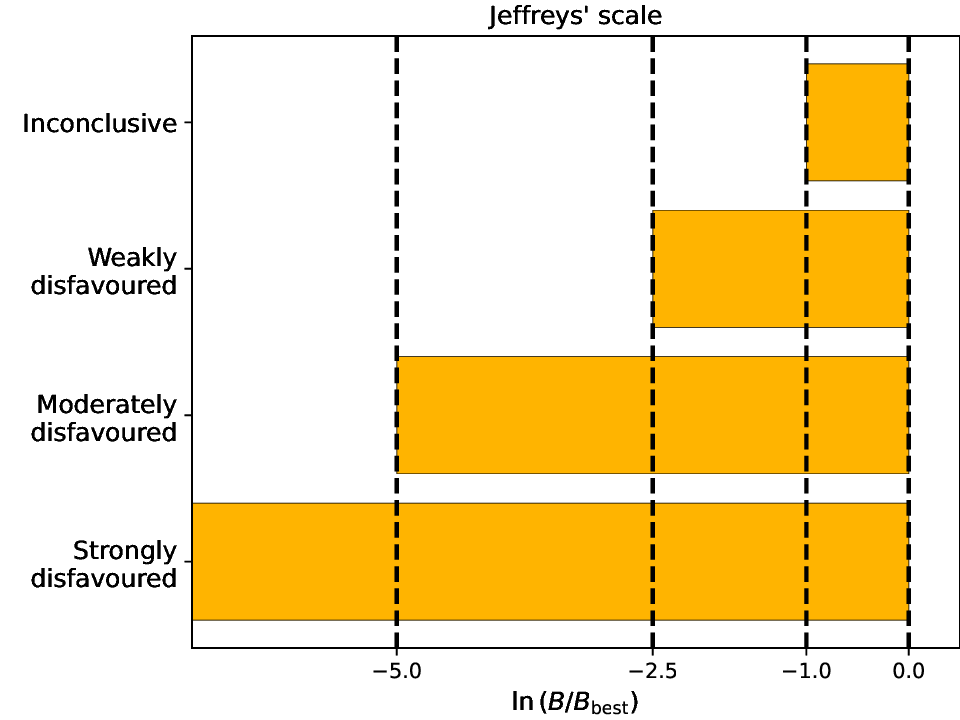}
\caption{Jeffreys' scale for measuring the strength of evidence
  when comparing a model having a Bayes factor $\BayesFactor{}{}$ to
  the best $\BayesFactor{}{\mathrm{best}}$. This scale appears as
  vertical gray lines in \cref{fig:bestevid_1,fig:bestevid_2}.}
\label{fig:jefscale}
\end{center}
\end{figure}

The posterior probability distributions
$\post{\bthetainf,\bthetareh}{\bdata}$ for a subset of the $\nnow$
tested models have been plotted in \cref{sec:modelposteriors}. We have
limited ourselves to the most interesting best models according to the
criteria discussed in this appendix. All the other posteriors will be
made available online~\cite{urlci}.

In \cref{fig:bestevid_1,fig:bestevid_2}, we have reported the natural
logarithm of the Bayes factors of all these inflationary models as
horizontal bars. The Bayes factors are in reference to the Bayesian
evidence of slow roll itself, i.e.,
\begin{equation}
\Bref{\calM} \equiv \dfrac{\evid{\bdata}{\calM}}{\evid{\bdata}{\calMref}}\,,
\end{equation}
where $\calMref=I$ denotes the so-called ``slow-roll model''. This one
is defined as the model for which the primordial power spectra are
given by \cref{eq:calPzsr3ast,eq:calPhsr3ast} with the strict (or,
possibly, extended) slow-roll prior $\vareps{i}\in[-0.2,0.2]$ and
$\ln\left(10^{10}\Pstar\right)\in [2.4,3.6]$. Its evidence is given by
the complete marginalization of the full posterior discussed in
\Sec{sec:srposteriors}
\begin{equation}
\evid{\bdata}{\calMref} = \int
\post{\bdata}{\bthetastd,\Pstar,\vareps{1},\vareps{2},\vareps{3},\vareps{4}}
\prior{\bthetastd,\Pstar,\vareps{1},\vareps{2},\vareps{3},\vareps{4}}
\, \ud \bthetastd
\ud \Pstar \ud \vareps{1} \ud \vareps{2} \ud \vareps{3} \ud \vareps{4},
\end{equation}
and we find (strict slow-roll prior)
\begin{equation}
\ln\left[\evid{\bdata}{\calMref}\right] = -4.525 \pm 0.008.
\label{eq:evidref}
\end{equation}
Assuming non-committal model priors $\prior{\calM_i}=1/\nnow$, the
Bayes factors immediately give us the probability of all models to
explain the data
\begin{equation}
\post{\calM}{\bdata} = \dfrac{\Bref{\calM}}{\sum_i \Bref{\calM_i}}\,.
\end{equation}
As such, left-extended bars in \cref{fig:bestevid_1,fig:bestevid_2}
are associated with models $\calM$ that are less probable than
$\calMref$ to explain the data sets $\bdata$ whereas models having a
bar extended to the right are more probable. We have also reported in
these figures the Jeffreys' scale, as gray vertical lines, evaluating
the strength of evidences between the model under scrutiny $\calM$ and
the best of all $\calMmax$. It belongs to the Exponential SUSY
Inflation scenarios and appears as $\mathrm{ESI}_\mathrm{o}$ in the
figures with
\begin{equation}
\ln\left[\evid{\calMmax}{\bdata}\right] = -3.49 \pm 0.01.
\end{equation}
The gray vertical lines are the boundaries of the four categories
associated with the Jeffreys' scale discussed in
\Refs{Trotta:2008qt,Gordon:2007xm} and summarized in
\cref{fig:jefscale}. All models having $\ln\left(\Bbest{\calM}\right) >
-1.0$, having an horizontal bar with an edge right of the gray line
labeled ``Best'', are as good as $\mathrm{ESI}_\mathrm{o}$ in
explaining the data sets $\bdata$. On the contrary, all models having
$\ln\left(\Bbest{\calM}\right) < -5.0$ are ``strongly disfavoured''
compared to the bests. Let us remark that the lowest Bayes factors
saturate at a relatively large negative value. This saturation is
artificial and related to our numerical limitation of having
$\ln(\calLeffmin)=-11$ for the effective likelihood. As such, it is
very well possible that these models would actually have even worse
Bayes factors.

As the figures show, quite a large number of models are landing in the
``strongly disfavoured'' category and it is safe to say that these are
ruled-out. However, there are also quite a significant number of
models being in the ``best'' (or inconclusive) region and, for those,
it is interesting to determine what it takes for them to fit the data so well.
In \Refa{Martin:2013nzq}, we had used the so-called
Bayesian Complexity~\cite{Kunz:2006mc} to assess how many effective
parameters were needed for each model to fit the data. In the next
section, in the same spirit, we derive their Bayesian dimensionality
as introduced in \Refa{Handley:2019pqx}.

\subsection{Bayesian dimensionality}
\label{sec:dim}

\begin{figure}
\begin{center}
  \includegraphics[width=\bigfigw]{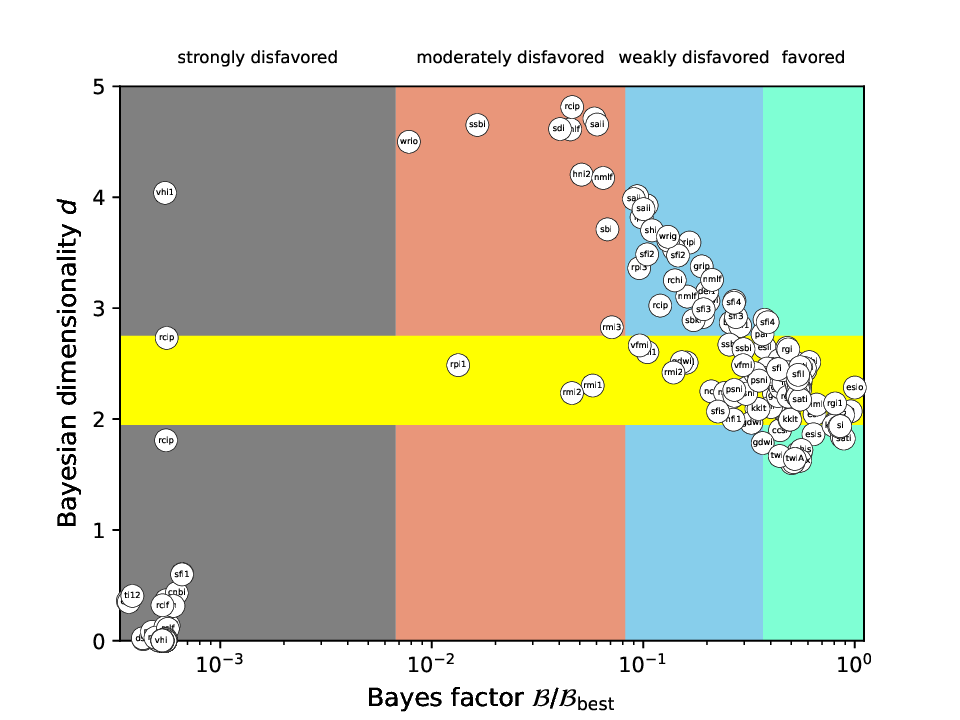}
  \includegraphics[width=\bigfigw]{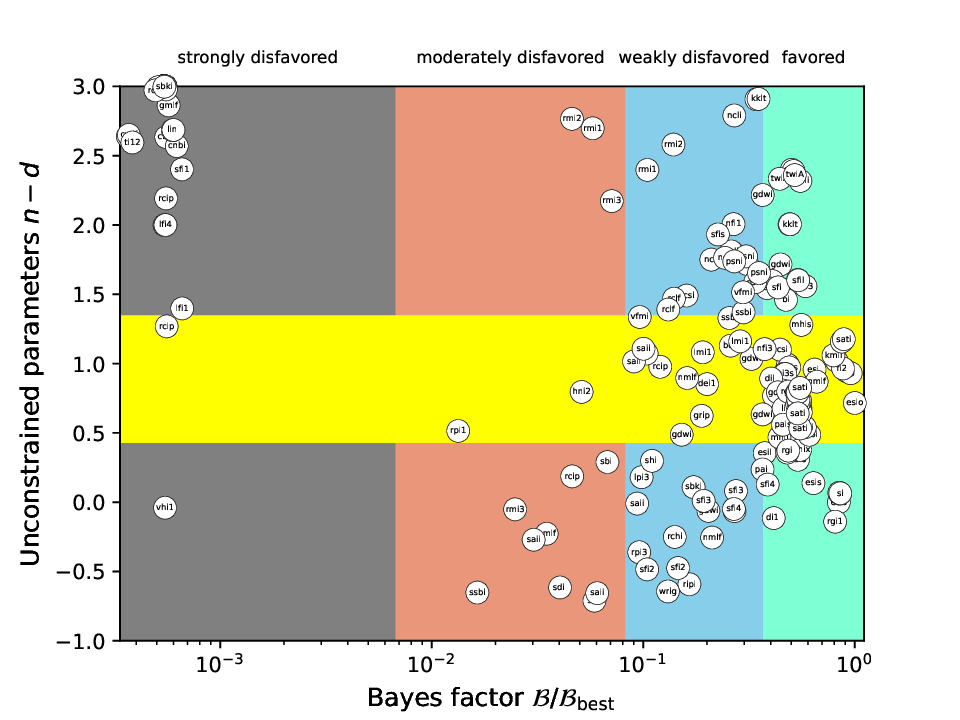}
\caption{Bayesian dimensionality $d$ (top panel) and number of
  unconstrained parameters $n-d$ (bottom panel), plotted against Bayes
  factors $\Bbest{}$ for the $\nnow$ models analyzed ($n$ is the number
  of model parameter). The yellow region is centered over the mean
  value of $d$ (top) and $n-d$ (bottom), with a width given by its
  standard deviation (weighted according to the model probability).}
\label{fig:dnfree}
\end{center}
\end{figure}

The Bayesian dimensionality is a measure of the number of parameters
that are constrained by the posterior probability
distribution. Introducing the total information gain $\DKLtot$, namely the
Kullback-Leibler divergence between the prior and the posterior
integrated over all parameters~\cite{kullback1951}, we have
\begin{equation}
\DKLtot = \int \post{\btheta}{\bdata} \ln\left[
\dfrac{\post{\btheta}{\bdata}}{\prior{\btheta}} \right] \ud \btheta,
\end{equation}
where we have used the shorthand notation $\btheta =
\{\bthetastd,\bthetainf,\bthetareh\}$. The Bayesian dimensionality is
defined by~\cite{Handley:2019pqx}
\begin{equation}
d = 2 \int \post{\btheta}{\bdata} \left\{ \ln\left[
  \dfrac{\post{\btheta}{\bdata}}{\prior{\btheta}} \right] - \DKLtot
\right\}^2 \ud \btheta.
\end{equation}
In terms of the Shannon's information
\begin{equation}
\calI =  \ln\left[ \dfrac{\post{\btheta}{\bdata}}{\prior{\btheta}}  \right],
\end{equation}
one sees that $\DKLtot = \mean{\calI}$ is the first moment of $\calI$
over the posterior while $d/2= \mean{\calI^2} - \mean{\calI}^2$ is
related to its variance. As discussed at length in
\Refa{Handley:2019pqx}, $d$ has many features wanted for being used as
a measure of the effective number of parameters. For instance, it has
almost no dependence on the prior distribution, it is equal to unity
for a one-dimensional Gaussian posterior and vanishes for a flat
posterior. It is also accurately computable by nested-sampling and we
have used the Python package {\ANESTHETIC} (incorporated within
{\INFDISTBAYES})~\cite{Handley:2019mfs} to determine the value of $d$
for the $\nnow$ models analyzed here.

In \cref{fig:dnfree}, all models have been represented by a circle in
the plane $(\Bbest{},d)$. There is a clear correlation between Bayes
factor and Bayesian dimensionality, the latter increasing for the most
disfavored models. This is expected as, for models that poorly fit the
data, the posterior is forced to peak in the small prior corners that
are closest to (but still far from) the favored regions. In this
sense, but in this sense only, they are well constrained. As such,
Bayesian dimensionality alone should be interpreted with care: data
sets can indeed be very constraining for models providing a very bad
fit. Let us also notice the few strongly disfavored models with Bayes'
factors slightly less than $10^{-3}$. Their Bayes' factors are not
accurate and their apparent vertical alignment comes from our flat
extrapolation of the effective likelihood along the $\vareps{3}$
direction for values outside the strict prior range $[-0.2,0.2]$. In
particular, we have checked that, using the $\calLeff$ associated with
the extended prior range $\vareps{3}\in[-1,1]$, these models end up
being displaced to much lower evidences, as expected.

For all models, we also keep a record of their number of free
parameters $n$, namely the total number of $\bthetainf$ and
$\bthetareh$. As such, $n-d$ is, for a given model, a measure of the
number of unconstrained parameters. Let is notice that, for sharply
peaked posteriors, $d$ can be greater than
$n$~\cite{Handley:2019pqx}. In these cases, the number of ``free
parameters'' can become negative and this simply signals that the
posteriors of the $n$ model parameters are sharper than Gaussian
distributions. In the bottom panel of \cref{fig:dnfree}, we have
positioned all models in the plane $(\Bbest{},n-d)$. Looking at the
most probable models, we find scenarios having typically one
unconstrained parameter, and scenarios having none; the latter being
the most economical models explaining well the data.

\subsection{Information gain on the reheating}
\label{sec:gain}

\begin{figure}
\begin{center}
  \includegraphics[width=\bigfigw]{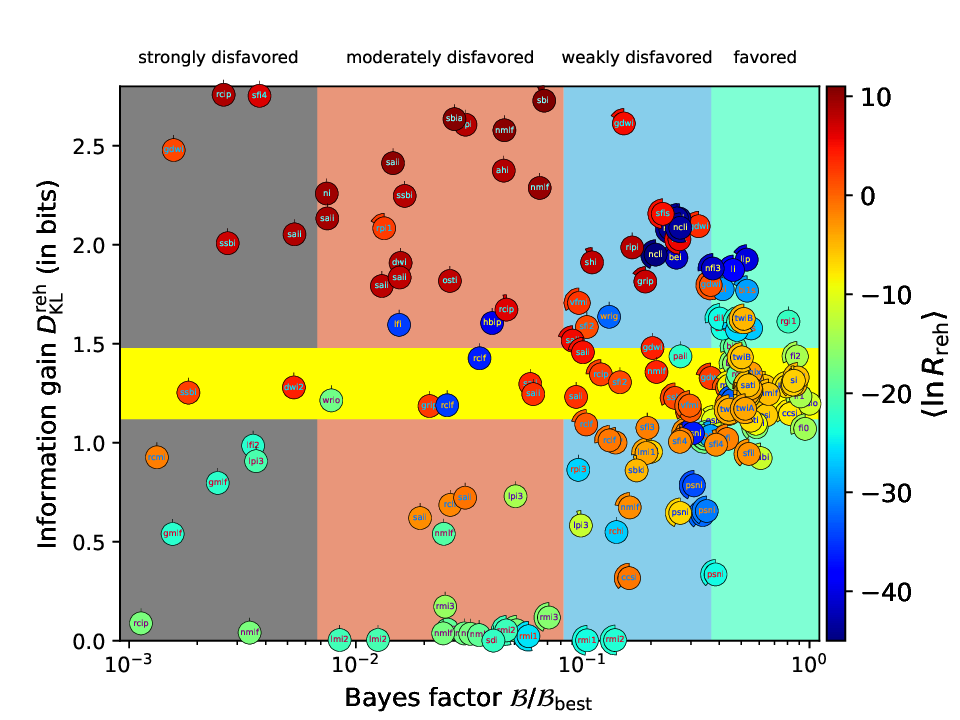}
  \includegraphics[width=\bigfigw]{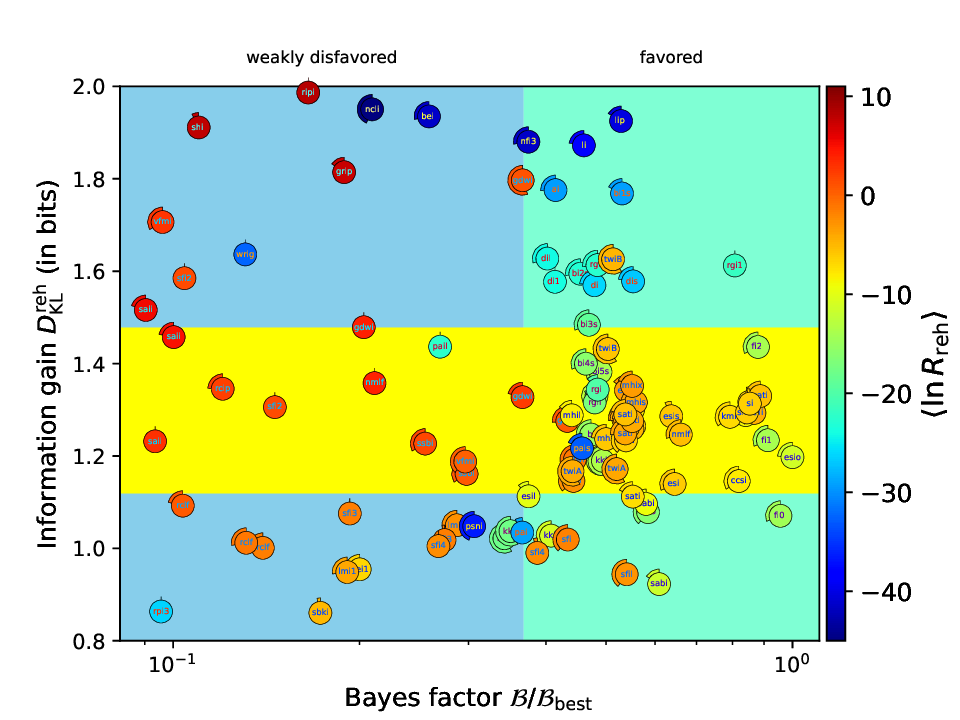}
\caption{Information gain on the rescaled reheating
  parameter $\ln\Rreh$ as a function of the Bayes' factors. The color
  scale displays the mean value of $\ln\Rreh$ over its posterior
  while the anti-clockwise gauge circling all models measures the
  fraction of unconstrained model parameters $\max\left[(n-d)/n,0\right]$.}
\label{fig:Dklreh}
\end{center}
\end{figure}

Many favored models have less than one unconstrained parameter and
this implies that the data are constraining the kinematics of the
reheating era~\cite{Martin:2010kz, Easther:2011yq, Martin:2014nya,
  Dai:2014jja}. This can be quantified by computing the information
gain, i.e., the Kullback-Leibler divergence, between the prior and the
posterior, of the reheating parameter, $\ln\Rreh$ here. Defining
\begin{equation}
\DKLreh = \int \post{\ln\Rreh}{\bdata} \ln\left[
\dfrac{\post{\ln\Rreh}{\bdata}}{\prior{\ln\Rreh}} \right] \ud \ln\Rreh,
\end{equation}
we have plotted in \Fig{fig:Dklreh} all models in the plane
$(\Bbest{},\DKLreh)$. Each scenario is represented by a colored
circle, the color of which gives the mean value $\mean{\ln\Rreh}$ over
its posterior $\post{\ln\Rreh}{\bdata}$. Each circle is enlarged by an
anti-clockwise gauge displaying $(n-d)/n$, i.e., the percentage of
unconstrained model parameters. As such, when this extra-shell
collapses to a single tick, it means that all model parameters are
constrained (or over-constrained with $n-d<0$). When the shell closes
around, it means that all model parameters are unconstrained. As
already noticed in \Sec{sec:dim}, models providing a ``bad fit'' to
the data have essentially all their model parameters constrained and
this gauge is most relevant for the best models.

The horizontal yellow band in \Fig{fig:Dklreh} is centered around the
mean value of $\DKLreh$, and has a width given by its standard
deviation. These have been calculated according to the posterior probability
of all models to explain the data and we get (in bits)
\begin{equation}
\mean{\DKLreh} = \sum_i \post{\calM_i}{\bdata} \DKLreh(\calM_i)
\simeq 1.3,
\end{equation}
with
\begin{equation}
\sqrt{\mean{\left(\DKLreh\right)^2} - \mean{\DKLreh}^2} \simeq 0.36.
\end{equation}
These figures have increased by more than a factor of two compared to
the first Planck data release, which is a significant
achievement. Indeed, according to \Refa{Martin:2016oyk}, one had
$\mean{\DKLreh} \simeq 0.55 \pm 0.14$ for Planck 2013, $\mean{\DKLreh}
\simeq 0.82 \pm 0.13$ for Planck 2015 with BICEP/KECK data. Because
the reheating parameter is constrained for all best models, by at
least one bit (see \Fig{fig:Dklreh}), inflationary model building is
no longer about deriving an inflationary potential only: the reheating
kinematics also needs to be specified.  In other words, at exact same
field potential and background evolution during inflation, two
different reheating histories will yield two different Bayesian
evidences: they can be distinguished by present cosmological data.

\section{Conclusion}
\label{sec:bests}

\begin{figure}
\begin{center}
\includegraphics[width=\onefigw]{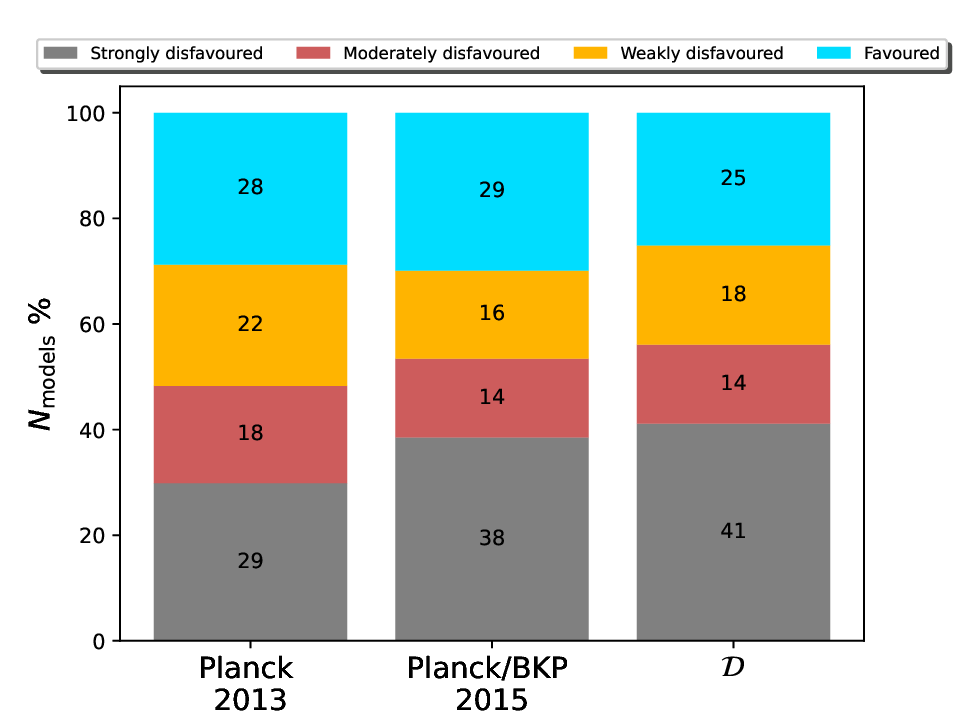}
\caption{Probability distribution of the models within the four
  Jeffreys' categories with respect to the three cosmological data
  sets used within the past decade in \Refa{Martin:2013nzq},
  \Refa{Martin:2016oyk} and the present work $\bdata$. The
  constraining power of data is winning against the increase of
  theoretical inflationary models.}
\label{fig:counting}
\end{center}
\end{figure}

We have presented a comprehensive Bayesian data analysis, and model
comparison, within the landscape of nearly 300 models of single-field
slow-roll inflation. Such an analysis produces a large amount of
information, the posterior probability distributions of all the
model's parameters have indeed been computed, a sub-sample of which
being presented in \Sec{sec:modelposteriors}. As such, we could very
well discuss, model per model, the theoretical consequences of the
preferred values of their parameters. Instead, in this paper, we have
chosen to discuss global features, determined by averaging over the
landscape various properties shared by all models and weighted by the
probability of each model to explain the data
$\post{\calM}{\bdata}$. One of the main result is the significant
boost in sensitivity of current cosmological data to the kinematics of
the reheating era. Most of the current proposed models of inflation
are loosely making predictions on how long, and with which equation of
state, the reheating era proceeded. As we have detailed in
\Sec{sec:rehparam}, the observable consequences of the reheating
kinematics can be fully encoded in the rescaled reheating parameter
$\Rreh$. In the absence of specific information on its value, starting
from the most uninformative prior (see \Sec{sec:modelspace}), we find
that the present data give us a posterior with $1.3$ bits of
information, more than a two-fold increase within a decade of CMB and
LSS measurements.

It is also informative to compare the distribution of models within
the four Jeffreys' categories of evidences. As represented in
\Fig{fig:counting}, in spite of the inclusion of $\nadd$ new models in
the present work, we find that the constraining power of the data is
slightly pushing more and more models into the disfavored
regions. Let us remark that many of these new $\nadd$ models have been
proposed \emph{after} the Planck 2013 data release. Still, many of
them are actually not able to explain the data better than various
simple models proposed well before. For one part, this comes from the
inclusion of new model parameters, that can penalize complex models by
the Occam's razor effects intrinsically present within Bayesian
evidences. For other parts, some of these models still ignore
reheating kinematics, and when these are properly taken into account, as we
have done here, the model predictions are landing out of the data
favored regions.

On another aspect, our results definitely demonstrate that theoretical
realizations of cosmic inflation are testable and can be falsified. As
\Fig{fig:counting} shows, more than $40\%$ of the proposed scenarios
are presently ruled-out. Let us stress than even if the current range
of wavenumbers probed by CMB and LSS data corresponds to a relatively
small window along the inflaton's potential, of about a few {\efolds},
our marginalization over all reheating histories is making this window
moving along the potential by tens of {\efolds}. The Bayesian evidence
for each model is thus sensitive to a large part of the inflationary
potential and strongly disfavoured scenarios are unlikely to be
resuscitated by tweaking their potential outside the CMB window with
the hope of slightly shifting its location. Nevertheless, one may have
a situation in which a strongly disfavoured model is able to fit the
data well but only within a fine-tuned region of its potential (see,
e.g., the $\rcipiTWO$ cases). Such a potential shape might be salvaged
but only within a new theoretical embedding making definite
predictions for the reheating era. The predicted reheating history
should, however, be such that it allows for the fine-tuned region of
the potential to land exactly onto the cosmological window, a
difficult task indeed.

After the first chapter concluded recently, inflation is therefore
ready for the second one where new high-accuracy data will help us to
constrain this scientific theory even more. All the data analysis
pipeline presented here is indeed readily applicable to other
cosmological data sets, and, in particular, we plan to use it for the
soon to be released Euclid LSS data. Clearly, probing the small length
scales is expected to provide tighter bounds on $\vareps{3}$, and this
will propagate into disfavoring models having too large running of the
spectral index. Finally, we hope the future LiteBIRD data will clarify
the curious remaining excess in $B$-modes observed in the present
work.

\section*{Acknowledgments}
We are indebted to Steven Gratton and Erik Rosenberg for having
provided us with their latest {\CAMSPEC} likelihood module for {\COSMOMC}.
This work is supported by the ESA Belgian Federal PRODEX Grant
$\mathrm{N^{\circ}} 4000143201$. Computing support has been provided
by the CURL cosmo development cluster and the Center for High
Performance Computing and Mass Storage (CISM) at UCLouvain.

\appendix

\section{Posteriors for the non-primordial parameters}
\label{sec:nuisances}

\begin{figure}
\begin{center}
\includegraphics[width=0.95\textwidth]{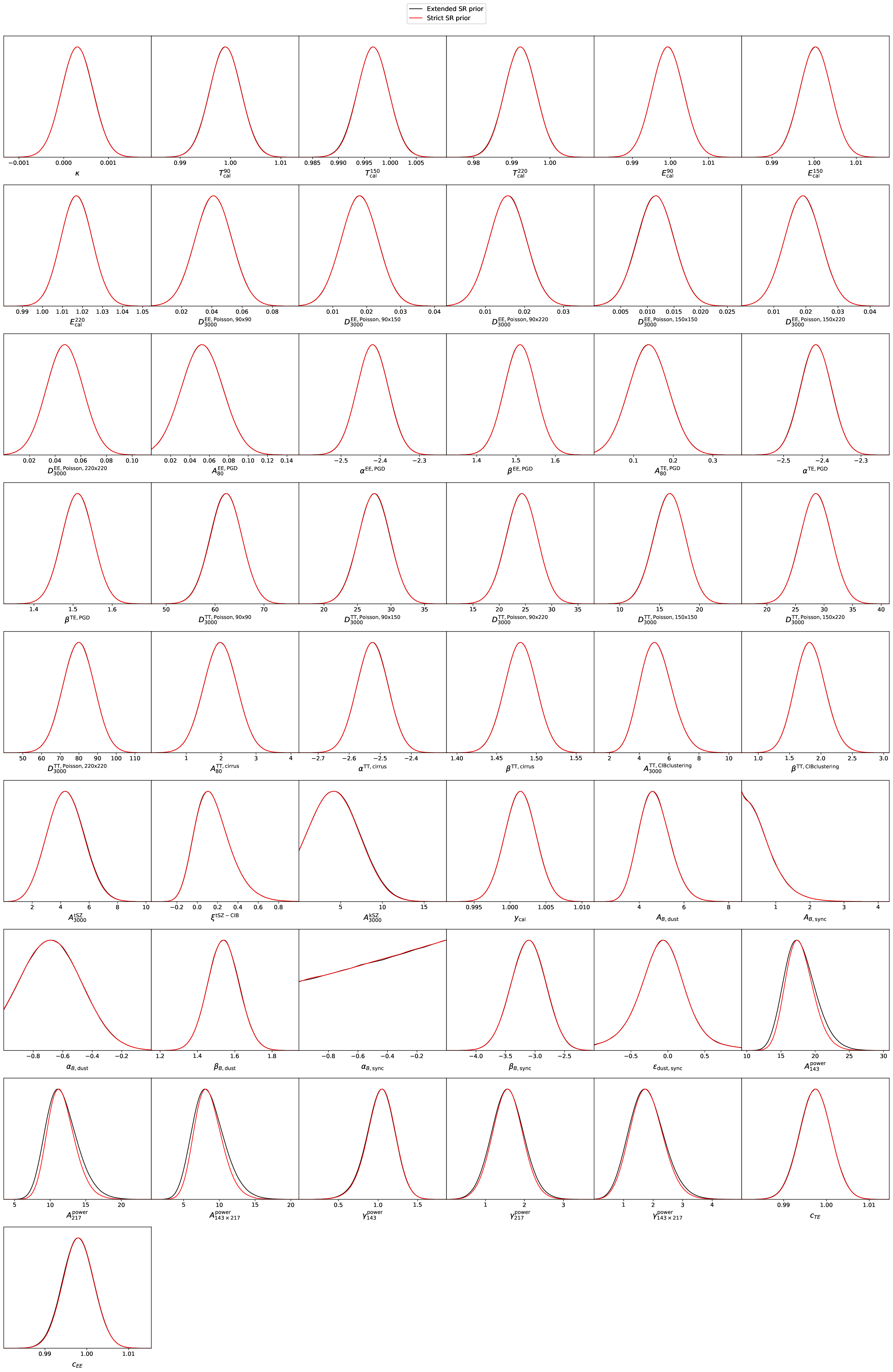}
\caption{One-dimensional posteriors for the $49$ non-cosmological, nor
  primordial, parameters for both the strict and extended slow-roll
  priors.}
\label{fig:nuisances_1D}
\end{center}
\end{figure}

The MCMC exploration of the full parameter space, as performed in
\Sec{sec:srposteriors}, is necessary to determine the effective
likelihood defined in \cref{eq:Leff}. In \Sec{sec:srposteriors}, we
have presented only the posterior probability distributions for the
cosmological and primordial parameters. These have been obtained by
marginalizing over the remaining $49$ astrophysical and instrumental
parameters, whose posteriors have been plotted in
\Fig{fig:nuisances_1D}. The definitions, priors, and notations for
these are identical to the ones of \Refs{Planck:2018lbu,
  Efstathiou:2019mdh, Planck:2020olo, Rosenberg:2022sdy}.

\section{Prior choices for the new models}
\label{sec:modelpriors}

In the following, we summarize the prior probability distributions
chosen for the primordial parameters, generically referred to as
$\cte{1}$, $\cte{2}$, $\cte{3}$\dots associated with the $\nadd$ new
inflationary models $\calM$ that were not part of the models
discussed in the Appendix A of \Refa{Martin:2013nzq}. As mentioned in
\Sec{sec:modelspace}, we have dropped $\nignore$ models from the
original batch of \Refa{Martin:2013nzq}. These are strongly disfavored
models associated with extreme fine-tuning: $\gmssmiopA$,
$\gmssmiomA$, $\gmssmiopB$, $\gmssmiomB$, $\gmssmiem$, $\gmssmiep$,
$\gripiopA$, $\gripiomA$, $\gripiopB$, $\gripiomB$, $\gripiep$,
$\gripiem$, $\tiem$, $\tiep$, $\tie$, and, $\rpiTWO$. We have also put
aside three favored, but quite peculiar models, that were exploring
slow-roll violating regions for some values of their parameters. They
are $\psniepA$, $\psniepB$ and $\psniepC$, the evidences of which were
analytically extrapolated based on three other slow-roll models (still
included): $\psnioA$, $\psnioB$ and $\psnioC$, respectively. As shown
in \Refa{Martin:2013nzq}, the evidences of the former are only slightly
reduced with respect to the latter. The interested reader may find the
analytical formulas to derive the evidences of $\psniepA$, $\psniepB$,
$\psniepC$ in the Appendix A of \Refa{Martin:2013nzq}, see Eq.~(A72). Let
us stress that the drop of these $\nignore$ models has been
consistently taken into account in the comparison made in
\Fig{fig:counting}.

The new models presented below are introduced with a minimal amount of
information to justify the prior choices: all the physical and
technical details, for each of these scenarios, can be found in the
``Opiparous Edition'' of {\EI} in \Refa{EIopiparous}.

\subsection{Axion Hilltop Inflation (AHI)}
\label{sec:ahiprior}

Axion Hilltop Inflation has a potential of the form
\begin{equation}
V(\phi) = M^4\left[\nu_0 - 2 \cos\left(\frac{\phi}{f}\right) +
  \left(\pi-\frac{\phi}{f}\right)
  \sin\left(\frac{\phi}{f}\right)\right],
\end{equation}
which involves one parameter $f$, a typical vacuum expectation value,
whose order of magnitude is unknown. We have therefore
chosen a Jeffreys' prior, namely an uniform prior of the logarithm of
$f$.
\begin{center}
  \begin{priortable}{1}
    $\ahi$ & $\log(f/\Mp) \in [-3,3]$  \\
  \end{priortable}
\end{center}

\subsection{Cublicly Corrected Starobinsky Inflation (CCSI)}
\label{sec:ccsiprior}

This is a modified gravity model in which a cubic power of the Ricci
scalar is added as a correction to the Starobinsky model. In the
Einstein frame, the potential reads
\begin{equation}
V(\phi) = \dfrac{\Mg^2 \mu^2}{2} \left(1 - e^{-y} \right)^2
\dfrac{1+\sqrt{1+3\alpha\left(e^y-1\right)} + 2\alpha \left(e^y-1
  \right)}{\left[1 + \sqrt{1 + 3\alpha\left(e^y - 1\right)}\right]^3}\,,
\end{equation}
where $y=\sqrt{2/3}\,\phi/\Mg$ and $\Mg \simeq \Mp$. There is one
small parameter $\alpha$, and setting $\alpha=0$ gives back the
Starobinsky model, see \Sec{sec:siprior}. Such a potential generates
three distinct inflationary regimes, $\ccsiONE$, $\ccsiTWO$,
$\ccsiTHREE$. One of them, $\ccsiTWO$, is quite fine-tuned as
requiring a very small coupling constant, an external mechanism to
stop inflation (for instance, a tachyonic instability), and some
tuning of the field value at which this mechanism occurs to be viable:
$\phiend > \phiendmin$ where $\phiendmin$ is set to ensure that
inflation lasts, at least, for $120$ {\efolds}. As such, we have
analyzed these three different scenarios separately.

\begin{center}
  \begin{priortable}{2}
    $\ccsiONE$ & $\log(\alpha)\in[-6,0]$ & -  \\
    $\ccsiTWO$ & $\log(\alpha)\in[-6,-3]$  & $\phiend/\phiendmin \in [1,5]$  \\
    $\ccsiTHREE$ & $\log(\alpha)\in[-6,0]$ & -  \\
  \end{priortable}
\end{center}

\subsection{Double Exponential Inflation (DEI)}
\label{sec:deiprior}
The potential of this phenomenological model is given by
\begin{equation}
V(\phi) = M^4 \left(\ee^{\beta\frac{\phi}{\phizero}}-\beta^2\ee^{\frac{1}{\beta}\frac{\phi}{\phizero}}\right),
\end{equation}
where the two exponential terms are set to generate a large hilltop
region. This is a two-parameter model with $\phizero>\Mp$, the order
of magnitude being unknown, and $0<\beta<1$.

\begin{center}
  \begin{priortable}{2}
    $\deiONE$ & $\beta \in [0,1]$ & $\log(\phizero/\Mp)\in[0,3]$  \\
  \end{priortable}
\end{center}

\subsection{Dual Inflation (DI)}
\label{sec:diprior}

Dual Inflation is a UV-complete and non-perturbative inflation model
coming from an $N=2$ Supersymmetric Yang-Mills theory. The potential has
no simple from, but can be parameterized as
\begin{equation}
\begin{aligned}
V(m) & = \dfrac{f_0^2 \Lambda^2}{\pi^2} \left\{1 + V_0 - 2 \dfrac{K-E}{m K}
- \dfrac{\pi}{m K K'} \nu^2(m) \Theta[\nu(m)] \right\},\\
\nu(m) & \equiv 1 - \dfrac{8 \sqrt{2}}{\pi^2} \dfrac{\Lambda}{f_0}
\dfrac{K (E'-K')^2}{m^{1/2}}\,,\\
\phi(m) & = \Lambda \dfrac{2\sqrt{2}}{\pi} \int_m^1
\dfrac{\sqrt{K(p^{1/2}) K'(p^{1/2})}}{p^{3/2}} \, \ud p,
\end{aligned}
\label{eq:dipot}
\end{equation}
where $\Theta(x)$ stands for the Heaviside function, $V_0$ is an order
unity uplifting constant (which has to be numerically determined). The
functions $E(x)$, $K(x)$, $E'(x)$ and $K'(x)$ are known as the first-
and second-kind complete elliptic integrals. The potential is
parameterized by $f_0$ and the typical vacuum expectation value
$\Lambda$. The internal consistency of the theory requires
$f_0<\Lambda$ and, as such, we work with the dimensionless parameter
$f=f_0/\Lambda<1$. Dual Inflation is also a model predicting the
amplitude of the CMB anisotropies (the mass scale of the potential is
$M \propto \sqrt{f_0\Lambda}$), $\Lambda(f)$ is thus not free. DI is
finally a one parameter model parameterized by $f$ only. We have
split it into various motivated priors as described in the
following table. 
\begin{center}
  \begin{priortable}{1}
    $\di$ & $\log(f) \in [-5,0]$  \\
    $\dis$ & $\log(f) \in [-5,-2]$  \\
    $\dil$ & $\log(f) \in [-2,0]$ \\
    $\diONE$ & $f=1$ \\
  \end{priortable}
\end{center}

\subsection{Double Well Inflation (DWI)}
\label{sec:dwiprior}

This models are a phenomenological UV-completion of the Small
Field quadratic inflation models ($\sfiTWO$) and have already been
considered in \Refa{Martin:2013nzq}. Their potential reads
\begin{equation}
V(\phi) = M^4 \left[\left(\dfrac{\phi}{\phizero}\right)^2 - 1\right]^2.
\end{equation}
They do significantly differ from $\sfiTWO$ in the super-Planckian
field regime and we have added a missing realization of this effect as
a motivated prior in $\dwiTF$.
  \begin{center}
  \begin{priortable}{1}
    $\dwiTF$ & $\phizero/\Mp=25$ \\
  \end{priortable}
\end{center}

\subsection{Exponential SUSY Inflation (ESI)}
\label{sec:esiprior}
The potential reads
\begin{equation}
V(\phi)=M^4\left(1-\ee^{-q\phi/\Mp}\right),
\end{equation}
with a unique model parameter $q$. This class of models was already
presented in \Refa{Martin:2013nzq}, but a new theoretical realization
was proposed since then, in which $q=\sqrt{8}$~\cite{EIopiparous}. We have
therefore introduced the new model $\esiEIGHT$.

\begin{center}
  \begin{priortable}{1}
    $\esiEIGHT$ & $q=\sqrt{8}$  \\
  \end{priortable}
\end{center}

\subsection{Fiber Inflation (FI)}
\label{sec:fiprior}

This is a model built upon String Theory, where the potential reads
\begin{equation}
  V(\phi)=M^4\left[\left(1+\frac{2}{3}\delta\right)
\ee^{-\frac{4}{\sqrt{3}}\frac{\phi}{\Mp}}
-4\left(1+\frac{\delta}{6}\right)
\ee^{-\frac{1}{\sqrt{3}}\frac{\phi}{\Mp}}
+\frac{\delta}{1+n}
\ee^{\frac{2(1+n)}{\sqrt{3}}\frac{\phi}{\Mp}}+3-\frac{\delta}{1+n}\right].
\end{equation}
Here, $n$ is an integer number and $\delta$ is a small positive
number, of unknown order or magnitude, but related to a string
coupling $q$ by $\delta \propto q^{4(1+n/3)}$. We have therefore
considered the three scenarios listed in the following table.

\begin{center}
  \begin{priortable}{2}
    $\fiZERO$ & $\log(\delta)\in[-8,-4]$ & $n=0$  \\
    $\fiONE$ & $\log(\delta)\in[-32/3,-16/3]$ & $n=1$  \\
    $\fiTWO$ & $\log(\delta)\in[-40/3,-20/3]$ & $n=2$  \\
  \end{priortable}
\end{center}

Some peculiar values of $\delta$ may lead to slow-roll violations and,
for the three scenarios, we have added a hard prior ignoring all
inflating regions in which $\vareps{1}>0.2$ or having a total
number of {\efolds} less than $120$.

\subsection{Generalized Double Well Inflation (GDWI)}
\label{sec:gdwiprior}

This is a phenomenological completion of the non-quadratic Small Field
Inflationary model $\sfi$ introduced in \Refa{Chowdhury:2019otk}. The
potential reads
\begin{equation}
V(\phi) = M^4 \left[ \left(\dfrac{\phi}{\phizero}\right)^{2p} - 1\right]^2,
\end{equation}
with two parameters, a power index $p$ and a vacuum expectation value
$\phizero$. It has been studied under the prior choices listed in the
table below.

\begin{center}
  \begin{priortable}{2}
    $\gdwiTWOs$ & $p=2$ & $\log(\phizero/\Mp)\in[-3,0]$  \\
    $\gdwiTWOl$ & $p=2$ & $\log(\phizero/\Mp)\in[0,3]$  \\
    $\gdwiTHREEs$ & $p=3$ & $\log(\phizero/\Mp)\in[-3,0]$  \\
    $\gdwiTHREEl$ & $p=3$ & $\log(\phizero/\Mp)\in[0,3]$  \\
    $\gdwiFOURs$ & $p=4$ & $\log(\phizero/\Mp)\in[-3,0]$  \\
    $\gdwiFOURl$ & $p=4$ & $\log(\phizero/\Mp)\in[0,3]$  \\
    $\gdwis$ & $p \in [1/10,10]$ & $\log(\phizero/\Mp)\in[-3,0]$  \\
    $\gdwil$ & $p \in [1/10,10]$ & $\log(\phizero/\Mp)\in[0,3]$  \\
  \end{priortable}
\end{center}

\subsection{Hyperbolic Inflation (HBI)}
\label{sec:hbiprior}
The potential is parameterized by two parameters, a power index $n$
and a vacuum expectation value $\mu$, as
\begin{equation}
V(\phi) = M^4 \sinh^n\left(\dfrac{\phi}{\mu}\right).
\label{eq:pothbi}
\end{equation}
The model describes a scalar field embedded within a perfect fluid. As
such, we have introduced two models, one, $\hbio$, describing this
very theoretical setup in which one has $n= 2\nphi/(\nphi-\nf)$ and
$\mu/\Mp= 2 \sqrt{\nphi}/(\nphi-\nf)$. In that case, the underlying
model parameter are $\nphi$ and $\nf$. In another model, named
$\hbip$, we consider \cref{eq:pothbi} as a phenomenological potential
and set the prior directly on $n$ and $\log(\mu/\Mp)$. For both
cases, we have also added a hard prior ignoring inflating regions in
which $\vareps{1}>0.2$.

\begin{center}
  \begin{priortable}{2}
    $\hbio$ & $\nf\in[0,1]$ & $\nphi-\nf \in [0.1,10]$  \\
    $\hbip$ & $ n\in[0,5]$ & $\log(\mu/\Mp)\in[0,3]$ \\
  \end{priortable}
\end{center}

\subsection{Higgs Inflation (HI)}
\label{sec:hiprior}

Higgs inflation is a non-minimal model of gravity built upon the
Standard Model Higgs field. In the Einstein frame, its potential is
parametrically given by
\begin{equation}
\begin{aligned}
  V(\phi) & = \dfrac{\Mp^4\lambda}{4\xi^2} \left(\frac{\barh^2 - \barv^2}{1+\barh^2}\right)^2,\\
  \dfrac{\phi}{\Mp} & =  \sqrt{\dfrac{1+6\xi}{\xi}} \ln \left[ \sqrt{1+(1+6\xi) \barh^2}
    + \sqrt{(1+6\xi) \barh^2} \right] \\ & + \sqrt{6} \ln
  \left[\frac{\sqrt{1+ \barh^2}}{\sqrt{1+(1+6\xi)\barh^2} + \sqrt{6\xi
        \barh^2}} \right],
\end{aligned}
\end{equation}
with $\lambda=0.13$ and $\barv = \sqrt{\xi} v/\Mp$ ($v=246\GeV$). As
such, Higgs inflation is also a zero-parameter model because $\xi$ is
fixed by the amplitude of the CMB anisotropies. As explained in {\EI},
at leading order $\si$ and $\hi$ are the same model, but we have
dropped this approximation here and consider them exactly, and
independent, as future data may possibly disambiguate between the two
scenarios.

\subsection{Hybrid Natural Inflation (HNI)}
\label{sec:hniprior}

This is an extension of Natural Inflation that incorporates a
tachyonic instability to stop inflation at a field value
$\phiend$. The potential is given by
\begin{equation}
  V(\phi) = M^4 \left[1 + \alpha \cos\left(\dfrac{\phi}{f}\right)\right],
\end{equation}
where the coupling $\alpha \in ]0,1[$ and $f$, the vacuum expectation
value, should be super-Planckian to allow for slow roll. With
$\phiend$, this is a three-parameter model. However, there are some regimes
in which inflation gracefully ends before the tachyonic
instability actually takes place, and, for these, $\phiend$ is not
relevant. We have accordingly split the model in three scenarios.

\begin{center}
  \begin{priortable}{3}
    $\hniONEf$ & $\alpha\in]0,1[$ & $\log(f/\Mp)\in[0,3]$ & -  \\
    $\hniTWOf$ & $\alpha\in]0,1[$ & $\log(f/\Mp)\in[0,3]$ & $\phiend/f
    \in ]0,\pi[$ \\
    $\hniTWOl$ & $\log(\alpha)\in [-3,0[$ & $\log(f/\Mp)\in[0,3]$ & $\phiend/f
    \in ]0,\pi[$ \\
  \end{priortable}
\end{center}

\subsection{Mutated Hilltop Inflation (MHI)}
\label{sec:mhi}

The potential of this model reads
\begin{equation}
V(\phi) = M^4\left[1-\sech\left(\dfrac{\phi}{\mu}\right)\right],
\end{equation}
and was included in the analysis of \Refa{Martin:2013nzq}. In the
present work, we have added one new realization having a deep sub-Planckian
vacuum expectation value $\mu$ as $\mhixs$.
\begin{center}
  \begin{priortable}{1}
    $\mhixs$ & $\mu=10^{-2}$  \\
  \end{priortable}
\end{center}

\subsection{Non-renormalizable Corrected Loop Inflation (NCLI)}
\label{sec:ncliprior}
This is an effective-field-theory model that can be seen as including
corrections from non-renormalizable operators onto the so-called Loop
Inflation model. The potential has three parameters and reads
\begin{equation}
V(\phi) = M^4\left[1+\alpha\ln\left(\frac{\phi}{\Mp}\right)
+\left(\frac{\phi}{\phizero}\right)^{4+2n}\right].
\end{equation}
We have considered five priors, motivated by the fact that $n$ is a
power index, $\phizero$ a sub-Planckian vacuum expectation value of unknown order of
magnitude and $\alpha$ a small coupling constant to keep corrections
under control. Moreover, we have added hard priors excluding inflating
regions with $\vareps{1}>0.2$ or the ones having a total number of
{\efolds} less than $120$.

\begin{center}
  \begin{priortable}{3}
    $\ncliONE$ & $\log(\alpha)\in[-6,-1]$ &
    $\log(\phizero/\Mp)\in[-3,0]$ & $n=1$ \\
    $\ncliTWO$ & $\log(\alpha)\in[-6,-1]$ &
    $\log(\phizero/\Mp)\in[-3,0]$ & $n=2$ \\
    $\ncliTHREE$ & $\log(\alpha)\in[-6,-1]$ &
    $\log(\phizero/\Mp)\in[-3,0]$ & $n=3$ \\
    $\ncliFOUR$ & $\log(\alpha)\in[-6,-1]$ &
    $\log(\phizero/\Mp)\in[-3,0]$ & $n=4$ \\
    $\ncli$ & $\log(\alpha)\in[-6,-1]$ & $\log(\phizero/\Mp)\in[-3,0]$ & $n\in[1,10]$ \\
  \end{priortable}
\end{center}

\subsection{N-Formalism Inflation (NFI)}
\label{sec:nfiprior}

This is a class of phenomenological models built to generate a
power-law behavior of the Hubble-flow functions at leading order in
slow roll. The potential is given by
\begin{equation}
V(\phi) = M^4 e^{-a x^b},
\end{equation}
with $x=\phi/\Mp$, the two parameters $a$ and $b$ being
arbitrary. Depending on their values, various inflationary regimes
appear, some of them requiring a third parameter to end inflation, a
field value $\xend=\phiend/\Mp$ assumed to be associated with a
tachyonic instability. As detailed in {\EI}, the consistency of the
scenario requires $\xend$ to be bounded by the two known functions
$\xendmin(a,b)$ and $\xendmax(a,b)$, and these ones have been required
to trigger, at least, $120$ {\efolds} of inflation. This leads to
several cases, to which specific priors can be associated, as listed
below.
\begin{center}
  \begin{priortable}{3}
    $\nfiONE$ & $a\in[0,10]$ & $b\in]1,10]$ & - \\       
    $\nfiTHREEn$ & $a\in[-10,0[$ & $b\in[0,1[$ & - \\
    $\nfiTHREEp$ & $a\in]0,10]$ & $b\in[-10,0[$ & - \\
    $\nfiTWO$ & $a\in[-10,0]$ &
            $b\in]1,10]$ & $\xend\in[\xendmin,\xendmax]$ \\
    $\nfiFOURp$ & $a\in[1,10]$ & $b\in]0,1[$ &
        $\xend\in[\xendmin,\xendmax]$ \\
    $\nfiFOURn$ & $a\in[-10,0[$ & $b\in[-10,0]$ &
            $\xend\in[\xendmin,\xendmax]$ \\
  \end{priortable}
\end{center}
In all these cases, a hard prior requiring inflation to verify
$\vareps{1}<0.2$ has been added.

\subsection{Non-Minimal Large Field Inflation (NMLFI)}
\label{sec:nmlfiprior}

These models correspond to Large Field Inflation where a non-minimal coupling to
the Ricci scalar, of strength $\xi$, is added. In the Einstein Frame,
the potential is a parametric function given by
\begin{equation}
  \begin{aligned}
    V(\barh) & = M^4 \frac{\barh^p}{1+ \barh^2}\,, \\
    \dfrac{\phi}{\Mp} & =  \sqrt{6 + 1/\xi} \ln \left[ \sqrt{1+(1+6\xi) \barh^2}
  + \sqrt{(1+6\xi) \barh^2} \right] \\ & + \sqrt{6} \ln
\left[\frac{\sqrt{1+ \barh^2}}{\sqrt{1+(1+6\xi)\barh^2} + \sqrt{6\xi
      \barh^2}} \right].
  \end{aligned}
\end{equation}
There are two parameters, a power index $p$, and the non-minimal
coupling constant $\xi$. Only for $p=4$ this potential has a
plateau. For all the other cases, it exhibits a hilltop and various
inflationary regimes can take place, some of them (NMLFI3) requiring a
new mechanism to end inflation~\cite{EIopiparous}. For them, we
require inflation to last, at least, $120$ {\efolds} and we demand, by
a hard prior, that it occurs with $\vareps{1}<0.2$. The scenarios are
summarized in the table below.
\begin{center}
  \begin{priortable}{3}
    $\nmlfiONEs$ & $\log(\xi)\in[-4,2]$ & $p\in[1/10,4]$ & - \\
    $\nmlfiONEl$ & $\log(\xi)\in[-4,2]$ & $p\in[4,8]$ & - \\
    $\nmlfiONEONE$ & $\log(\xi)\in[-4,2]$ & $p=1$ & - \\
    $\nmlfiONETWO$ & $\log(\xi)\in[-4,2]$ & $p=2$ & - \\
    $\nmlfiONETHREE$ & $\log(\xi)\in[-4,2]$ & $p=3$ & - \\
    $\nmlfiONEFOUR$ & $\log(\xi)\in[-4,2]$ & $p=4$ & - \\
    $\nmlfiONEFIVE$ & $\log(\xi)\in[-4,2]$ & $p=5$ & - \\
    $\nmlfiONESIX$ & $\log(\xi)\in[-4,2]$ & $p=6$ & - \\
    $\nmlfiTHREEs$ & $\log(\xi)\in[-4,2]$ & $p\in[0.1,0.5]$ &
    $\log(\barhend)\in[-1,2]$ \\
    $\nmlfiTHREEl$ & $\log(\xi)\in[-4,2]$ & $p\in[0.6,3.5]$ &
    $\log(\barhend)\in[-1,2]$ \\
    $\nmlfiTHREEONE$ & $\log(\xi)\in[-4,2]$ & $p=1$ &
    $\log(\barhend)\in[-1,2]$ \\
    $\nmlfiTHREETWO$ & $\log(\xi)\in[-4,2]$ & $p=2$ &
    $\log(\barhend)\in[-1,2]$ \\
    $\nmlfiTHREETHREE$ & $\log(\xi)\in[-4,2]$ & $p=3$ &
    $\log(\barhend)\in[-1,2]$ \\
  \end{priortable}
\end{center}

\subsection{Pure Arctan Inflation (PAI)}
\label{sec:paiprior}
A pure phenomenological model with a potential given by
\begin{equation}
V(\phi) = M^4 \arctan\left(\dfrac{\phi}{\mu}\right),
\end{equation}
with a vacuum expectation value $\mu$ or unknown order of
magnitude. Three priors have been devised.
\begin{center}
  \begin{priortable}{1}
    $\pai$ & $\log(\mu/\Mp)\in[-3,3]$  \\
    $\pais$ & $\log(\mu/\Mp)\in[-3,0]$  \\
    $\pail$ & $\log(\mu/\Mp)\in[0,3]$  \\
  \end{priortable}
\end{center}

\subsection{Radiatively Corrected Inflection Point Inflation (RCIPI)}
\label{sec:rcipiprior}

This scenario is motivated by effective-field-theory considerations in
which loop corrections, tuned to generate an inflection point, are
corrected by higher-order operators preventing the inflection point
to appear. The potential has three parameters and reads
\begin{equation}
V(\phi) = M^4 \left(\frac{\phi}{\Mp}\right)^p
\left[1 + \alpha \ln\left(\frac{\phi}{\Mp}\right) + \beta
  \ln^2\left(\frac{\phi}{\Mp}\right) \right],
\end{equation}
where $p$ is a power index, $\alpha$ and $\beta$ are coupling
constants associated with loops and higher-order operators. In this
theoretical framework $\alpha$ and $\beta$ are not arbitrary and
should take values close to the ones leading to an exact inflection
point. At given $\beta$ and $p$, as discussed in \Refa{EIopiparous},
the inflection point appears for $\alpha=\pm\alphazero$ where $\alphazero
=2 \sqrt{\beta} \sqrt{1-\beta/p^2}$. A theoretical prior is that
$\alpha-\alphazero$ should be a small quantity and these models will
be referred to as ``tuned''. Various inflationary regimes appear, as
the inflection point can become either a hilltop or lead to a monotonous
potential. We have also considered RCIPI as a phenomenological potential
in which one is allowed to be far from the inflection point. In that
case, we have set priors in which $\alpha$ can be quite different from
$\alphazero$. The resulting models are listed below.

\begin{center}
  \begin{priortable}{3}
    $\rcipiONEtuneTWOp$ & $p=2$ & $\log(\alpha-\alphazero) \in
    [-5,-2]$ & $\log(\beta)\in[-2,\log(4)]$ \\
    $\rcipiONEtuneFOURp$ & $p=4$ & $\log(\alpha-\alphazero) \in [-5,-2]$ & $\log(\beta)\in[-2,\log(16)]$ \\
    $\rcipiONEtuneTWOm$ & $p=2$ & $\log(-\alpha-\alphazero) \in
    [-5,-2]$ & $\log(\beta)\in[-2,\log(4)]$ \\
    $\rcipiONEtuneFOURm$ & $p=4$ & $\log(-\alpha-\alphazero) \in
    [-5,-2]$ & $\log(\beta)\in[-2,\log(16)]$ \\
    $\rcipiONEm$ & $p \in [2,4]$ & $(2\sqrt{\beta}+\alpha)/(2\sqrt{\beta}-\alphazero) \in
    [0,1]$ & $\log(\beta)\in[-2,\log(4)]$ \\
    $\rcipiTWOTWO$ & $p=2$ & $\alpha/\alphazero \in [-1,1]$ &
    $\log(\beta)\in[-2,\log(4)]$ \\
    $\rcipiTWOTHREE$ & $p=3$ & $\alpha/\alphazero \in [-1,1]$ &
    $\log(\beta)\in[-2,\log(9)]$ \\
    $\rcipiTWOFOUR$ & $p=4$ & $\alpha/\alphazero \in [-1,1]$ &
    $\log(\beta)\in[-2,\log(16)]$ \\
    $\rcipiTWOFIVE$ & $p=5$ & $\alpha/\alphazero \in [-1,1]$ &
    $\log(\beta)\in[-2,\log(25)]$ \\
    $\rcipiTWOSIX$ & $p=5$ & $\alpha/\alphazero \in [-1,1]$ &
    $\log(\beta)\in[-2,\log(36)]$ \\
    $\rcipiTWO$ & $p\in[2,6]$ & $\alpha/\alphazero \in [-1,1]$ & $\log(\beta/\sqrt{p})\in[-2,0]$ \\    
  \end{priortable}
\end{center}

\subsection{Radiatively Corrected Large Field Inflation (RCLFI)}
\label{sec:rclfiprior}

The potential is given by
\begin{equation}
V(\phi) = M^4 \left[ \left(\frac{\phi}{\mu}\right)^p + \alpha
  \left(\frac{\phi}{\mu}\right)^4 \ln \left(\frac{\phi}{\mu}\right) \right],
\end{equation}
and models loop corrections to the Large Field Inflation
potential (monomial term). There are three parameters, the power index
$p$, the coupling constant $\alpha$ and a renormalization mass scale
$\mu$. In this theoretical framework, one may expect the loop
corrections to be small, and this perturbative regime is referred to
as RCLFI4. As detailed in {\EI}, when loop corrections become
sizable, various novel inflationary regimes appear, named RCLFI1,
RCLFI2, RCLFI2, as the potential may develop a local maximum for some
parameter values. The separatrix of these different regimes involve
some functions, derived and explained in \Refa{EIopiparous}, which are
$\alphazero(p)=-e(p-4)$ and $\alphaone(p) = -p(p-4) e^{2-p/4}/4$. In
some cases, inflation can occur only at the hilltop and there is a
minimal value of $\mu=\mumin$ for which the total number of {\efolds}
realized without an extreme tuning can exceed an acceptable value
which has been set to $120$ here. The precise value of $\mumin$ does
not really matter as these situations are actually strongly
disfavored. However notice that, tuning these cases even more, by
allowing for smaller $\mumin$, would lower their evidences. There are
also some fine-tuned cases in which the power index $p$ is forced to
remain smaller than a numerically computed value $\pmax$.

\begin{center}
  \begin{priortable}{3}
    $\rclfiONEpm$ & $p\in[4.1,6]$ & $\alpha/\alphaone\in ]1,10[$ &
            $\log(\mu/\mumin)\in[0,3]$ \\            
    $\rclfiONEmm$ & $p\in[0.1,3.9]$ & $\alpha \in [-4,-0.1[$ &
                $\log(\mu/\mumin)\in[0,3]$ \\
    $\rclfiONEmp$ & $p/\pmax\in[0.1,3.9]$ & $\alpha/\alphaone \in
              ]1,10]$ & $\log(\mu/\mumin)\in[0,3]$ \\
    $\rclfiTWOpm$ & $p \in[4.1,8]$ & $\alpha/\alphazero \in ]1,10]$ &
        $\log(\mu/\mumin)\in[0,3]$ \\
    $\rclfiTWOmm$ & $p \in[0.1,3.9]$ & $\alpha \in [-2,-0.2]$ &
        $\log(\mu/\mumin)\in[0,3]$ \\
    $\rclfiTWOmp$ & $p/\pmax \in[0.1,0.9]$ & $\alpha/\alphazero \in ]1,10]$ &
        $\log(\mu/\mumin)\in[0,3]$ \\
    $\rclfiTHREEpp$ & $p \in[4.1,8]$ & $\log(\alpha) \in [-2,2]$ &
        $\log(\mu/\Mp)\in[-3,3]$ \\
    $\rclfiTHREEpm$ & $p \in[4.1,8]$ & $\alpha/\alphazero \in ]1,10]$ &
        $\log(\mu)\in[-3,3]$ \\
    $\rclfiTHREEmp$ & $p \in[0.1,3.9]$ & $\alpha/\alphazero \in ]1,10]$ &
        $\log(\mu/\Mp)\in[-3,3]$ \\
    $\rclfiFOURp$ & $p \in[4.1,8]$ & $\log(\alpha/\alphaone) \in [-3,0]$ &
        $\log(\mu/\Mp)\in[-2,4]$ \\
    $\rclfiFOURm$ & $p \in[0.1,3.9]$ & $\log(\alpha/\alphaone) \in [-3,0]$ &
        $\log(\mu/\Mp)\in[-2,4]$ \\        
  \end{priortable}
\end{center}

\subsection{String Axion Inflation I (SAII)}
\label{sec:saiiprior}

This is a String Theory motivated model emerging from geometric
compactifications with a two-parameter potential given by
\begin{equation}
V(\phi) = M^4 \left[1 - \cos\left(\frac{\phi}{\mu}\right) +
  \alpha \frac{\phi}{\mu} \sin\left(\frac{\phi}{\mu}\right)\right].
\end{equation}
The coupling $\alpha$ does not need to be small while the vacuum
expectation value $\mu$ needs to be super-Planckian for slow-roll
inflation to take place. The potential having a hilltop, there are two
possible inflationary regimes, SAII1 and SAAI2. A hard prior requires
inflation to last, at least, $120$ {\efolds}. We have therefore
considered the priors listed below.

\begin{center}
  \begin{priortable}{2}
    $\saiiONEn$ & $\log(-\alpha)\in[-3,3]$ & $\log(\mu/\Mp)\in[0,3]$ \\
    $\saiiONEp$ & $\log(\alpha)\in[-3,3]$ & $\log(\mu/\Mp)\in[0,3]$  \\
    $\saiiONEf$ & $\alpha \in[-10,10]$ & $\log(\mu/\Mp)\in[0,3]$ \\
    $\saiiTWOn$ & $\log(-\alpha)\in[-3,3]$ & $\log(\mu/\Mp)\in[0,3]$ \\
    $\saiiTWOp$ & $\log(\alpha)\in[-3,3]$ & $\log(\mu/\Mp)\in[0,3]$  \\
    $\saiiTWOf$ & $\alpha \in[-10,10]$ & $\log(\mu/\Mp)\in[0,3]$  \\
  \end{priortable}
\end{center}

\subsection{String Axion Inflation II (SAIII)}
\label{sec:saiiiprior}

This model shares the same String Theory motivations as SAII in
\Sec{sec:saiiprior} but includes additional higher-order corrections. Its potential reads
\begin{equation}
V(\phi) = M^4 \left\{1 - \cos\left(\frac{\phi}{\mu}\right) +
  \alpha \left[ \frac{\phi}{\mu} \sin\left(\frac{\phi}{\mu}\right) +
    \frac{1}{2} \beta \left(\frac{\phi}{\mu} \right)^2 \right]\right\},
\end{equation}
where the amplitude of the added terms are set by the new parameter
$\beta$, which verifies $\alpha \beta >0$. This is a three-parameter
model that gives rise to three distinct inflationary regimes labeled
as SAIII1, SAIII2 and SAIII3. The parameter separatrices between these
regimes are given by some irrational numbers labeled as $\betazero$,
$\betatwo$ and $\betathree$ derived in \Refa{EIopiparous}. A hard
prior has been added to enforce that inflation lasts, at least, $120$
{\efolds}. We have analyzed all corresponding scenarios with the
priors listed in the following table.

\begin{center}
  \begin{priortable}{3}
    $\saiiiONEp$ & $\alpha\in]0,3]$ & $\beta\in]0,3]$ &
            $\log(\mu/\Mp)\in[-1,2]$ \\
    $\saiiiONEn$ & $\alpha\in[-3,0[$ & $\beta\in[-3,0[$ &
                    $\log(\mu/\Mp)\in[-1,2]$ \\
    $\saiiiTWOp$ & $\alpha\in]0,3]$ & $\beta\in]0,\betatwo]$ &
            $\log(\mu/\Mp)\in[-1,2]$ \\
    $\saiiiTWOn$ & $\alpha\in[-3,0[$ & $\beta\in[\betathree,0[$ &
                    $\log(\mu/\Mp)\in[-1,2]$ \\
    $\saiiiTHREEp$ & $\alpha\in]0,3]$ & $\beta\in[\betazero,3]$ &
            $\log(\mu/\Mp)\in[-1,2]$ \\
    $\saiiiTHREEn$ & $\alpha\in[-3,0[$ & $\beta\in[-3,-1]$ &
                    $\log(\mu/\Mp)\in[-1,2]$ \\                    
  \end{priortable}
\end{center}

\subsection{Superconformal $\alpha$-Attractor B Inflation (SABI)}
\label{sec:sabiprior}

These scenarios are based on Supergravity in which the conformal
symmetry associated with a highly symmetrical theory is broken. Their
potential reads
\begin{equation}
V(\phi) = M^4 \left(1 - e^{-\sqrt{\frac{2}{3\alpha}} x}\right)^{2n},
\end{equation}
with $x=\phi/\Mg$ and $\Mg \simeq \Mp$. There are two parameters, a
power index $n$ and a coupling constant $\alpha$ (associated with a
K\"ahler potential). We have considered various scenarios as listed
below. Let us notice that the case $n=1$ is a model referred to as
Superconformal $\alpha$-Attractor A Inflation in \Refa{EIopiparous}.

\begin{center}
  \begin{priortable}{2}
    $\sabiONE$ & $\log(\alpha)\in[-3,3]$ & $n=1$  \\
    $\sabiONETHREE$ & $\log(\alpha)\in[-3,3]$ & $n=1/3$  \\
    $\sabiONETWO$ & $\log(\alpha)\in[-3,3]$ & $n=1/2$  \\
    $\sabiTHREETWO$ & $\log(\alpha)\in[-3,3]$ & $n=3/2$  \\
    $\sabiFIVETWO$ & $\log(\alpha)\in[-3,3]$ & $n=5/2$  \\    
    $\sabiTHREE$ & $\log(\alpha)\in[-3,3]$ & $n=3$  \\
  \end{priortable}
\end{center}

\subsection{Superconformal $\alpha$-Attractor T Inflation (SATI)}
\label{sec:satiprior}

These models are another realization of conformal symmetry breaking
(see \Sec{sec:sabiprior}) in a highly symmetrical Supergravity theory
and are described by the potential
\begin{equation}
V(\phi) = M^4 \tanh^{2n}\negthinspace\left(\frac{\phi}{\Mp\sqrt{6\alpha}}\right).
\end{equation}
There are two parameters, a power index $n$ and the K\"ahler coupling
$\alpha$. The case $\alpha=1$ is also known as T-model inflation. The
priors chosen for the SATI class are enumerated in the following
table.

\begin{center}
  \begin{priortable}{2}
    $\satione$ & $\alpha=1$ & $n\in[1,10]$  \\
    $\satiONETHREE$ & $\log(\alpha)\in[-3,3]$ & $n=1/3$  \\
    $\satiONETWO$ & $\log(\alpha)\in[-3,3]$ & $n=1/2$  \\
    $\satiONE$ & $\log(\alpha)\in[-3,3]$ & $n=1$ \\
    $\satiTHREETWO$ & $\log(\alpha)\in[-3,3]$ & $n=3/2$  \\
    $\satiTWO$ & $\log(\alpha)\in[-3,3]$ & $n=2$  \\
    $\satiFIVETWO$ & $\log(\alpha)\in[-3,3]$ & $n=5/2$  \\
    $\satiTHREE$ & $\log(\alpha)\in[-3,3]$ & $n=3$  \\    
  \end{priortable}
\end{center}

\subsection{Symmetry Breaking K\"ahler Inflation (SBKI)}
\label{sec:sbkiprior}

This scenario is based on embedding quadratic Large Field Inflation
($\lfiTWO$) in a Supergravity context where the shift symmetry gets
broken by a new term in the K\"ahler potential. The field potential is
given by
\begin{equation}
V(\phi) = M^4\left(\frac{\phi}{\Mp}\right)^2
\exp\left[\alpha\left(\frac{\phi}{\Mp}\right)^2
+\frac{\alpha^2}{6}\left(\frac{\phi}{\Mp}\right)^4\right],
\end{equation}
and involves one new parameter $|\alpha|\ll 1$ representing a vacuum
expectation value in Planck units. We have analyzed the two models
listed below.

\begin{center}
  \begin{priortable}{1}
    $\sbkin$ & $\log(-\alpha)\in[-4,-1]$   \\
    $\sbkip$ & $\log(\alpha)\in[-4,-1]$  \\
  \end{priortable}
\end{center}

\subsection{S-Dual Inflation (SDI)}
\label{sec:sdiprior}

This model is phenomenological and motivated by S-duality in String
Theory where the inflaton field is a dilaton. The potential is
parameterized by a vacuum expectation value $\mu$ and reads
\begin{equation}
V(\phi) = \dfrac{M^4}{\cosh\left(\dfrac{\phi}{\mu}\right)}\,.
\end{equation}
As shown in \Refa{EIopiparous}, slow-roll inflation occurs only for
super-Planckian values of $\mu > \Mp/\sqrt{2}$ and a hard prior has
been added to ensure that it lasts more than $120$ {\efolds}.
\begin{center}
\begin{priortable}{1}
    $\sdi$ & $\log(\mu/\Mp) \in ]1/\sqrt{2},3]$  \\
  \end{priortable}
\end{center}

\subsection{Smeared Higgs Inflation (SHI)}
\label{sec:shiprior}
This scenario arises from an extension of the Colemann-Weinberg model
into the context of Grand Unified Theory within a SU(5) symmetry. The
potential is parameterized by two parameters, a vacuum expectation
value $\phizero$ and a coupling constant $\alpha$, both of them being
unconstrained. It reads
\begin{equation}
V(\phi) = M^4\left\lbrace
\left[1-\left(\frac{\phi}{\phizero}\right)^2\right]^2 +
\alpha\left(\frac{\phi}{\phizero}\right)^4
\left[\ln\left(\frac{\phi}{\phizero}\right)-\frac{1}{4}\right]
+\frac{\alpha}{4}\right\rbrace,
\end{equation}
and describes a hilltop. We have considered this model in the
slow-roll regime only by adding a hard prior enforcing
$\vareps{2}<0.2$.
\begin{center}
  \begin{priortable}{2}
    $\shi$ & $\log(\alpha)\in[-3,3]$ & $\log(\phizero/\Mp) \in [-2,2]$  \\
  \end{priortable}
\end{center}

\subsection{Starobinsky Inflation (SI)}
\label{sec:siprior}

Starobinsky inflation is a non-minimal model of gravity having, in the
Einstein frame, a potential given by
\begin{equation}
V(\phi)=M^4\left(1-\ee^{-\sqrt{2/3}\, \phi/\Mp}\right)^2.
\end{equation}
It is a zero-parameter model. However, as discussed in {\EI}, there
are theoretical realizations of this potential that live in the
Jordan frame, and, as such, they correspond to a minimally coupled
inflaton field. Even though both frameworks have exactly the same
potential, the presence of minimal and non-minimal couplings induces
slightly different reheating histories that have been taken into
account within two different scenarios, $\si$ and $\simc$.

\subsection{Mukhanov Inflation (VFMI)}
\label{sec:vfmiprior}

This model provides a hydrodynamical description of the background
evolution during inflation by fixing the evolution equation for the
equation of state parameter $w(N) =
\beta/[(3\beta/2)^{1/\alpha}+\Nend-N)^\alpha$, $N$ being the number of
{\efolds}. As discussed in {\EI}, this is strictly equivalent to
the field potential
\begin{equation}
V(\phi) = M^4\left[1 - \dfrac{\beta}{2
    \left(1+\dfrac{2-\alpha}{2\sqrt{3\beta}}\dfrac{\phi}{\Mp}\right)^{\frac{2\alpha}{2-\alpha}}}
  \right] \exp\left\{ \dfrac{3 \beta}{1-\alpha}
\left[\left(1+\dfrac{2-\alpha}{2\sqrt{3\beta}}\dfrac{\phi}{\Mp}\right)^{\frac{2(1-\alpha)}{2-\alpha}}
  - 1 \right] \right\}.
\end{equation}
It is parameterized by the two parameters $\alpha$ and $\beta$, the
order of magnitude of which is unspecified. We have therefore
chosen to analyze the model under the following prior choices.

\begin{center}
  \begin{priortable}{2}
    $\vfmi$ & $\alpha\in[1/10,10]$ & $\beta \in[1/10,10]$ \\
    $\vfmis$ & $\alpha\in[1/4,4]$ & $\beta \in [1/4,4]$  \\
  \end{priortable}
\end{center}

\section{Posterior distributions for the best models}
\label{sec:modelposteriors}

In this section, we report the marginalized posterior distributions
for the primordial parameters associated with the best models, i.e., the ones
landing in the favored zone of evidences having
$\ln\left(\Bref{}/\Bbest{}\right) > -1$. In order to avoid
redundancies, when several models share the same potential, the same
theoretical framework, but only differ by their priors, we have only
represented, when it exists, the model having the prior encompassing
the others. In all figures, the intrinsic theoretical model parameters
are enumerated as $\cte{1}$, $\cte{2}$, $\cte{3}$, \dots and we have added various
derived primordial parameters as the energy scale of the potential
$M$, the energy density at the end of inflation $\rhoend$, the
slow-roll parameters $\vareps{i}$, the spectral index $\nS$, the
tensor-to-scalar ratio $\reps$, the scalar running $\alphaS$ (all
computed at second order in slow-roll) and the reheating parameter
$\Rrad$.

\begin{figure}
\begin{center}
\includegraphics[width=\onefigw]{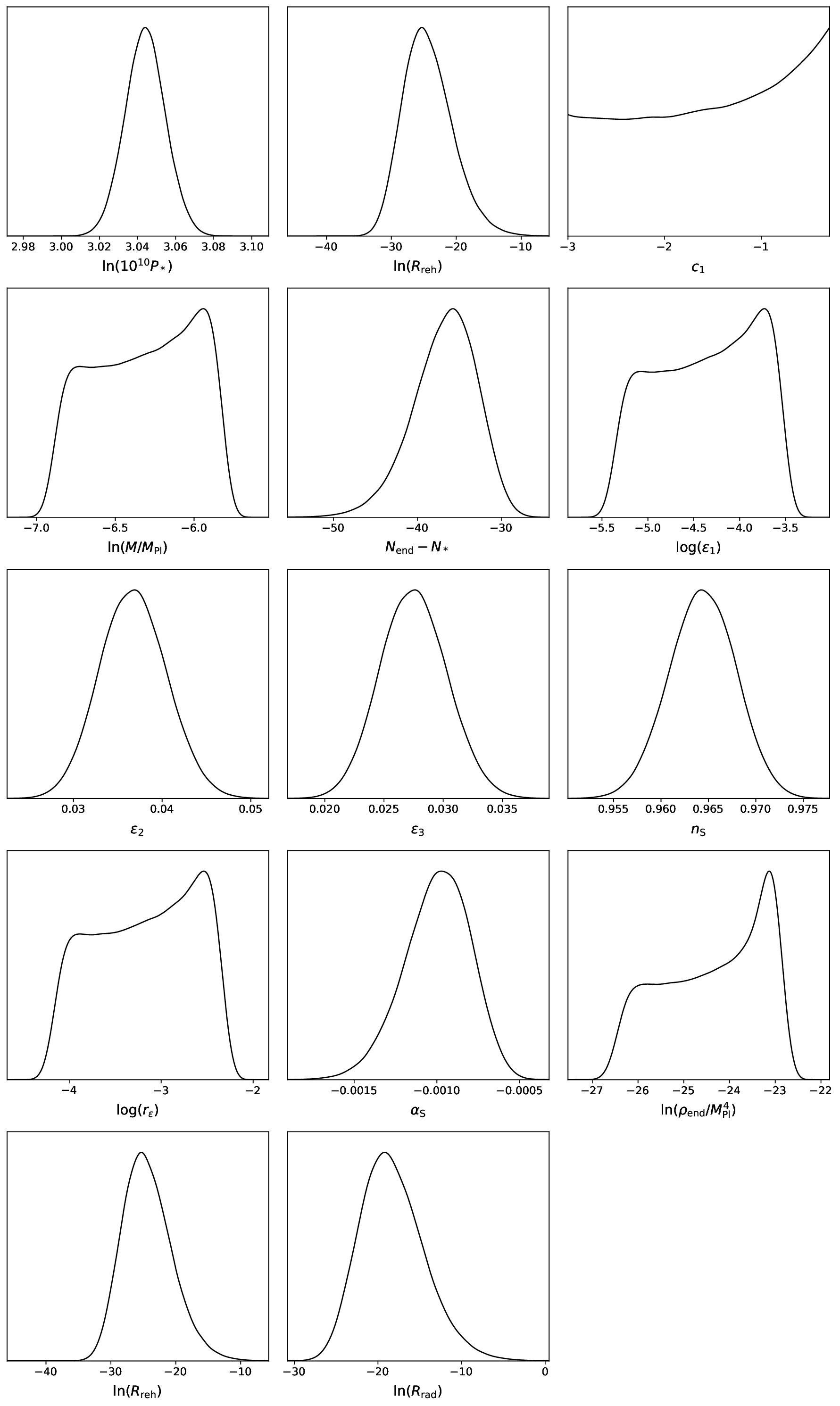}
\caption{One-dimensional posterior distributions for Arctan
  Inflation ($\ai$). This is a phenomenological model having only one parameter,
  a typical vacuum expectation value $\cte{1}=\log(\mu/\Mp)$.}
\label{fig:ai}
\end{center}
\end{figure}

\begin{figure}
\begin{center}
\includegraphics[width=\onefigw]{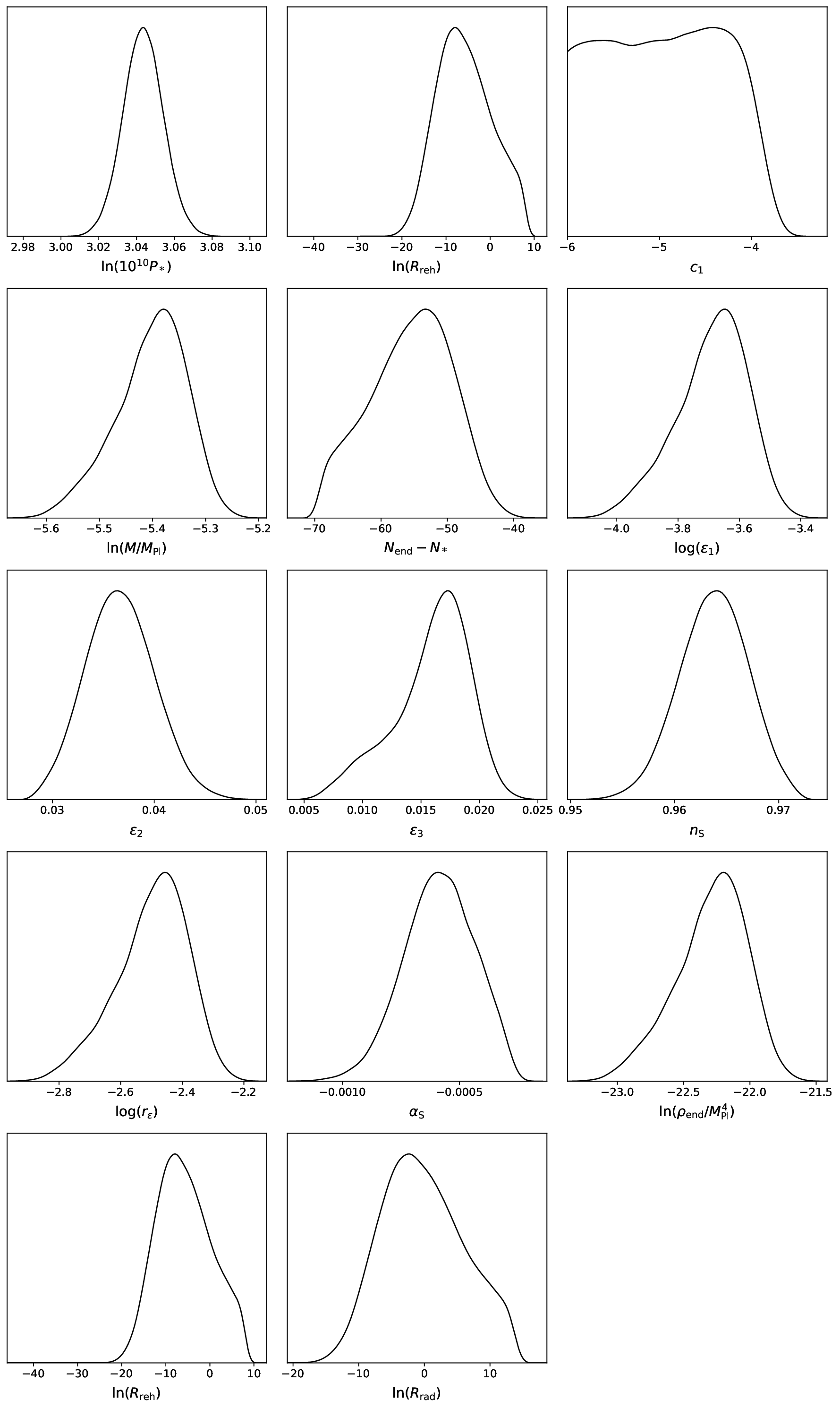}
\caption{One-dimensional posterior distributions for Cublicly
  Corrected Starobinsky Inflation 1 ($\ccsiONE$). It has one parameter
  $\cte{1}=\log(\alpha)$, where $\alpha$ is a small positive coupling constant.}
\label{fig:ccsi1}
\end{center}
\end{figure}

\begin{figure}
\begin{center}
\includegraphics[width=\onefigw]{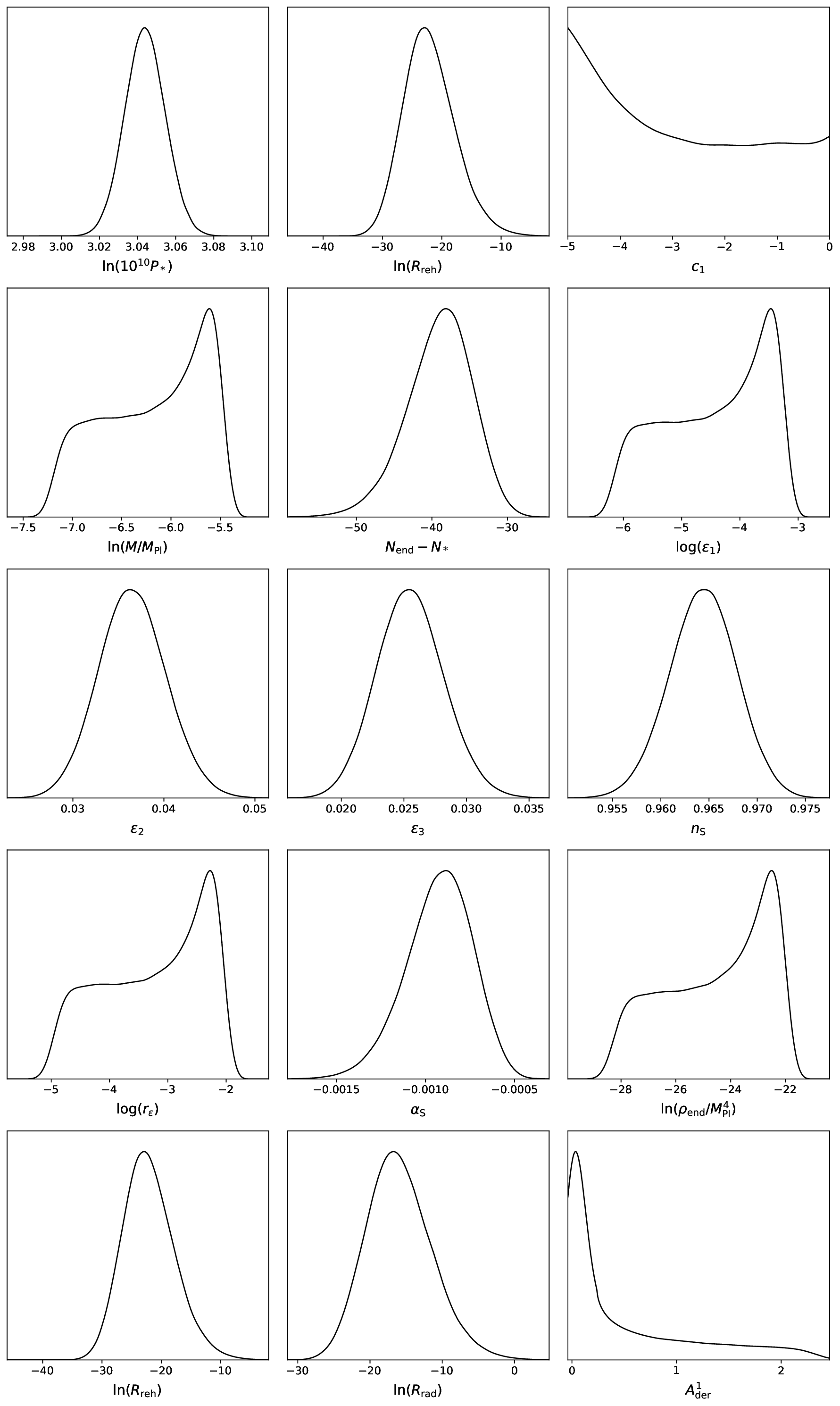}
\caption{One-dimensional posterior distributions for Dual Inflation
  ($\di$). It has one theoretical parameter $\cte{1}=\log(f)$ where
  $f<1$ is a ratio of two vacuum expectation values, which has to be
  less than unity for the underlying model to be well defined. The
  derived parameter $\Ader{1}=\Lambda/\Mp$ involves $\Lambda$, another
  typical vacuum expectation value fixing the amplitude of the CMB
  anisotropies.}
\label{fig:di}
\end{center}
\end{figure}

\begin{figure}
\begin{center}
\includegraphics[width=\onefigw]{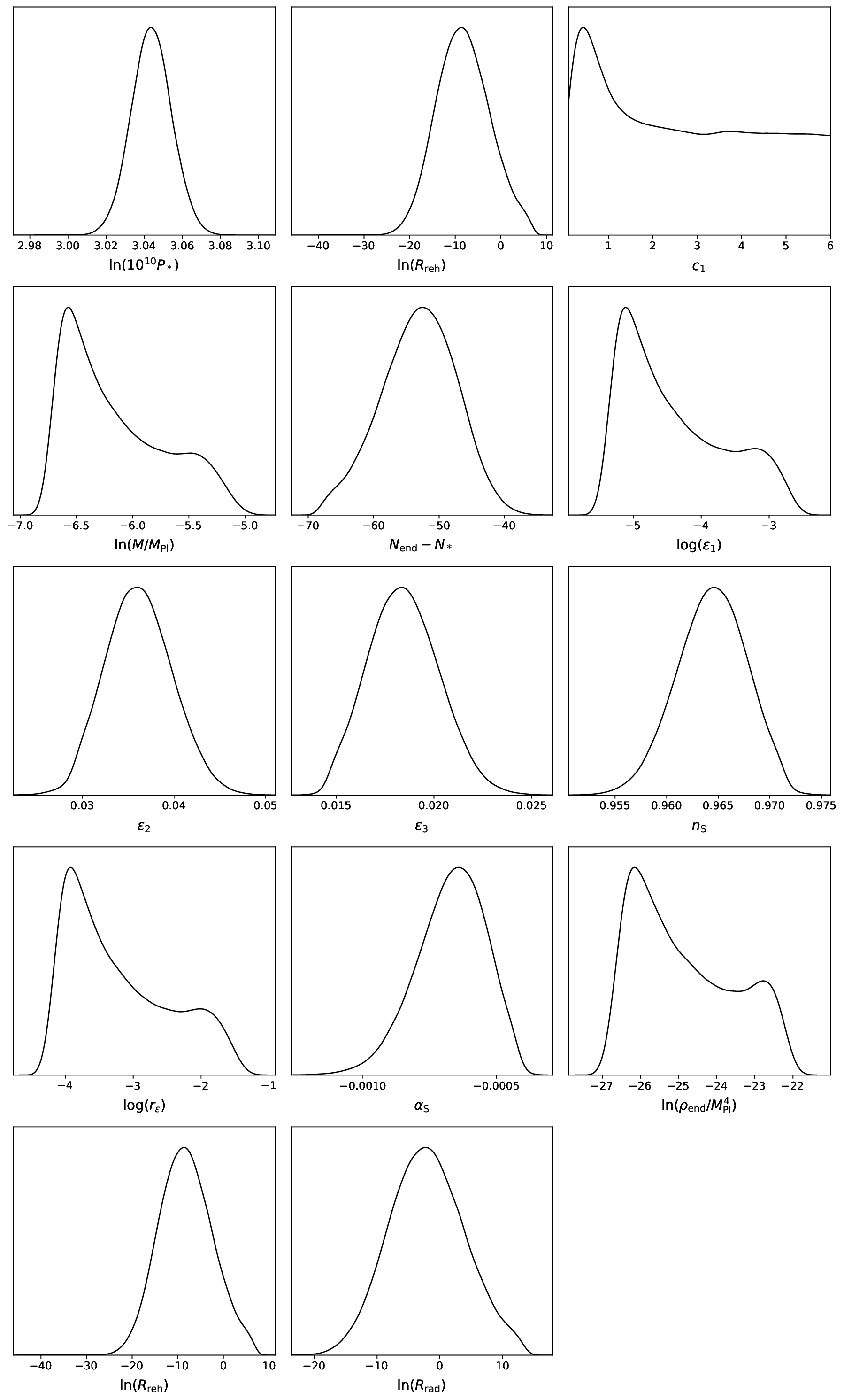}
\caption{One-dimensional posterior distributions for Exponential SUSY
  Inflation ($\esi$). There is one theoretical parameter, a coupling
  constant of order one $\cte{1}=q$.}
\label{fig:esi}
\end{center}
\end{figure}

\begin{figure}
\begin{center}
\includegraphics[width=\onefigw]{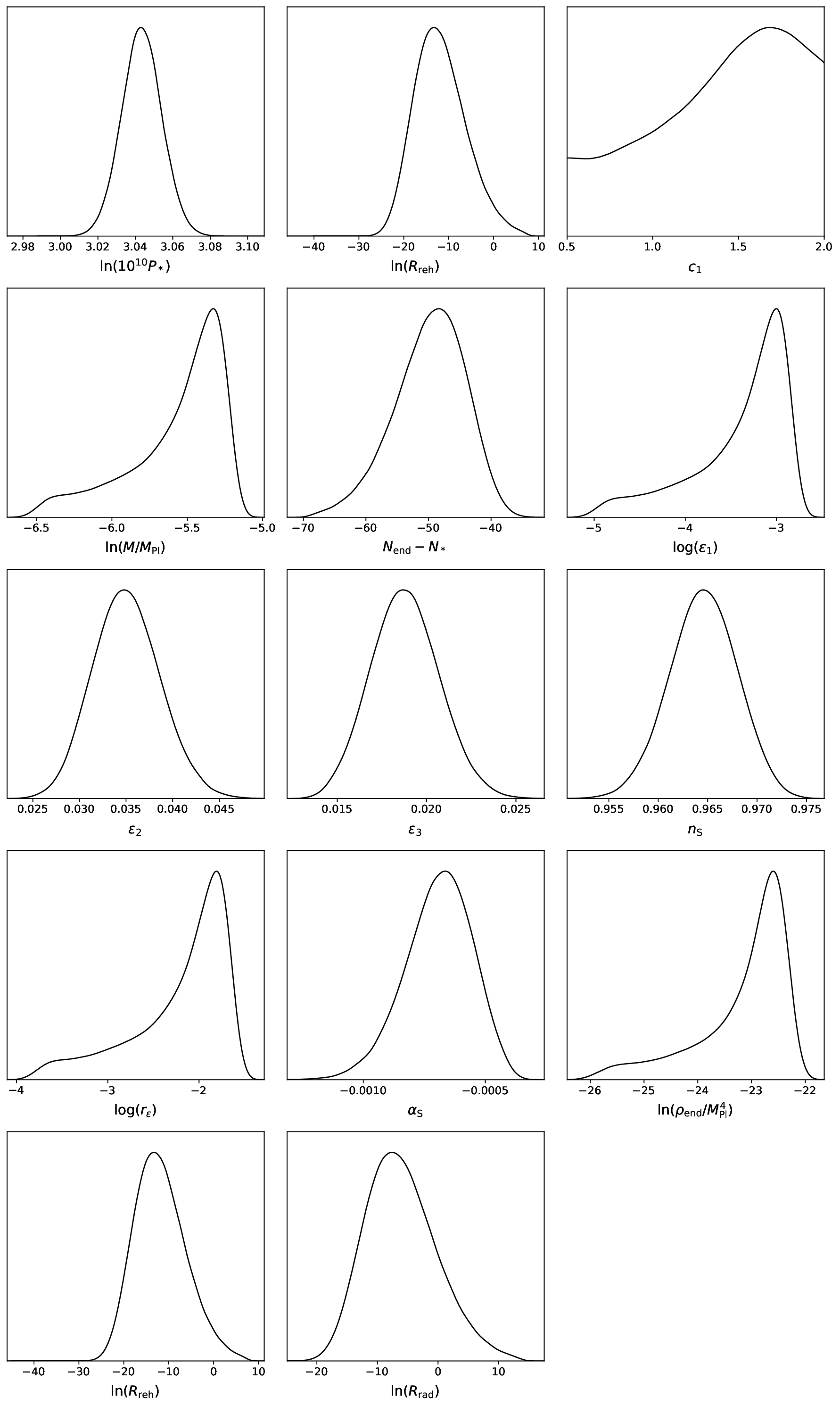}
\caption{One-dimensional posterior distributions for Exponential SUSY
  Original Inflation ($\esio$). There is one theoretical parameter
  $\cte{1}=\sqrt{2/\beta}$ where $\beta$ is the normalization of a
  K\"ahler potential in the range centered around unity,
  $\beta\in[1/2,2]$. It differs from \Fig{fig:esi} by the theoretically
  motivated prior being flat in $\beta$ as opposed to being flat in
  $\cte{1}$. This is the best model of this analysis.}
\label{fig:esio}
\end{center}
\end{figure}

\begin{figure}
\begin{center}
\includegraphics[width=\onefigw]{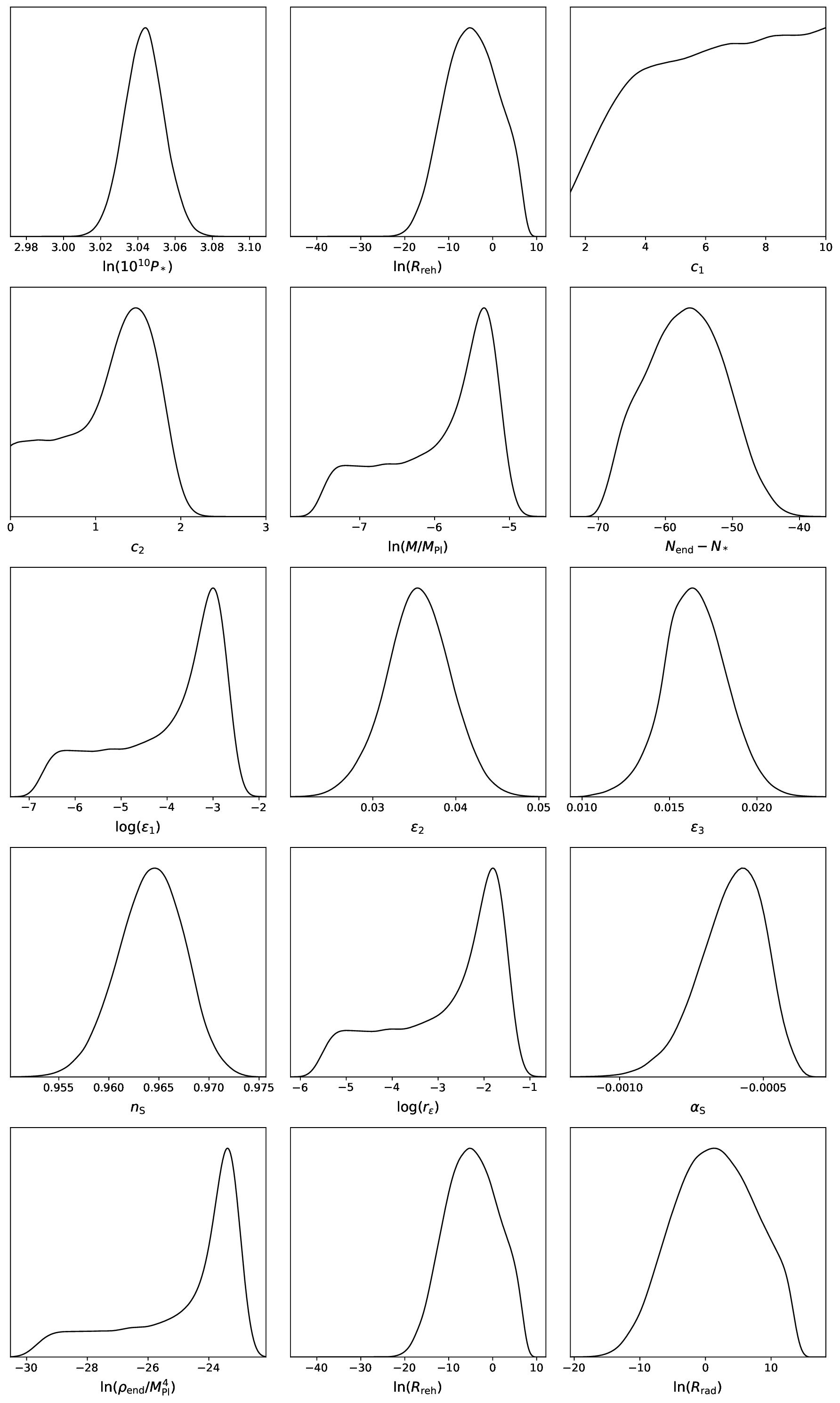}
\caption{One-dimensional posterior distributions for Generalized
  Double Well Inflation at large field values ($\gdwil$). There are two theoretical
  parameters, a power index $\cte{1}=p$ and a super-Planckian vacuum
  expectation value $\cte{2}=\log(\mu/\Mp)$.}
\label{fig:gdwil}
\end{center}
\end{figure}

\begin{figure}
\begin{center}
\includegraphics[width=\onefigw]{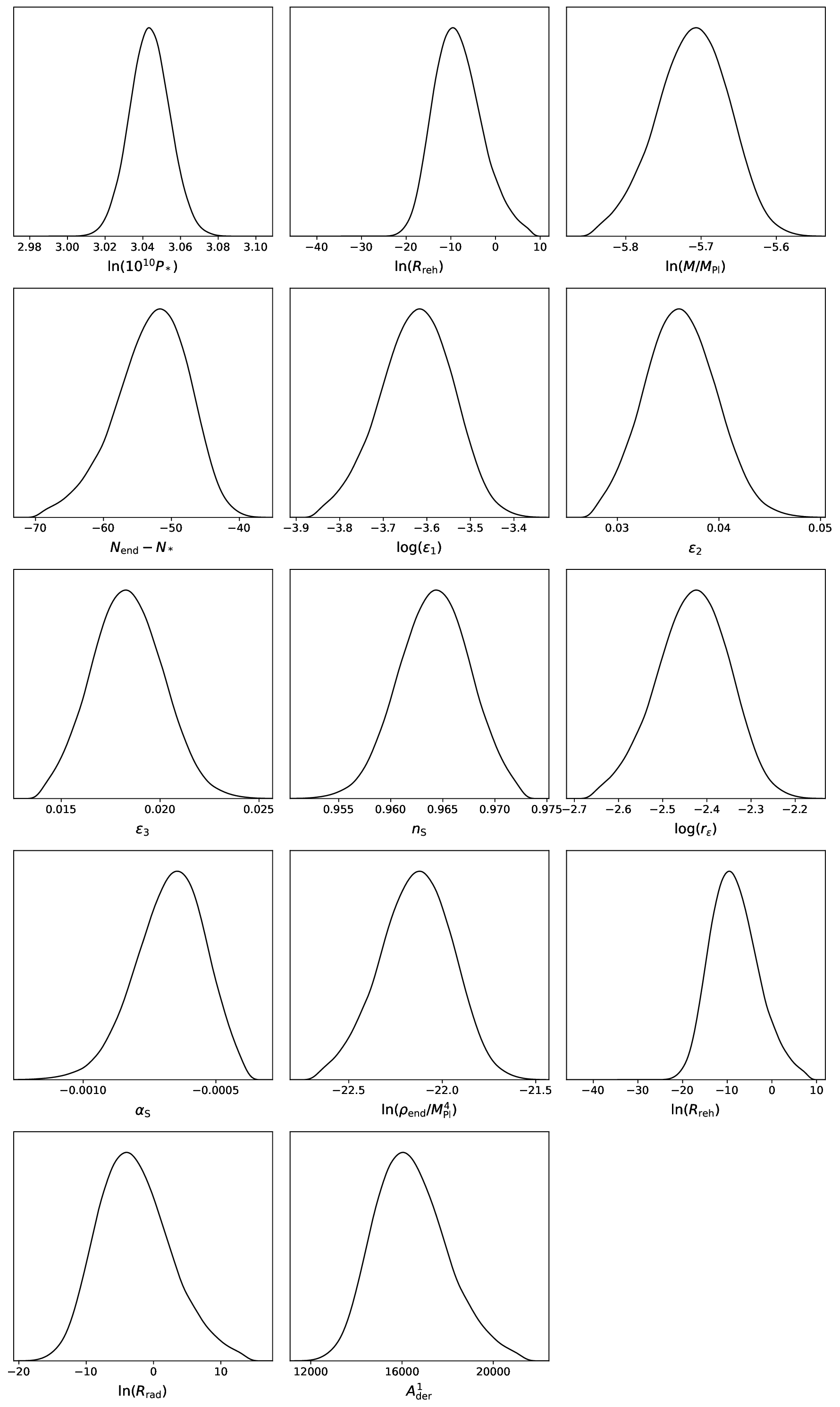}
\caption{One-dimensional posterior distributions for Higgs Inflation
  ($\hi$). There is no model parameter, but we have represented the
  posterior of the derived parameter $\Ader{1}=\xi$, the
  non-minimal coupling of the Higgs field.}
\label{fig:hi}
\end{center}
\end{figure}

\begin{figure}
\begin{center}
\includegraphics[width=\onefigw]{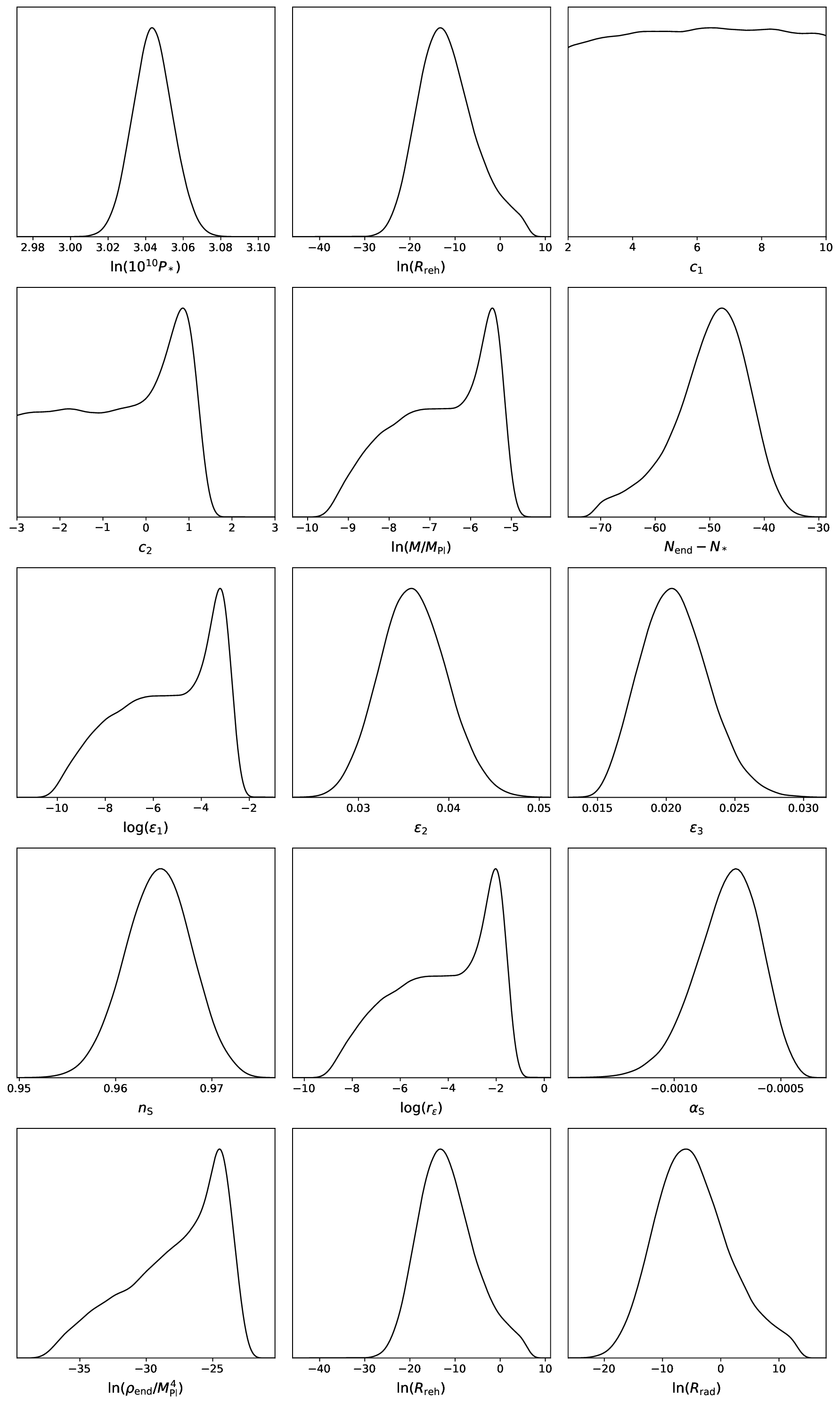}
\caption{One-dimensional posterior distributions for KKLT Inflation
  ($\kklti$). There are two model parameters $\cte{1}=p$, a power index, and
  $c_2=\log(\mu/\Mp)$, a typical vacuum expectation value.}
\label{fig:kklti}
\end{center}
\end{figure}

\begin{figure}
\begin{center}
\includegraphics[width=\onefigw]{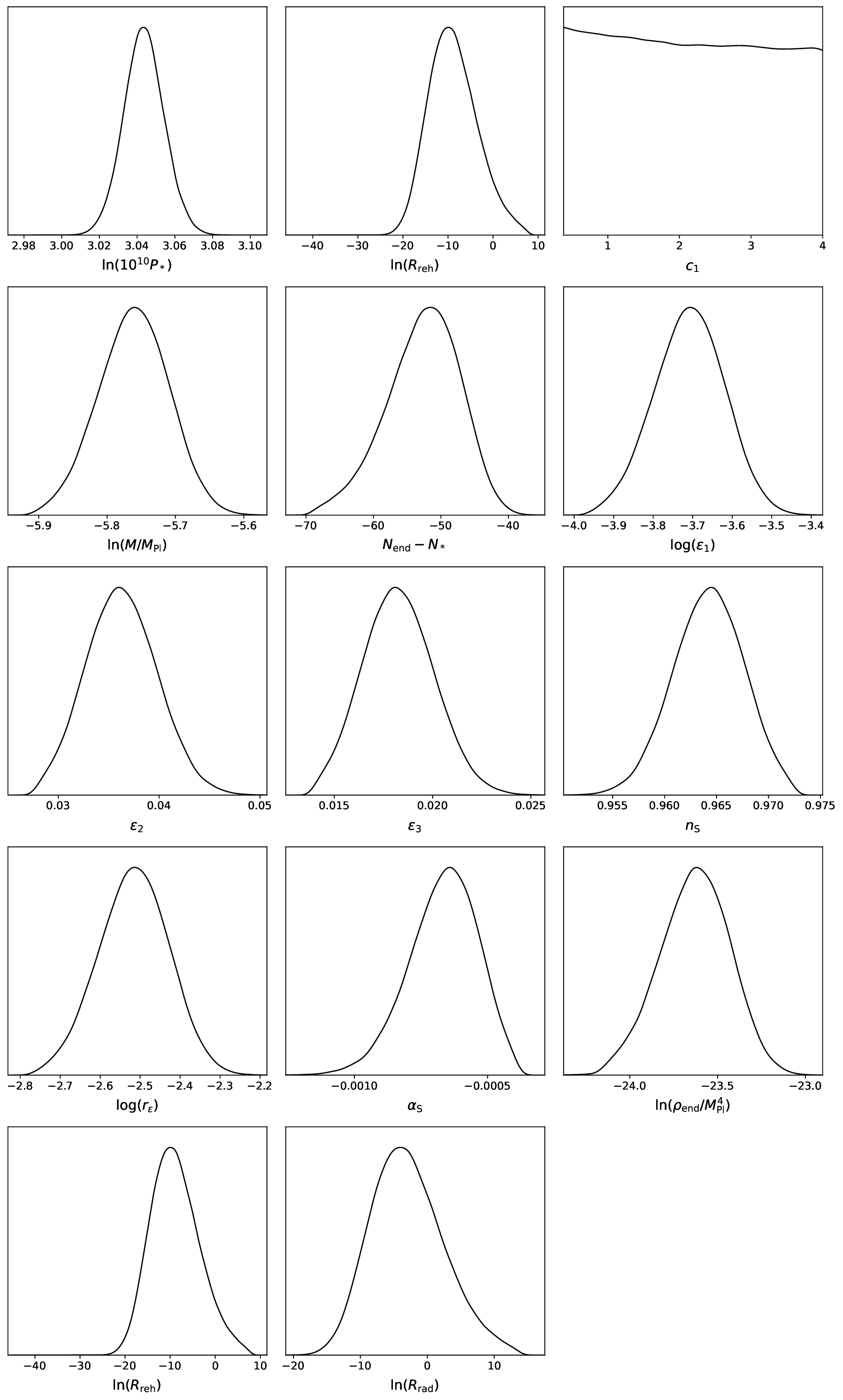}
\caption{One-dimensional posterior distributions for K\"ahler Moduli
  Inflation I ($\kmii$). The unique model parameter is a coupling
  constant $\cte{1}=\log(\alpha)\in[0.382,4]$, this prior range being set by
  the underlying theory~\cite{Martin:2013nzq}.}
\label{fig:kmii}
\end{center}
\end{figure}

\begin{figure}
\begin{center}
\includegraphics[width=\onefigw]{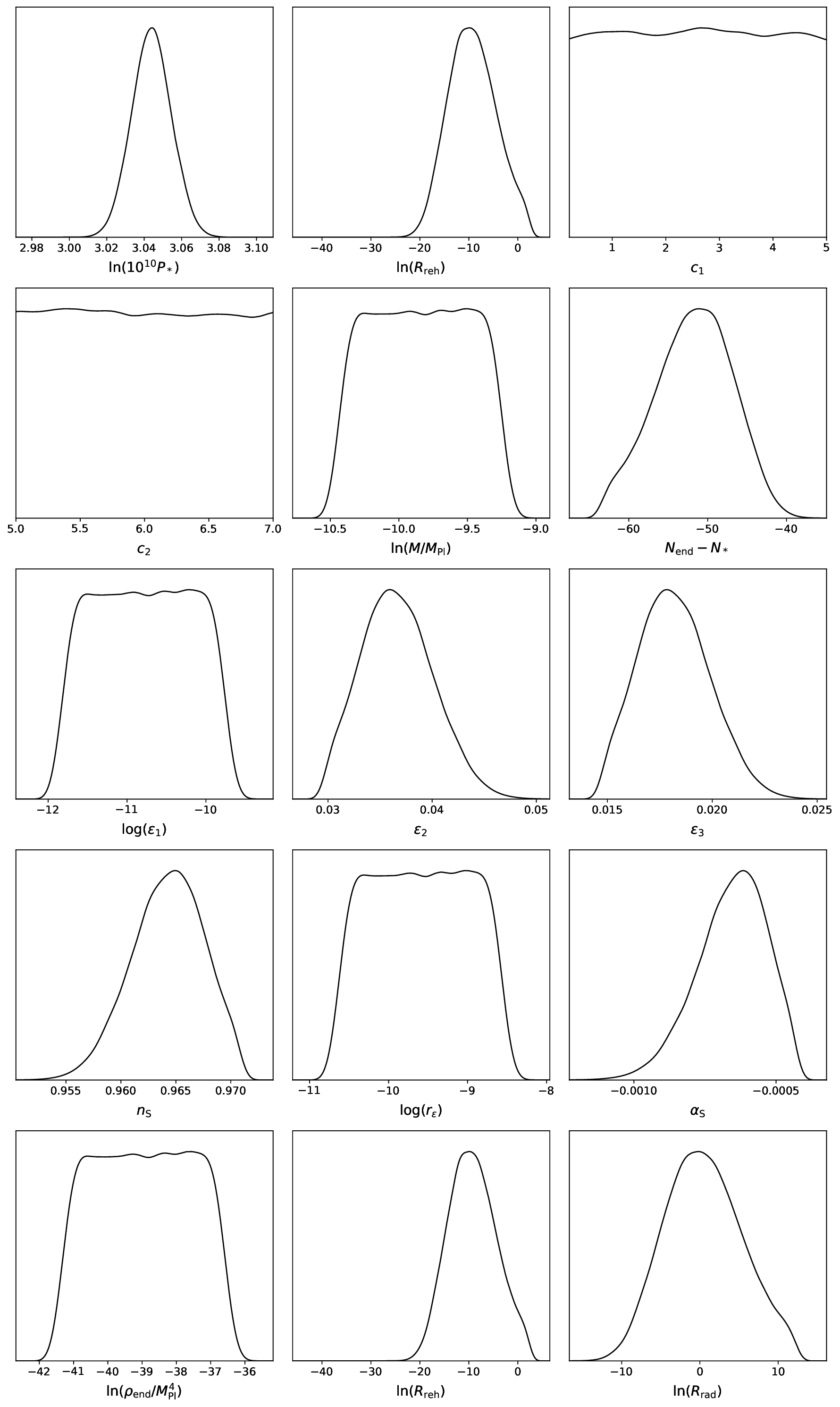}
\caption{One-dimensional posterior distributions for K\"ahler Moduli
  Inflation II ($\kmiii$). It has two model parameters $\cte{1}=
  \alpha/(\beta\calV)$, an order one quantity made of the ratio between
  coupling constants, and $\cte{2}=\log(\calV)$ where $\calV$ is a
  dimensionless compactification volume, whose order of magnitude lies
  within a decade of $10^6$~\cite{Martin:2013nzq}.}
\label{fig:kmiii}
\end{center}
\end{figure}

\begin{figure}
\begin{center}
\includegraphics[width=\onefigw]{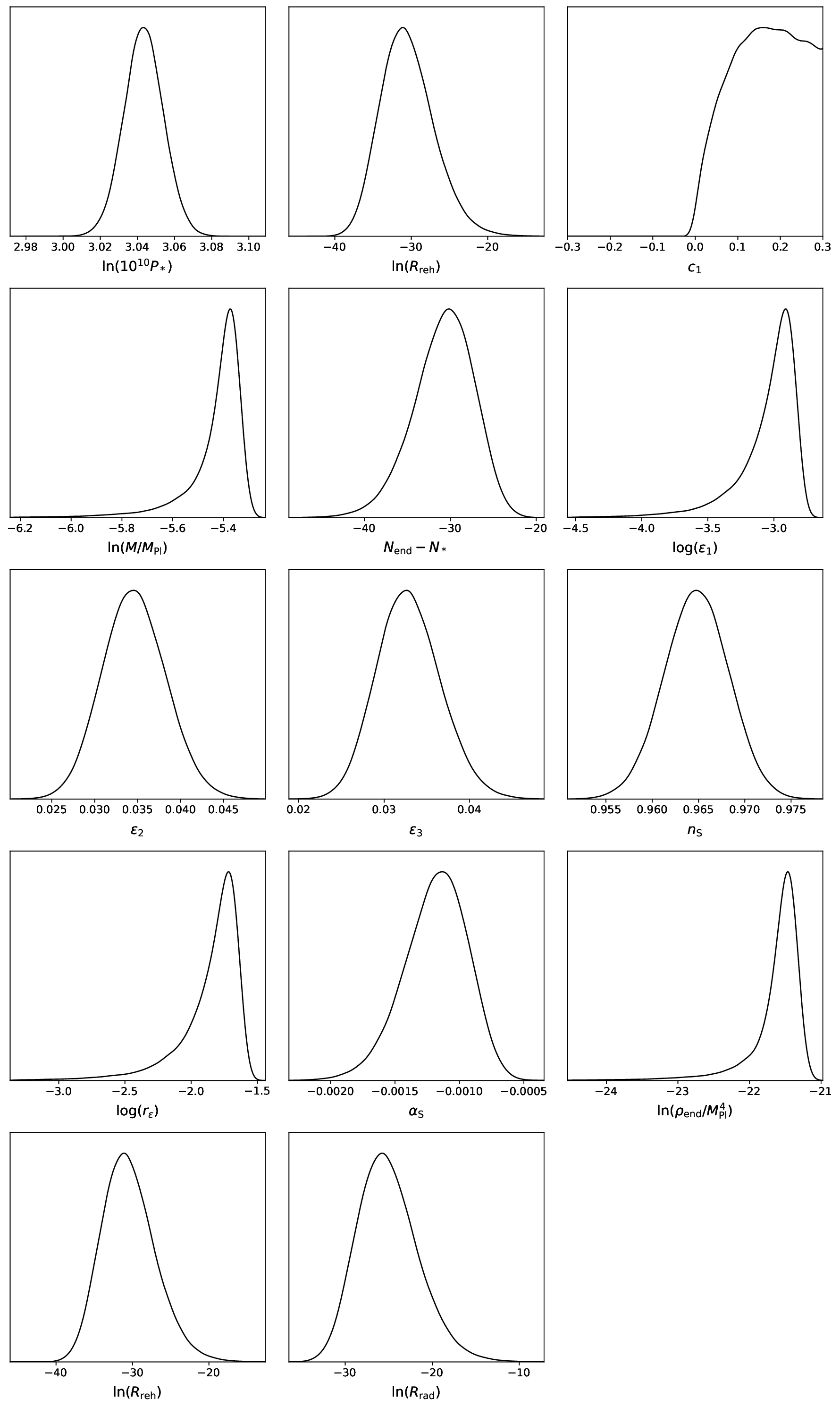}
\caption{One-dimensional posterior distributions for Loop Inflation
  ($\li$). There is a unique model parameter $\cte{1}=\alpha$, a
  small coupling constant setting the amplitude of quantum loop
  corrections and living in a disconnected set
  $\alpha\in[-0.3,-0.1]\cup[0,0.3]$~\cite{Martin:2013nzq}.}
\label{fig:li}
\end{center}
\end{figure}

\begin{figure}
\begin{center}
\includegraphics[width=\onefigw]{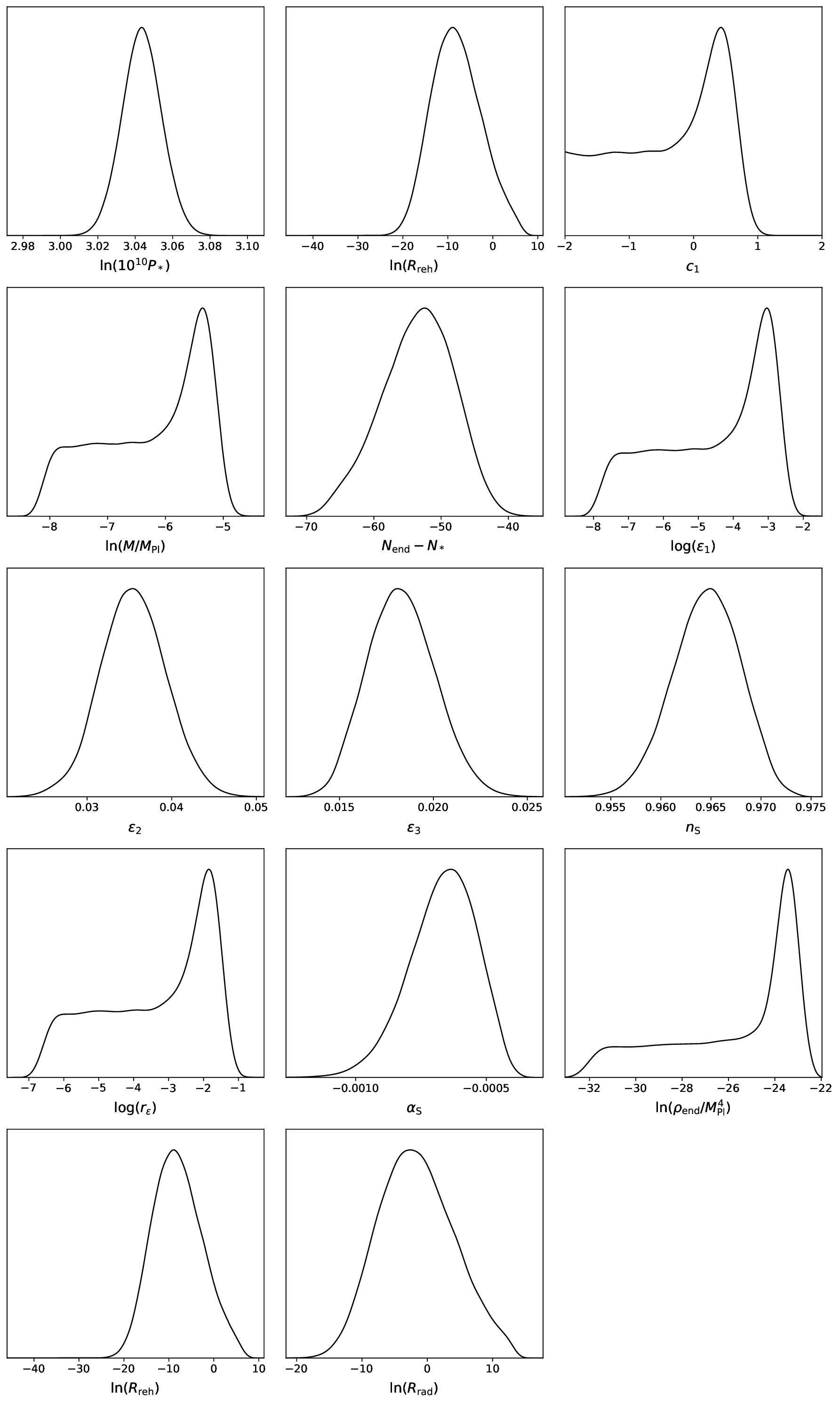}
\caption{One-dimensional posterior distributions for Mutated Hilltop
  Inflation ($\mhi$). There is a unique model parameter
  $\cte{1}=\log(\mu/\Mp)$, where $\mu$ is the inflaton typical vacuum
  expectation value.}
\label{fig:mhi}
\end{center}
\end{figure}

\begin{figure}
\begin{center}
\includegraphics[width=\onefigw]{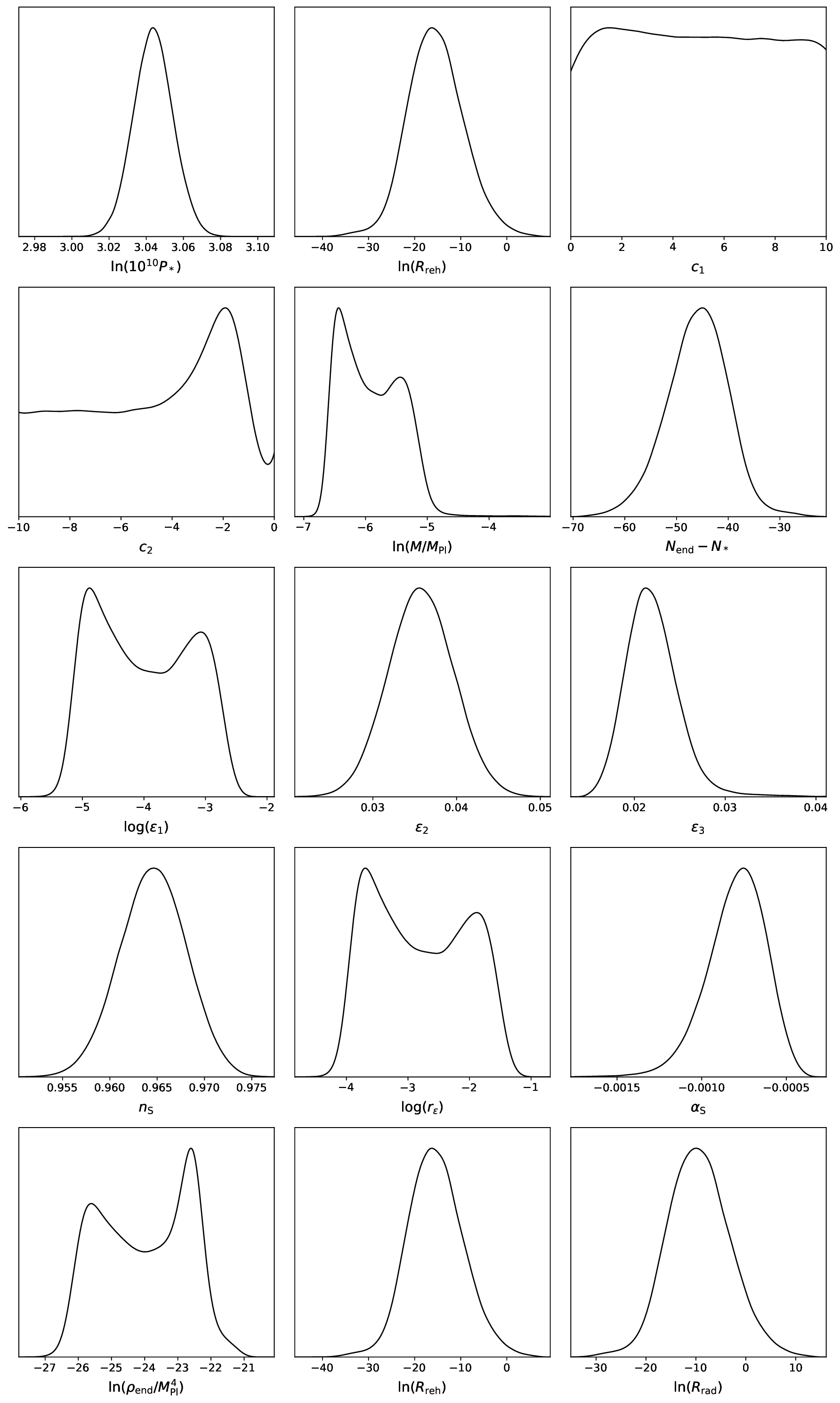}
\caption{One-dimensional posterior distributions for N-Formalism
  Inflation in the regime $a>0$ and $b<0$ ($\nfiTHREEp$, see \Sec{sec:nfiprior}).
  The two model parameters are
  $\cte{1}=a$ and $\cte{2}=b$, and are constrained to be smaller than $10$ in absolute value
  (phenomenological model).}
\label{fig:nfi3p}
\end{center}
\end{figure}

\begin{figure}
\begin{center}
\includegraphics[width=\onefigw]{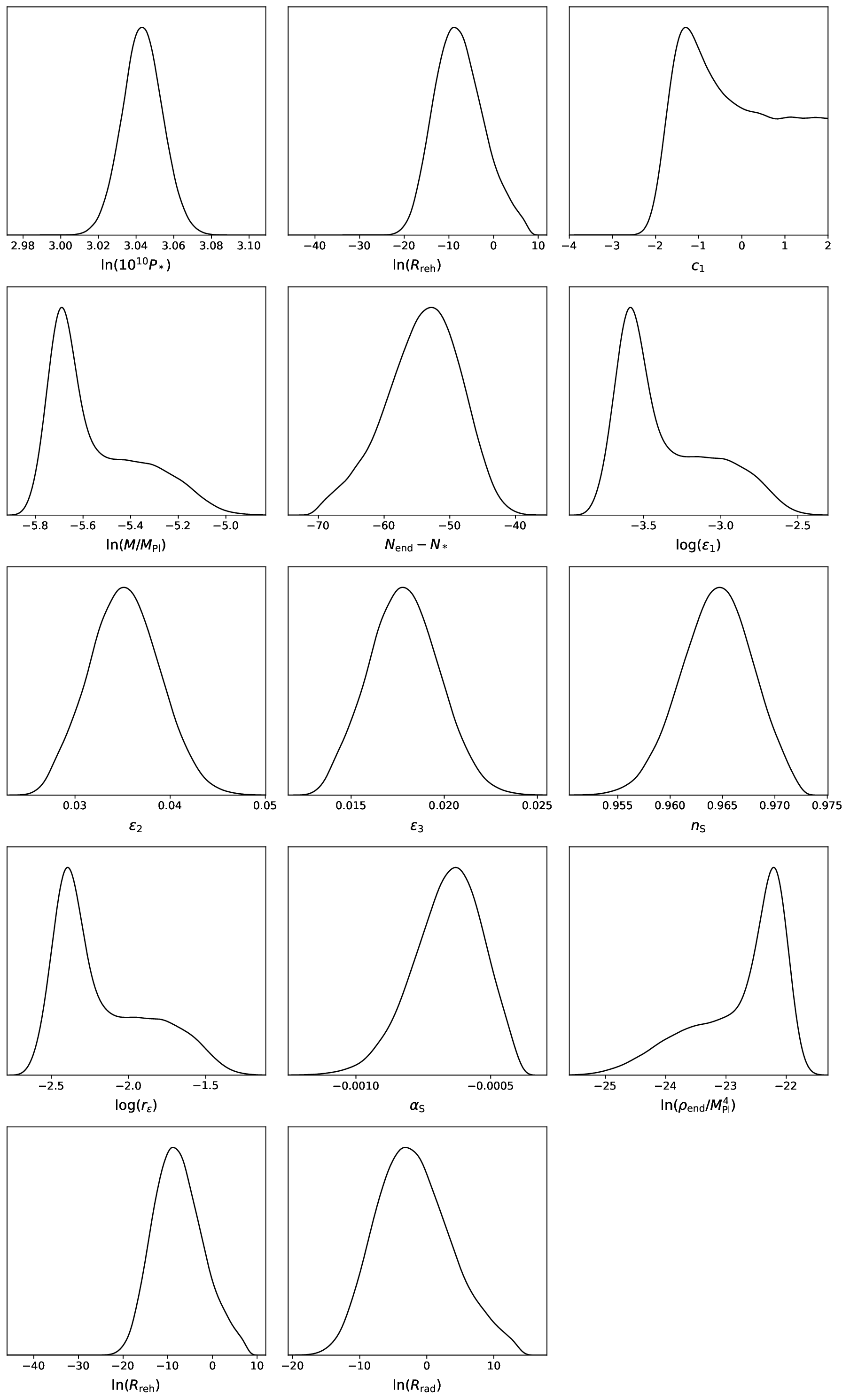}
\caption{One-dimensional posterior distributions for Non-Minimally
  Coupled Large Field Inflation, with a quartic power and occurring at
  decreasing field values ($\nmlfiONEFOUR$). The unique parameter is
  $\cte{1}=\log(\xi)$, where $\xi$ is the non-minimal coupling to gravity.}
\label{fig:nmlfi14}
\end{center}
\end{figure}

\begin{figure}
\begin{center}
\includegraphics[width=\onefigw]{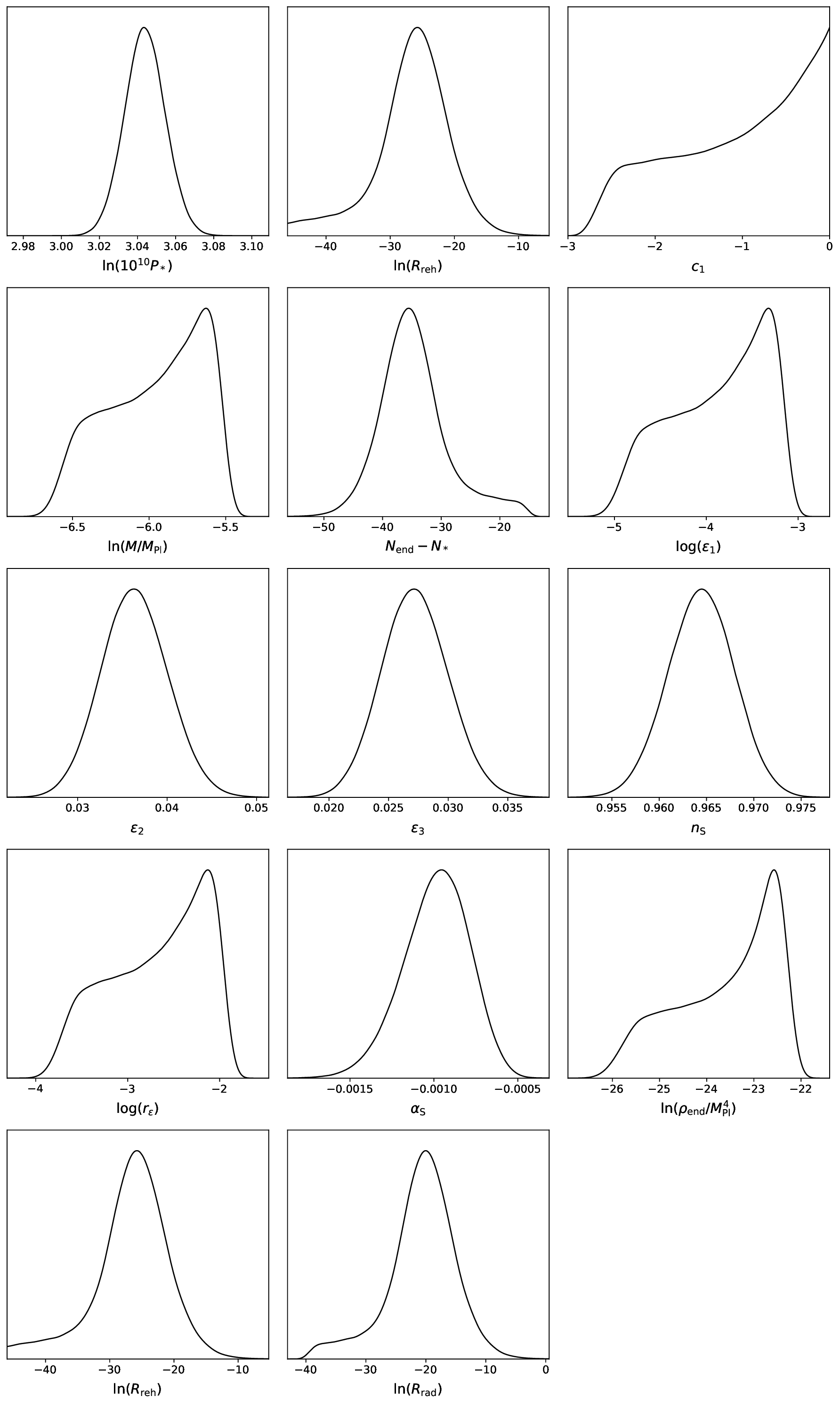}
\caption{One-dimensional posterior distributions for Pure Arctan
  Inflation with small vacuum expectation value ($\pais$). This is a
  phenomenological model having only one parameter
  $\cte{1}=\log(\mu/\Mp)$, $\mu$ is a sub-Planckian vacuum
  expectation value.}
\label{fig:pais}
\end{center}
\end{figure}

\begin{figure}
\begin{center}
\includegraphics[width=\onefigw]{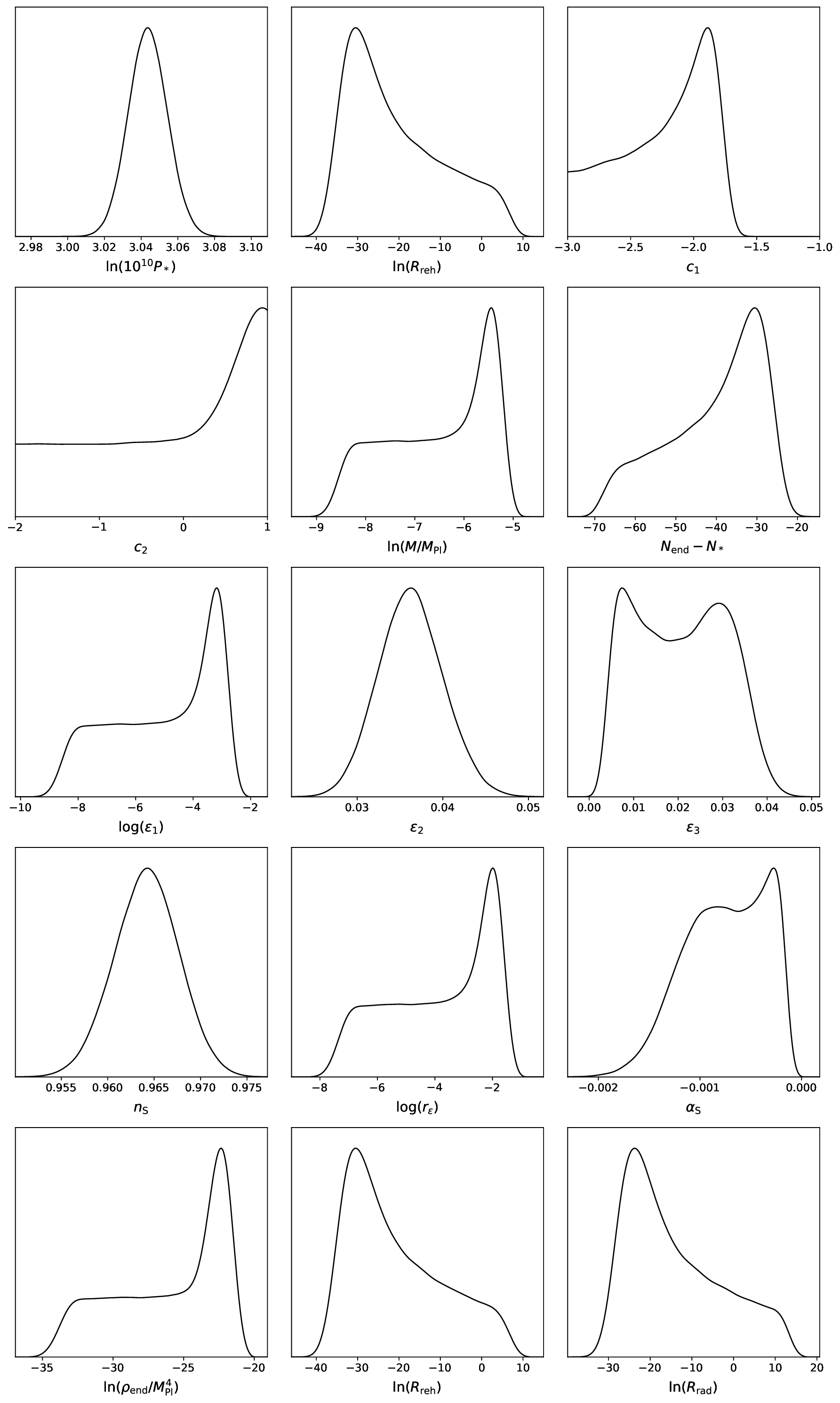}
\caption{One-dimensional posterior distributions for Pseudo Natural
  Inflation with a type 2 prior choice ($\psniftTWO$). This is a
  semi-phenomenological model having two-parameters
  $\cte{1}=\log(\alpha\Mp^2/f^2)$ and
  $\cte{2}=\log(f/\Mp)$~\cite{Martin:2013nzq}. The other versions of
  this model differ only by different motivated choices for the prior
  distributions.}
\label{fig:psnift2}
\end{center}
\end{figure}

\begin{figure}
\begin{center}
\includegraphics[width=\onefigw]{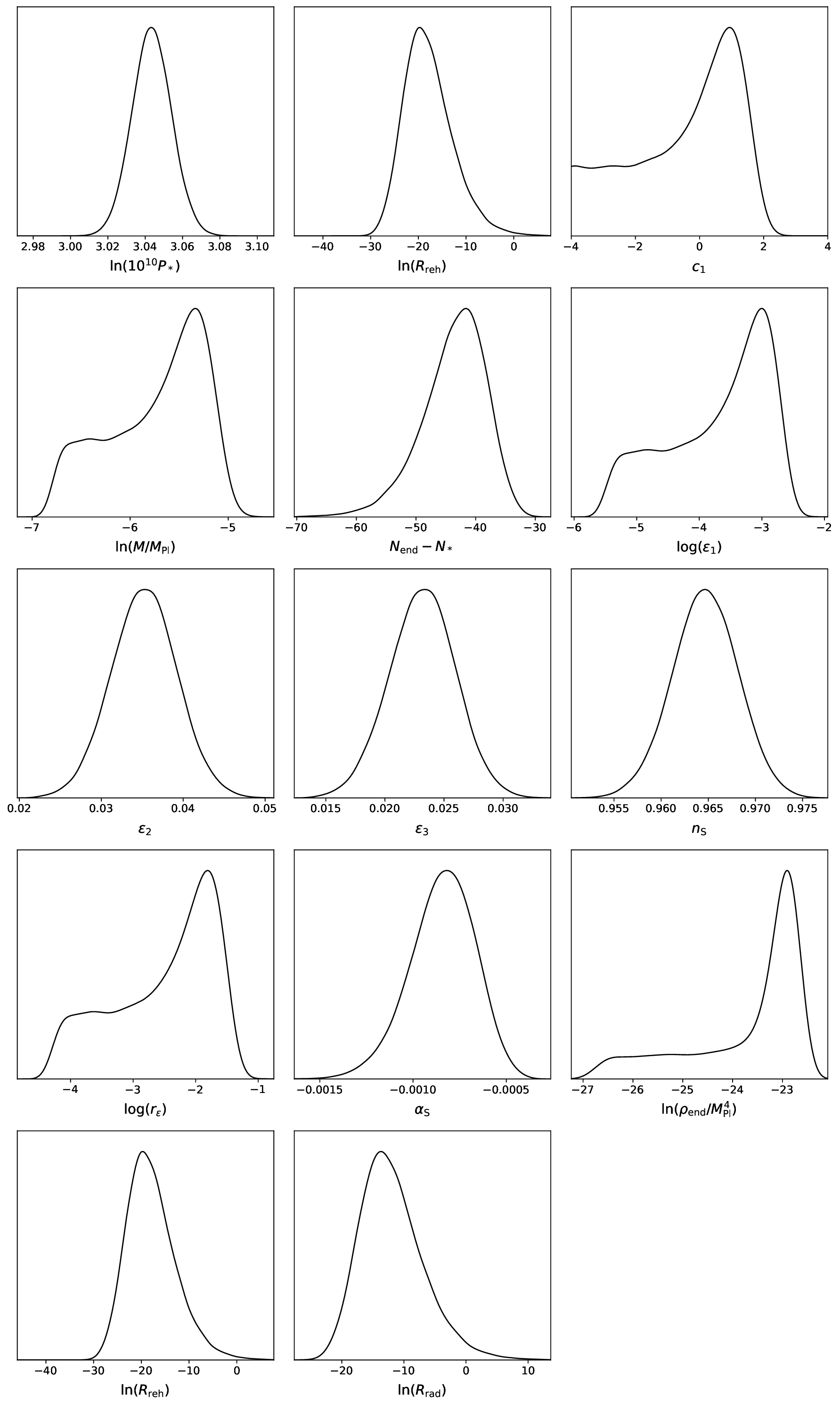}
\caption{One-dimensional posterior distributions for Radion Gauge
  Inflation ($\rgi$). There is only one parameter
  $\cte{1}=\log(\alpha)$, $\alpha$ being of unknown order of
  magnitude for the phenomenological version of this scenario~\cite{Martin:2013nzq}.}
\label{fig:rgi}
\end{center}
\end{figure}

\begin{figure}
\begin{center}
\includegraphics[width=\onefigw]{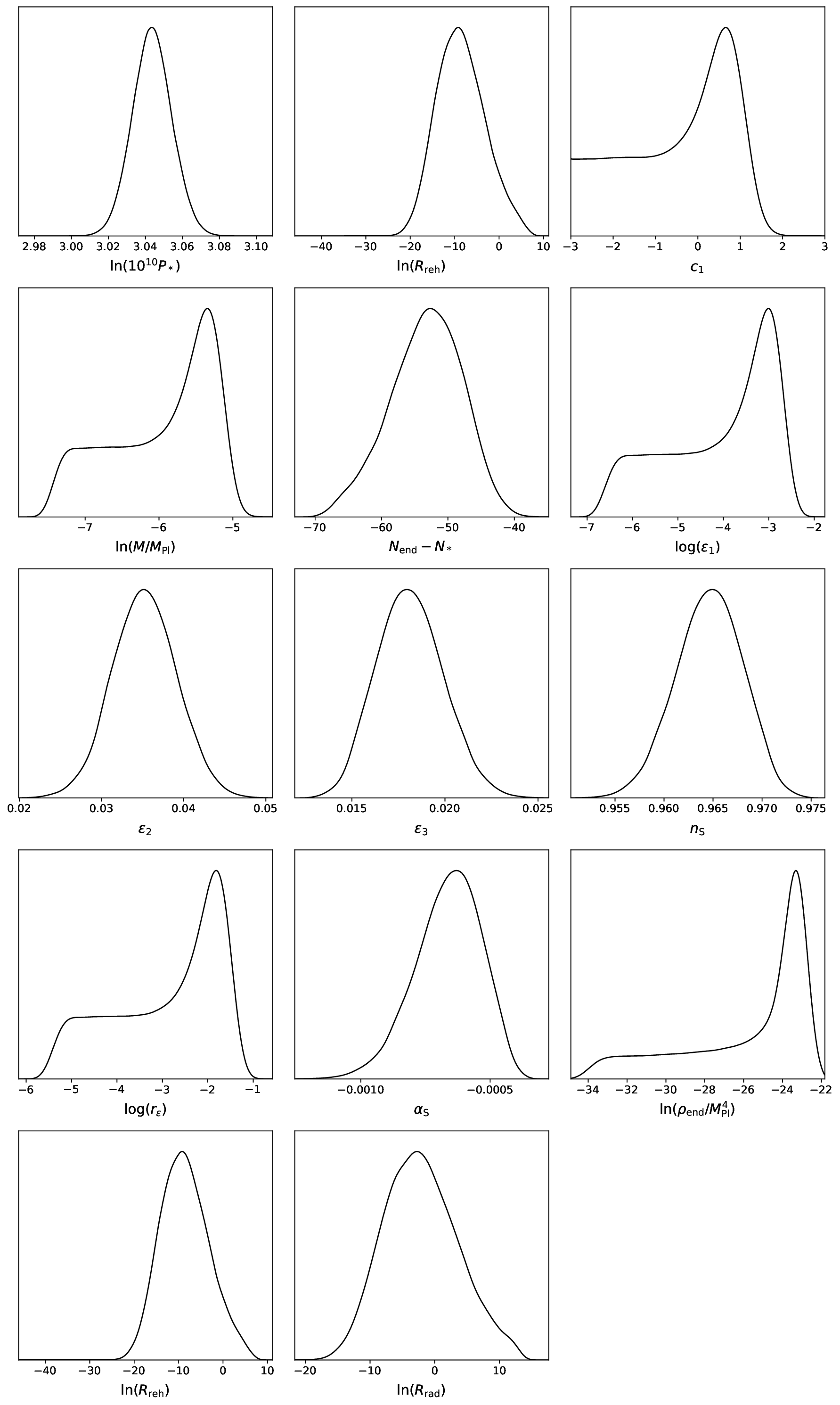}
\caption{One-dimensional posterior distributions for Superconformal
  $\alpha$-Attractor A Inflation ($\sabiONE$). There is one
  parameter $\cte{1}=\log(\alpha)$, $\alpha$ being of unknown order of
  magnitude.}
\label{fig:saai}
\end{center}
\end{figure}

\begin{figure}
\begin{center}
\includegraphics[width=\onefigw]{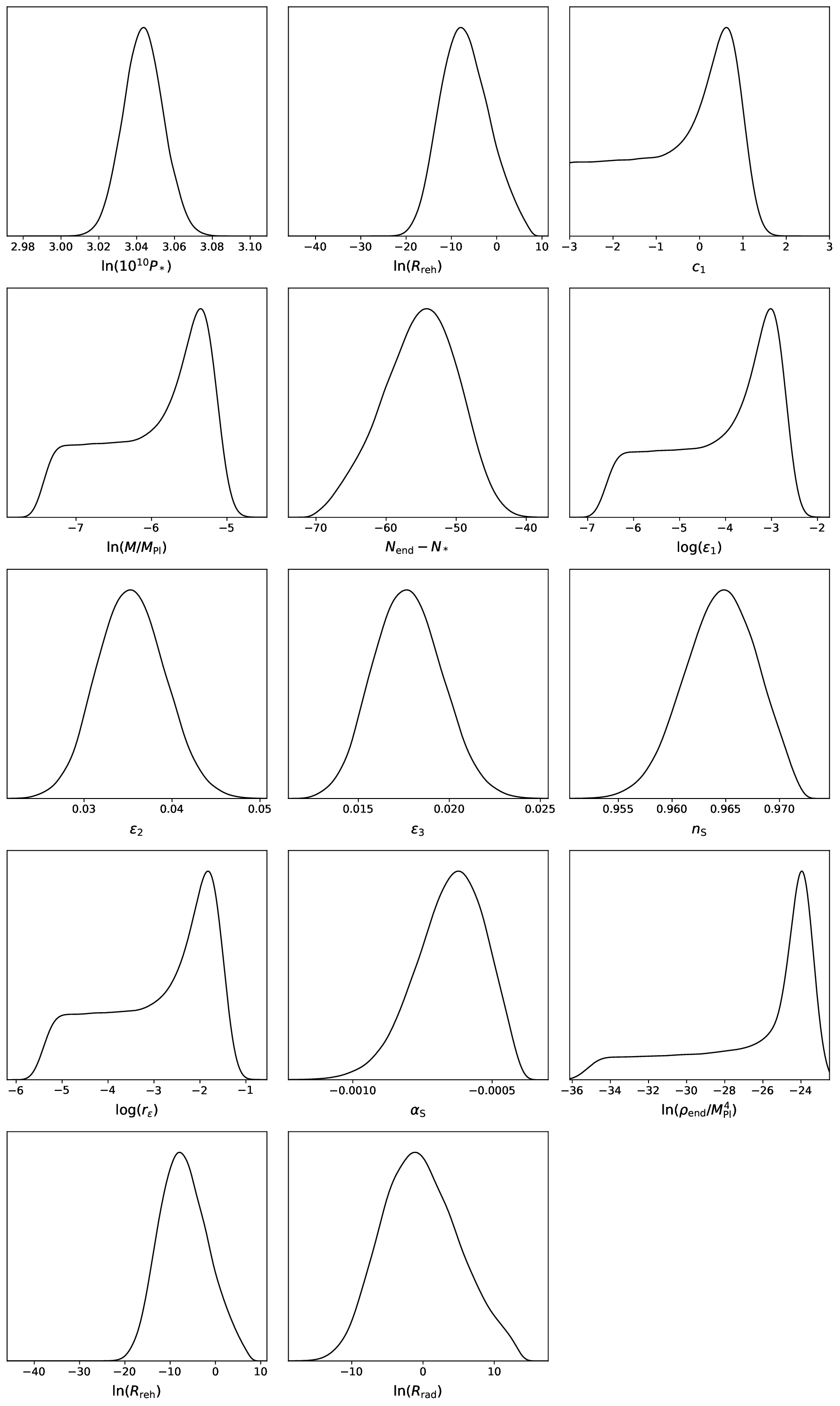}
\caption{One-dimensional posterior distributions for Superconformal
  $\alpha$-Attractor T Inflation with quadratic power ($\satiONE$). There is one
  parameter $\cte{1}=\log(\alpha)$, $\alpha$ being of unknown order of
  magnitude.}
\label{fig:sati1}
\end{center}
\end{figure}

\begin{figure}
\begin{center}
\includegraphics[width=\onefigw]{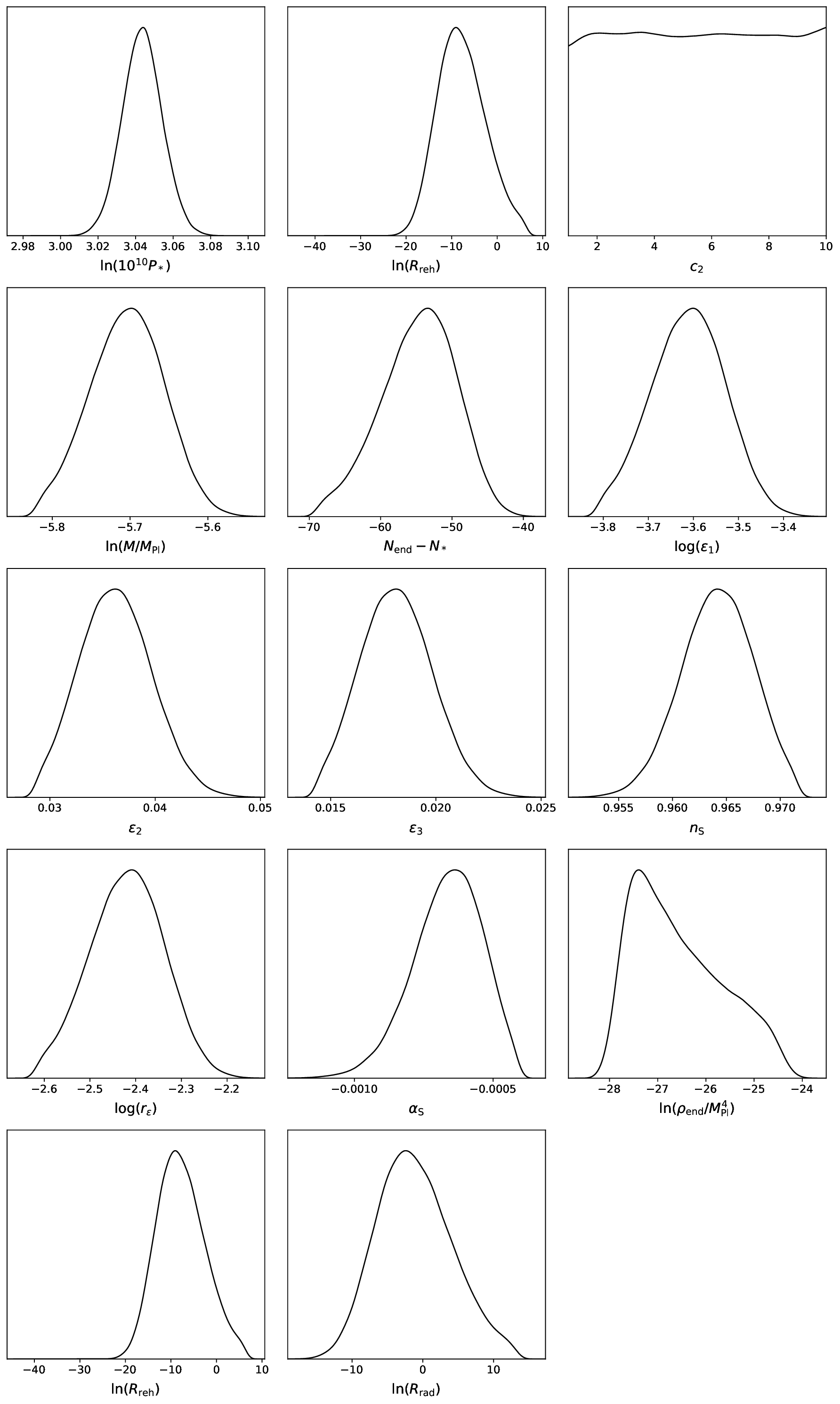}
\caption{One-dimensional posterior distributions for T-Model Inflation
  ($\satione$). There is one parameters $\cte{2}=n$, a power index.}
\label{fig:tmi}
\end{center}
\end{figure}

\begin{figure}
\begin{center}
\includegraphics[width=\onefigw]{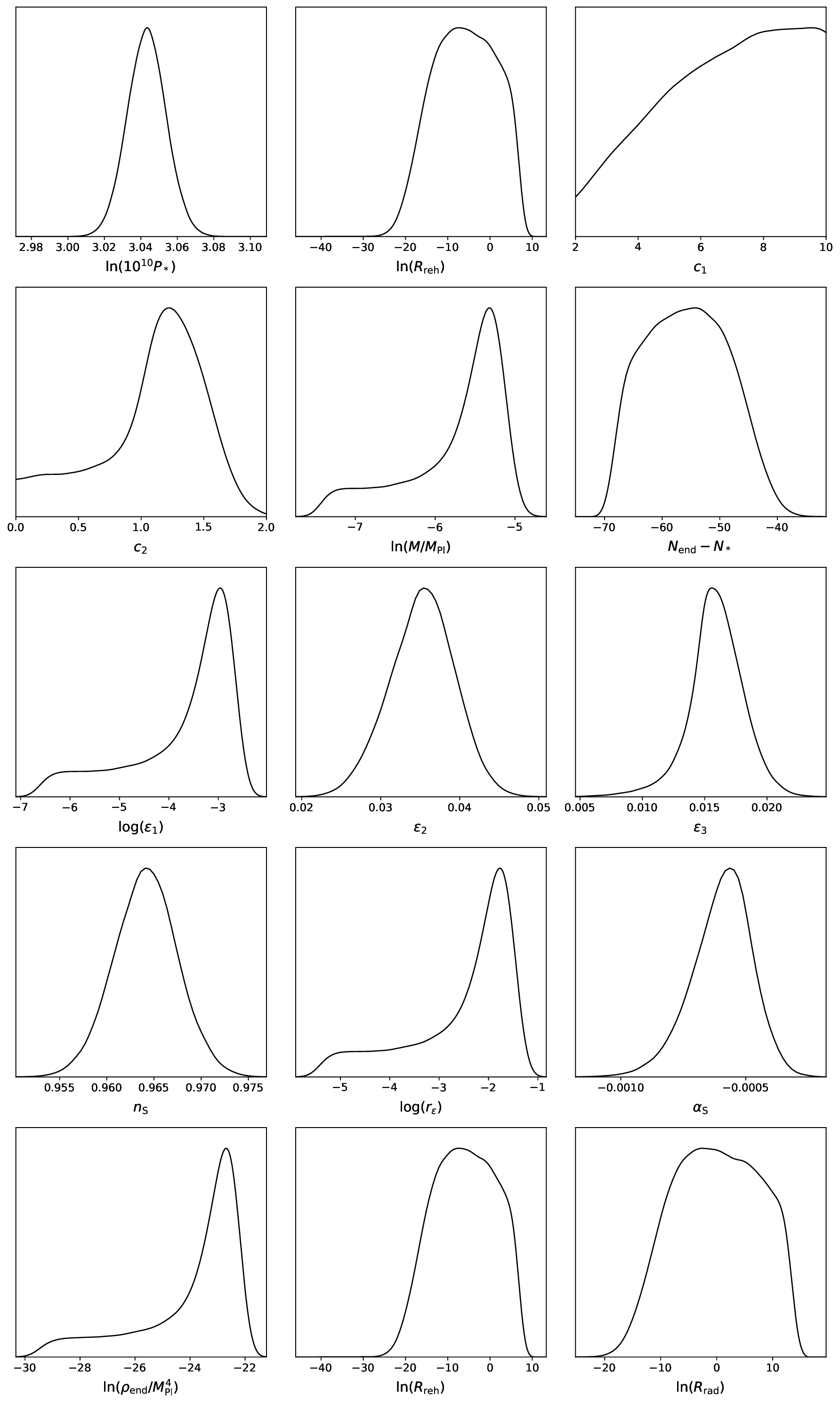}
\caption{One-dimensional posterior distributions for Small Field
  Inflation at large vacuum expectation values ($\sfil$). There are two
  parameters: $\cte{1}=p$, a power index, and $\cte{2}=\log(\mu/\Mp)$, a
  super-Planckian vacuum expectation value.}
\label{fig:sfil}
\end{center}
\end{figure}

\begin{figure}
\begin{center}
\includegraphics[width=\onefigw]{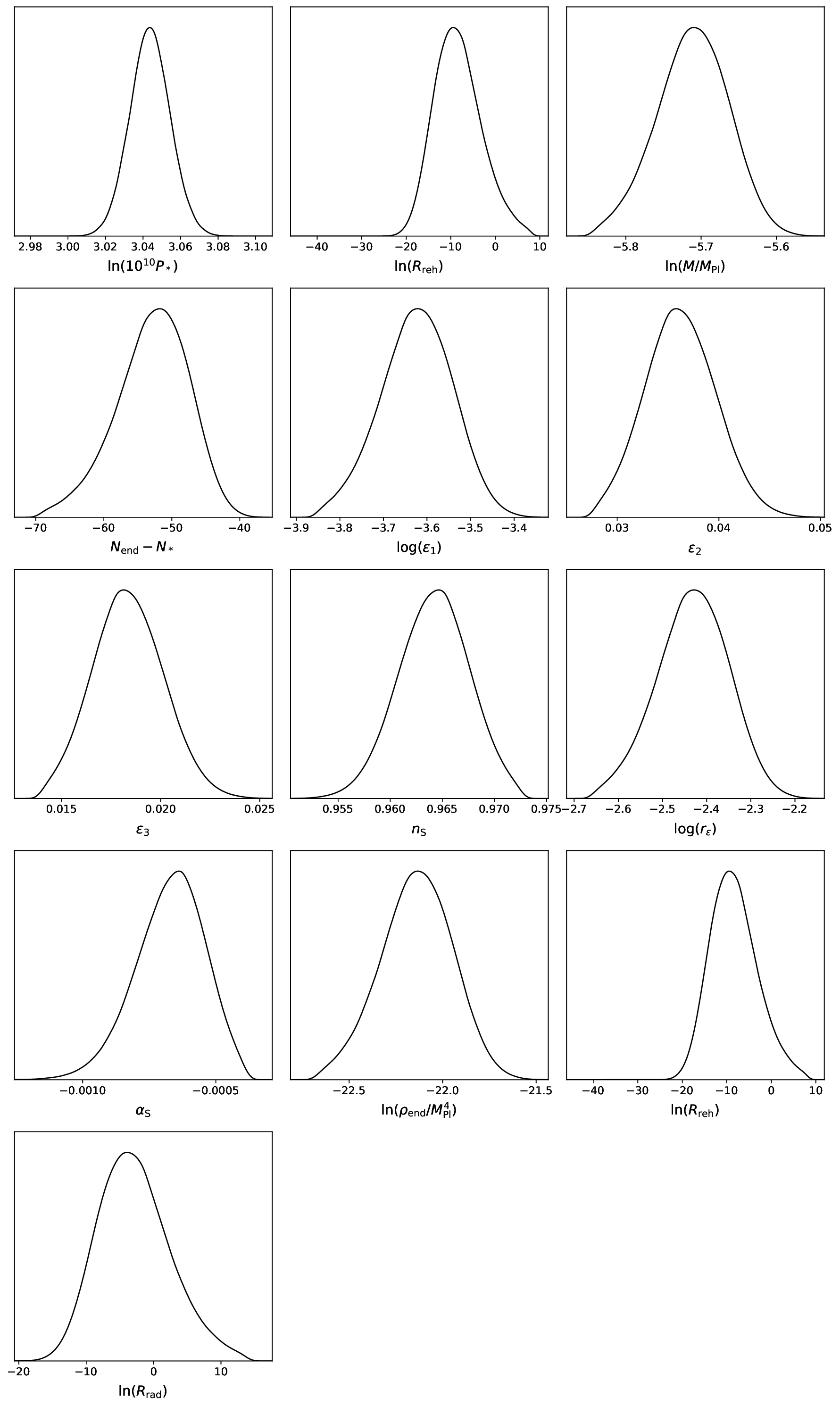}
\caption{One-dimensional posterior distributions for Starobinsky
  Inflation ($\si$). There is no model parameter. As can be checked in
  \Fig{fig:hi}, the data do not yet allow disambiguation between
  $\si$, $\simc$ and $\hi$.}
\label{fig:si}
\end{center}
\end{figure}

\begin{figure}
\begin{center}
\includegraphics[width=\onefigw]{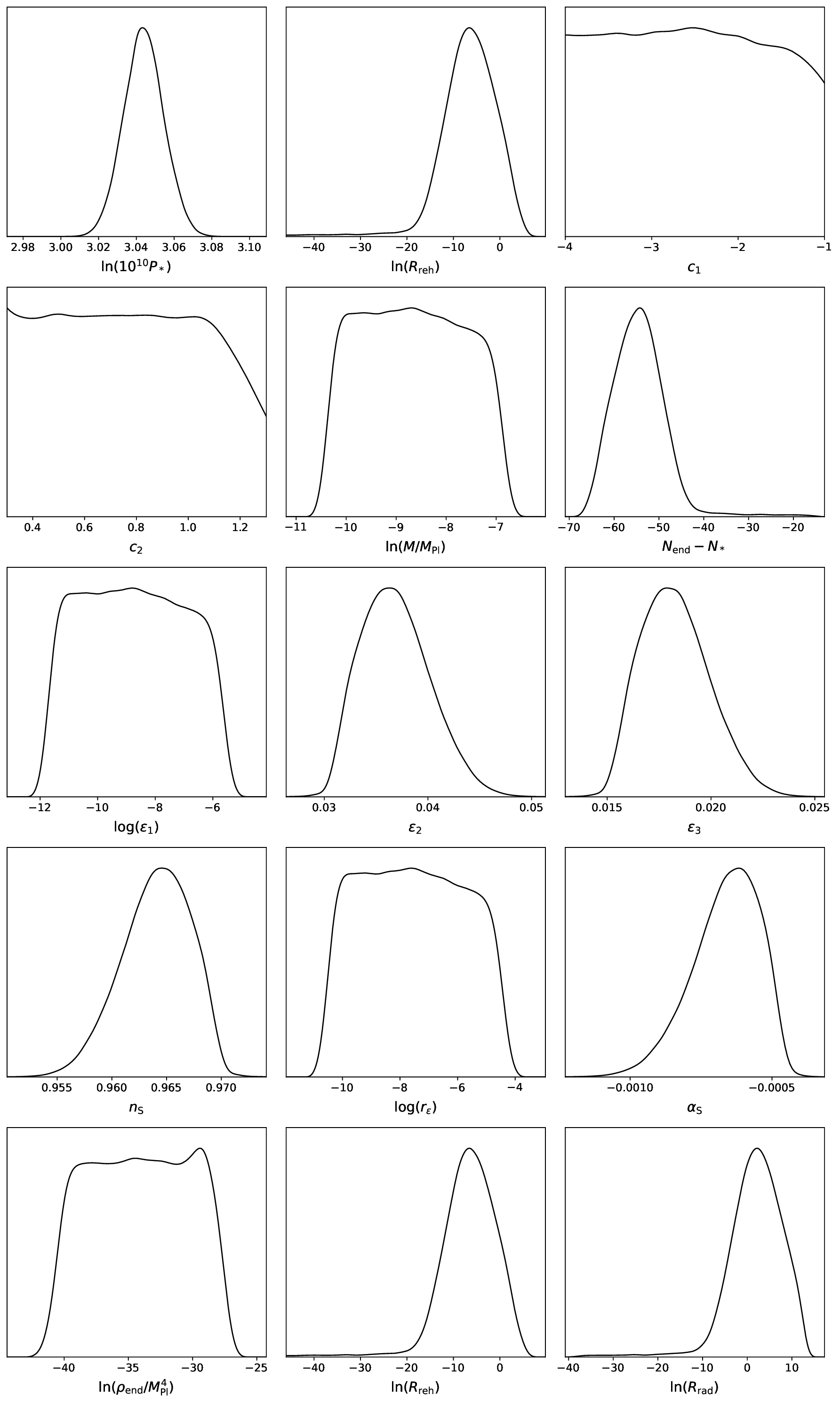}
\caption{One-dimensional posterior distributions for Twisted Inflation
  ($\twiAONE$). This is a two-parameter model with
  $\cte{1}=\log(\phizero/\Mp)$, a vacuum expectation value, and
  $\cte{2}=\log(\phiend/\phizero)$, the field value at which inflation
  ends (in unit of $\phizero$)~\cite{Martin:2013nzq}.}
\label{fig:twiA1}
\end{center}
\end{figure}

\newpage

\bibliographystyle{JHEP}

\bibliography{biblio}

\end{document}